\documentclass[useAMS,usenatbib]{mn2e}
\usepackage{amsmath,fleqn,graphicx,amssymb}
\usepackage{multirow}
\def\df{\textsc{df}}
\def\bolth{\mbox{\boldmath$\theta$}}

\def\Myr{\,\mathrm{Myr}}
\def\kpc{\,\mathrm{kpc}}
\def\mix{\mathrm{mix}}
\def\res{\mathrm{res}}
\def\cut{\mathrm{cut}}
\def\const{\mathrm{const}}
\def\kms{\,\mathrm{km\,s}^{-1}}
\def\masyr{\,\mathrm{mas\,yr}^{-1}}

\def\bolJ{\mbox{\boldmath{$J$}}}
\def\bolx{\mbox{\boldmath{$x$}}}
\def\bolv{\mbox{\boldmath{$v$}}}
\def\vsol{\mbox{\boldmath{$v_\odot$}}}
\def\pc{\,\mathrm{pc}}

\def\d{\mathrm{d}}
\def\Rc{R_\mathrm{c}}

\title[Extending the Hyades]
{Extending the Hyades}

\author[P.~J.~McMillan]{
  Paul~J.~McMillan\thanks{E-mail: p.mcmillan1@physics.ox.ac.uk}\\  
  Rudolf Peierls Centre for Theoretical Physics, 1 Keble Road,
  Oxford, OX1 3NP, UK
}

\begin{document}
\maketitle

\begin{abstract}
  We explore the implications of models of the Hyades moving group in
  which it has a resonant origin, for regions of the Galaxy beyond the
  Solar neighbourhood. We show that while models associated with
  different resonances can produce nearly identical substructure in
  the local velocity distribution, the velocity distribution away from
  the Solar neighbourhood has different properties for different
  models. In particular there is a variation between different models
  of where in Galactocentric radius the observed Hyades signal in velocity
  space is strongest, at a given Galactic azimuth. We note, however,
  that the uncertainties in currently available data, primarily due to
  uncertain distances to stars, hide these signatures rather
  effectively, meaning we are not yet able to determine which
  resonance is the cause of the Hyades.
\end{abstract}


\section{Introduction} \label{sec:intro} It is well established that
the velocity distribution in the Solar neighbourhood is far from
smooth. The existence of ``moving groups'' of stars with similar
velocities in the Solar neighbourhood has been recognised for over a
hundred years \citep[][and references therein]{Ka1905,Eg96}, but it
was generally believed that these groups are composed of stars from
dissolved clusters, which retain similar velocities (at a given point
in the Galaxy) while spreading in space.

Using observations by the \emph{Hipparcos} satellite
\citep{Hipparcos}, \cite{WD98} showed that these moving groups
dominate the distribution of blue ($B-V<0.4$) stars, \emph{and}
contain a substantial proportion of the redder stars in the Solar
Neighbourhood (beyond Parengo's discontinuity, $B-V>0.61$), with the
same moving groups being seen in the velocity distributions of the
different subsamples of Solar Neighbourhood stars.  This suggested
that a single moving group contains stars of many different ages, a
result confirmed by \cite{Faea05}, who added to the Hipparcos data by
determining radial velocities for a sample of Hipparcos stars,
allowing them to directly associate a given star with a given moving
group. Comparison of these stars to isochrones in the
Hertzsprung-Russell diagram then revealed the wide range of ages in
each moving group.
This is a very strong argument that moving groups must have dynamical
origins. 
More recently, \cite{Poea11} used chemical ``tagging'' to show that a large fraction of the stars in the Hyades moving group can not have originated from the Hyades cluster.

The dynamical processes that shape the velocity distribution of the
Solar neighbourhood in to moving groups is still unclear. \cite{Ka91}
suggested that the Hyades and Sirius moving groups could be associated
with an outer Lindblad resonance with the Galactic bar,\footnote{In
  the discussion section of these conference proceedings,
  \citeauthor{Ka91} (questioned by Lindblad) concedes that the same
  structure in the velocity space at a single point could be produced
  by an inner Lindblad resonance with some unknown perturbation, but
  notes ``I could tell the difference if I knew the stellar velocity
  distribution in a large enough region around the sun. But I will
  never know that.''  } while Dehnen (\citeyear{WD99:Bar,WD00:OLR})
used test-particle simulations to show how this resonance with the bar
could instead produce the Hercules moving group (which is at
velocities further from the circular velocity than the Hyades or
Sirius groups, and was not fully recognised when \citeauthor{Ka91} was
writing). \cite{Fu01} came to a similar conclusion, and more recently
\cite{Miea10} argued that the moving groups at velocities near to the
circular velocity may also be due to the influence of the bar's OLR.
Other work, beginning with \cite*{DSea04}, has looked at the possible
influence of spiral structure, and shown that it is capable of
producing a variety of substructures in velocity space, including
those observed in the Solar Neighbourhood
\citep[e.g][]{QuMi05,Anea11,Poea11}. Other studies have sought to explain at
least some of the moving groups as the response of the disc to the
accretion of a satellite galaxy \citep{Quea09,Miea09} or as the debris
of an accreted satellite \citep[][though this is only plausible for
the higher velocity moving groups]{Heea06}.  All of these works used
some form of test particle integration or N-body simulation to model
the effects of these perturbations and their resonances.

More recent work has used angle-action coordinates to show that the
Hyades moving group\footnote{ Throughout this paper we will refer to
  dynamical structure associated with the Hyades moving group as ``the
  Hyades'' in the interests of brevity. This should not be interpreted
  as meaning the Hyades star cluster, which is part of the moving
  group, but certainly doesn't extend beyond the Solar neighbourhood!}
is associated with trapping at a Lindblad resonance
\citep{Se10}. However, it is not clear which resonance is responsible,
as very different resonances are associated with very similar
structures in the velocity distribution of the Solar neighbourhood
\citep[][henceforth M11]{PJM11:Hyades}. It is this approach that we
pursue in this study, because it allows us to create easily tuneable
models which make no assumptions about the nature of the perturber.

Recently it has become feasible to examine the velocity distribution
of stars some distance away from the Solar neighbourhood. Notably
\cite[][henceforth A12]{Anea12} used observations by the RAdial Velocity Experiment
\citep[RAVE: ][]{RAVE1_short} to investigate the velocity distribution
in fields up to $\sim1.5\kpc$ from the Sun, in the anti-rotation
direction. In a spatial bin located around the Sun's Galactocentric
radius, but at least $700\pc$ away in the Galactic azimuthal
direction, they found a structure in velocity space at a similar
position to that occupied by the Hyades in the Solar neighbourhood, as
well as other overdensities at other points in velocity space. In
other spatial bins which lie at similar Galactic azimuths, but at
least $300\pc$ inside or outside the Solar radius, they find no such
structure at the velocity of the Hyades, but still find other
substructure in velocity space.

In this paper we extend the models used by M11 to explore the
implications of the expected velocity distributions away from the
Solar neighbourhood. In Section~\ref{sec:aa} we discuss the nature of
the resonances considered and the use of angle-action coordinates in
describing them. In Section~\ref{sec:num} we give the numerical
details of the Hyades models that we test, before describing their
properties away from the Solar neighbourhood in
Section~\ref{sec:beyond}.

\section{Angle-action coordinates and Lindblad resonances}
\label{sec:aa}

Three actions $J_i$ and three conjugate angle coordinates $\theta_i$
provide exceptionally convenient coordinates for objects orbiting in a
stationary or slowly evolving gravitational potential (and natural coordinates for perturbation theory). 
The actions are conserved quantities and the
angles increase linearly with time, $\theta_i(t) =
\theta_i(0)+\Omega_i(\bolJ)t$, where $\Omega_i$ is a frequency. 
This means
that $\bolJ$ can be thought of as labeling an orbit, and $\bolth$ as
describing a point on that orbit. The usual phase space coordinates
$\bolx,\bolv$ are $2\pi$-periodic in each angle coordinate $\theta_i$.
For a phase-mixed distribution function (\df) the distribution of
stars is uniform in angle and therefore the \df \ $f=f(\bolJ)$.

The conversion between the coordinate systems 
$\bolth,\bolJ$ and $\bolx,\bolv$ is only known
analytically for a few types of gravitational potential.
However we can use the ``torus fitting'' method \citep[e.g.][and
references therein]{PJMJJB08} to approximately find this conversion in
an axisymmetric Galactic potential. 
This is done numerically, on an 
orbit-by-orbit basis -- i.e. for a given $\bolJ$ the torus fitting 
algorithm finds an expression giving the 
values of $\bolx,\bolv$ for \emph{any} $\bolth$, but a new
torus fit has to be performed to know this relationship 
($\bolx,\bolv$ for any $\bolth$) for a different $\bolJ$.

In this study we work under the approximation that we can ignore the
influence of the non-axisymmetric structure of the disc \emph{except}
for their trapping stars around resonances. This means that we can use
the angles and actions calculated in the underlying axisymmetric
potential, but that the \df\ is reshaped by resonant trapping, and is
not simply $f(\bolJ)$ 
but also depends on the angles. 
This reshaping produces the Hyades.

\cite{Se10} explored the distribution of the Solar neighbourhood stars
in angle-action coordinates (under similar approximations), using data
from the Geneva Copenhagen Survey \citep*[GCS, ][]{GCS09}. The
distribution of stars in action showed a clear indication that the
Hyades moving group was associated with a resonance between the radial
and azimuthal orbital frequencies of the stars ($\Omega_r$ and
$\Omega_\phi$, both of which are functions of $\bolJ$) associated with
a perturbation whose pattern speed $\Omega_p$ satisfies
\begin{equation}\label{eq:res_om}
  l\Omega_r(\bolJ)+m\Omega_\phi(\bolJ) = m\Omega_p.
\end{equation}
$l$ defines the type of resonance, with $l=+1$ for an outer Lindblad
resonance (OLR), $l=-1$ for an inner Lindblad resonance (ILR), $l=0$
at corotation, and $l=\pm \frac{1}{2}$ for ultra-harmonic
resonances. (Throughout this paper we will refer to a Lindblad
resonance for any $m$ as an $m$:1 ILR or OLR.)\footnote{
Note that in these models there is nothing to distinguish the effects of a Lindblad resonance with an $m=4$ pattern from the effects of an ultra-harmonic resonance with an $m=2$ pattern.
}


The condition given by equation~\ref{eq:res_om} is satisfied along
planes in action space, and to a good approximation these can be
thought of as a relationship between the radial action $J_r$ and the
azimuthal action $J_\phi$ (which is the angular momentum about the
symmetry axis of an axisymmetric potential), with the third action
$J_z$ having a negligible effect on the relevant frequencies. It
therefore makes sense to think about resonant lines in the
$J_r$-$J_\phi$ plane and ignore changes with $J_z$. For the values of
$J_\phi$ that are relevant for disc stars near the Solar
neighbourhood, changing $\Omega_p$ shifts the zero intercepts of these
lines (the value of $J_\phi$ for $J_r=0$, i.e. the angular momentum of
the relevant circular orbit), but has a negligible effect of the slope
of the line in the $J_\phi-J_r$ plane.

\begin{figure}
  \centerline{
    \includegraphics[width=\hsize]{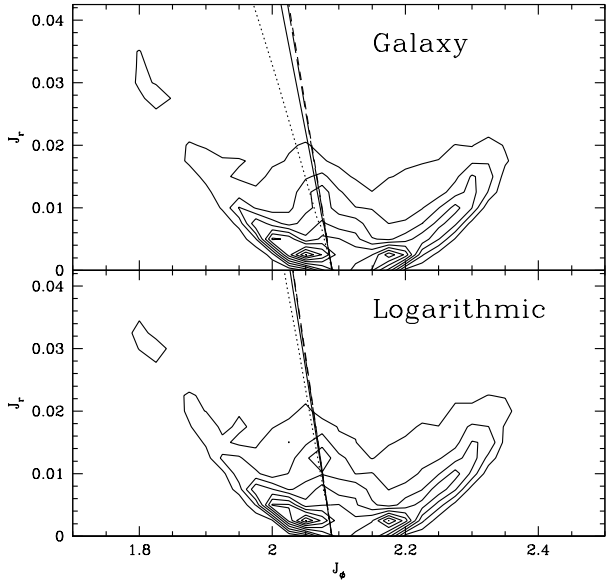}
  }
  \caption{ Figures showing the lines in action space associated with
    an 2:1 ILR (\emph{dotted}), 4:1 ILR (\emph{solid}), 2:1 OLR
    (\emph{short-dashed}), and a 4:1 OLR (\emph{long-dashed}) in the
    Galactic potential used in this study (upper) and a logarithmic
    potential (lower), for perturber pattern speeds $\Omega_p$ chosen
    such that the resonance includes the point $J_r=0, J_\phi=2.09$
    (note that in all cases except the 2:1 ILR this is the resonant
    line used in the Hyades model). In each  figure the  
    distribution in action of stars observed by the GCS is shown as a
    contour plot, both to give a sense of scale and to show the Hyades
    in action space (the broad overdensity near the resonance lines at
    higher $J_r$). Changed $\Omega_p$ for any of the resonances has
    the effect of moving the resonance line in $J_\phi$ while having a
    negligible effect on the gradient of the line.  This figure
    illustrates the fact that the resonant lines in action space are
    all very similar to one another, and can change significantly with
    a change of assumed potential.
\label{fig:RL}
}
\end{figure}
The resonant lines for different ILRs or OLRs that go through the
Solar neighbourhood have very similar slopes, and the stars associated
with the Hyades clearly lie around one of these, but it is impossible
to tell from the distribution in $\bolJ$ which values of $l$ and $m$
(and thus $\Omega_p$) define the resonance (see
Figure~\ref{fig:RL}). It is also worth noting that the slopes of the
lines in $\bolJ$ associated with the various resonances can be
seriously affected by relatively minor changes in the assumed
potential. The two panels of Figure~\ref{fig:RL} illustrate this by
showing the difference (most notably for the 2:1 ILR resonance)
between the lines assuming a logarithmic potential (and therefore a
flat rotation curve) and assuming a typical Galactic potential
\citep[taken from][]{PJM11:mass} which has a circular speed which
varies by $\sim4\kms$ in the $2\kpc$ either side of the Solar
radius. It may at some point be possible to use these differences to
apply observations of resonant structures to accurately determine the
shape of the Galactic potential \citep{JJB05}, but that is well beyond
the scope of this study.

It is also clear that the stars associated with the Hyades are not
uniformly distributed in angle -- most clearly because they are all
moving radially outwards in the disk whereas, if they were uniform in
angle, there would be as many stars moving out as in. This can also be
explained in terms of resonant trapping \citep[e.g.][section
3.7.2]{GDII}.  If we define $\tilde{\theta}_\phi =
\theta_\phi-\Omega_p\,t$ (effectively the $\phi$ angle coordinate in
the frame rotating with pattern speed $\Omega_p$), then it is clear
from eq.~\ref{eq:res_om} that for orbits near resonance, the value of
$ l\theta_r+m\tilde{\theta}_\phi$ will evolve slowly -- this is known
as the ``slow angle'' -- and the dynamical effect of the
non-axisymmetric perturbation is dominated by forces that affect only
the slow angle (as other forces rapidly average to zero). Resonantly
trapped orbits are ones in which the slow angle librates about a fixed
point, with an amplitude that can be of order unity \citep{GDII}.

Because of this effect, \cite{Se10} pointed out that one should also
expect to observe an overdensity in angle for any set of stars trapped
at resonance about the line in angle space
\begin{equation}\label{eq:res_th}
  l\theta_r+m\theta_\phi = \const ,
\end{equation}
where $l$ and $m$ take the same values as in
eq.~\ref{eq:res_om}.
\citeauthor{Se10} argued that the Hyades, as seen by the GCS, was
associated with an overdensity about one of these lines with $l=-1$
and $m=2,3$ or $4$, which would be a clear indication that it is a inner
Lindblad (or ultra-harmonic) resonance.

However, M11 showed that the selection effects associated with
sampling a finite volume (such as the Solar neighbourhood as observed
by the GCS) play a complicated role in shaping the observed
distribution in angle space. This means it is too difficult to
disentangle the true signal associated with some distribution in
$\bolth$ described by eq.~\ref{eq:res_th} (from which one might
determine the values $l$ \& $m$) from the signal associated with the
distribution in $\bolJ$ combined with the selection effects. This
conclusion was supported by \cite*{HaSePr11} who investigated the
Hyades in the Solar neighbourhood (within $200\pc$) as observed by
RAVE and the Sloan Digital Sky Survey \citep{SDSS7_short}.

None the less, it is clear that the different distributions in angle
associated with eq.~\ref{eq:res_th} are the key to discriminating
between the different resonances that may be associated with the
Hyades, as the resonances are associated with nearly identical areas
in action space. It is, however, necessary to look beyond the
immediate Solar neighbourhood.

\section{Numerical details} \label{sec:num}

\begin{table}
  \caption{Parameters of the phase mixed \df, $f_{\mix}$.}\label{tab:df}
  \begin{center}
    \begin{tabular}{l|ccccc}
      Disc & $R_\d$ & $\sigma_{r0}$ & $\sigma_{z0}$ &
      $L_0$ & $q$\\
      & $(\hbox{kpc})$ & $(\kms)$  & $(\kms)$  & $(\kpc\kms)$ & \\
      \hline
      Thin & 3.0 & 27 & 20 & 10 & 0.45 \\
      Thick& 3.5 & 48 & 44 & 10 & 0.45 \\
    \end{tabular}
  \end{center}
\end{table}

\begin{table*}
  \caption{Parameters of the resonant component of the \df, $f_{\res}$. 
    These are the parameters used in eqs.~\ref{eq:alpha},
    \ref{eq:beta} and \ref{eq:res_theta}. 
    $\Omega_p$ and $R_{\mathrm{corotation}}$ are the
    pattern speed and corotation radius of the perturbation modelled in
    each case. 
    Actions are in units of $\kpc^2\Myr^{-1}$.}\label{tab:dfres}
  \begin{center}
    \begin{tabular}{l|ccccccccc}
      Resonance & $\Omega_p (\Myr^{-1})$ & $R_{\mathrm{corotation}} (\kpc)$ & $J_{\phi,res}$ & 
      $\Delta_{J,res}$ & $B$ & $J_{R,cut}$ & $\Delta_{R,cut}$ & 
      $C_\theta$ & $\Delta_{\theta,res}$\\
      \hline
      OLR 2:1 & 0.058 & 4.5 & $2.09-3\,J_R/2$ & 0.05 & 0.3 & 0.004 & 0.001 & -1.7 & 0.3\\
      OLR 4:1 & 0.051 & 6.0 & $2.09-8\,J_R/5$ & 0.05 & 0.3 & 0.004 & 0.001 & -2.0 & 0.3\\
      ILR 2:1 & 0.028 & 28 & $2.11-3\,J_R$   & 0.05 & 0.3 & 0.004 & 0.001 & 1.3 & 0.3\\
      ILR 4:1 & 0.036 & 13 & $2.09-9\,J_R/5$ & 0.05 & 0.3 & 0.004 & 0.001 & 1.0 & 0.3\\
    \end{tabular}
  \end{center}
\end{table*}

In all cases we use the ``convenient'' model Galactic potential given
by \cite{PJM11:mass}. This model consists of a bulge component, thin
and thick exponential discs, and a \cite*{NFW96} halo.  This sets the
solar radius $R_0=8.5\kpc$ and the circular velocity at the Sun (the
local standard of rest) $v_0=244.5\kms$. When we need to transform
observational data into Galactocentric velocity measurements we assume
that the velocity of the Sun with respect to the local standard of
rest is the best-fitting value found by \cite*{SBD10}
\begin{equation}\label{eq:vsol}
  \vsol  =   (U_\odot,V_\odot,W_\odot) =  (11.1,12.24,7.25)\kms ,
\end{equation}
with $U$ being the velocity towards the Galactic Centre, $V$ being the
velocity in the direction of Galactic rotation, and $W$ the velocity
towards the north Galactic pole.

We use the torus-fitting method to find the values for $\bolx,\bolv$
for any values of $\bolth,\bolJ$.  All actions are quoted in units of
$\kpc^2\,\Myr^{-1}$.\footnote{N.B.  $1\kpc^2\,\Myr^{-1} \approx 978
  \kms\kpc$. The angular momentum of a circular orbit at the Solar
  radius is $\sim 2.13\kpc^2\,\Myr^{-1} = 2080 \kms\kpc$. } The
zero-points for the angles can be defined arbitrarily (provide the
same convention is applied for all orbits), and for clarity we follow
the conventions used by M11, so each component of $\theta_i$ lies in
the range $[-\pi,\pi]$, we define the zero point of $\theta_\phi$ such
that at apocentre $\theta_\phi=\phi$.  and we take $\theta_r=0$ at
apocentre, and therefore $\theta_r=\pm\pi$ at pericentre. The
Galactocentric coordinates are aligned such that the Sun is at a
position in real space with Galactocentric coordinate $\phi=0$.  Note
that for small $J_r,J_z$, where it is appropriate to use the epicycle
approximation, the value of $\theta_\phi$ corresponds to the position
(in $\phi$) of the guiding centre.

The dynamical models we use have two elements -- a phase mixed
component with a \df\ $f_\mix(\bolJ)$, and a component that is
associated with the resonance which produces the Hyades with \df\
$f_{\res}(\bolth,\bolJ)$. Both are based around the
``pseudo-isothermal'' \df s which have been used in a number of recent
papers to describe disc dynamics in the Milky Way
\cite[][M11]{JJB10,JJBPJM11:dyn,JJB12:dfs}. These are of the form
\begin{equation} \label{eq:totalDF}
  f(J_r,J_\phi,J_z)=f_{\sigma_r}(J_r,J_\phi)\times \frac{\nu_z}
  {2\pi\sigma_z^2}\,\mathrm{e}^{-\nu_z J_z/\sigma_z^2},
\end{equation}
where
\begin{equation} \label{eq:planeDF}
  f_{\sigma_r}(J_r,J_\phi)\equiv\frac{\Omega_\mathrm{c}\Sigma}{\pi\sigma_r^2\kappa}\bigg|_{\Rc}
  [1+\tanh(J_\phi/L_0)]\mathrm{e}^{-\kappa J_r/\sigma_r^2}.
\end{equation}
Here $\kappa(J_\phi)$ and $\nu(J_\phi)$ are the radial and vertical
epicycle frequencies respectively and
$\Sigma(J_\phi)=\Sigma_0\mathrm{e}^{-(\Rc-R_0)/R_\d}$ is the
(approximate) radial surface-density profile, where $\Rc(J_\phi)$ is
the radius of the circular orbit with angular momentum $J_\phi$. The
factor $1+\tanh(J_\phi/L_0)$ in equation (\ref{eq:planeDF}) is there
to effectively eliminate stars on counter-rotating orbits and the
value of $L_0$ is unimportant for these examples provided it is small
compared to the angular momentum of circular orbits at the radii of
interest. $\sigma_r$ and $\sigma_z$ are both functions of $J_\phi$
which control the radial and vertical velocity dispersions, with
\begin{eqnarray}\label{eq:sigmas}
  \sigma_r(J_\phi)&=&\sigma_{r0}\,\mathrm{e}^{q(R_0-\Rc)/R_\d}\nonumber\\
  \sigma_z(J_\phi)&=&\sigma_{z0}\,\mathrm{e}^{q(R_0-\Rc)/R_\d},
\end{eqnarray}
where $\sigma_{r0},\sigma_{z0}$ and $q$ are constants.

The phase mixed component with \df\ $f_\mix(\bolJ)$ is identical to
that used by M11. It is the sum of two ``pseudo-isothermal'' \df s,
corresponding to the thin and thick discs, normalised such that at the
Sun the surface density of thick-disc stars is 23 per cent of the
total stellar surface density. Table \ref{tab:df} lists the parameters
of each component of the \df.

The resonant component of the \df , $f_{\res}(\bolth,\bolJ)$, is of a
similar form to that used by M11, and can be written as
\begin{equation} \label{eq:resdf} f_{\res}(\bolth,\bolJ) = A_\res
  f_\mix(\bolJ) \times \alpha(\bolJ) \times \beta(\bolth)
\end{equation}
$A_\res$ is a normalisation constant, chosen such that in the Solar
neighbourhood (taken to be a sphere of radius $200\pc$ about the Sun),
8 per cent of the stars are associated with the resonance.  The term
$\alpha(\bolJ)$ describes how the stars are trapped around the
resonance line in action space. \cite{Se12} shows that for particles
at an ILR in an idealised $N$-body disc this trapping leads to an
excess of particles near the resonant line at high $J_R$, and a dearth
of particles at low $J_R$. We assume that this is what we would expect
for stars affected by any of the resonances considered. This leads us
to use
\begin{eqnarray} \label{eq:alpha} \alpha(\bolJ) & = &
  \exp\left(-\frac{(J_\phi-J_{\phi,\res}(J_r))^2}
    {\Delta_{J,\res}^2}\right)\times\\\nonumber &
  &\left(-B+\frac{1}{2}(1+B)
    (1+\tanh((J_R-J_{R,\cut})/\Delta_{J,\cut}))\right)
\end{eqnarray}
which is a Gaussian in $J_\phi$ of width $\Delta_{J,res}^2$ centred on
$J_{\phi,res}(J_r)$, multiplied by a function which runs from $1$ at
$J_R \gg J_{R,cut}$ to $-B$ at $J_R \ll J_{R,cut}$ with the transition
being over a range controlled by the parameter $\Delta_{J,cut}$. The
line $J_\phi=J_{\phi,res}(J_r)$ is, in each case, a good approximation
to the line described by the condition on orbital frequency
(equation~\ref{eq:res_om}) for a given perturber pattern speed
$\Omega_p$. We are careful to ensure that in all models
$f_\mix(\bolJ)+f_\res(\bolJ)>0$ for all $\bolJ$.

The term $\beta(\bolth)$ in equation~\ref{eq:resdf} is
\begin{equation} \label{eq:beta} \beta(\bolth) =
  \exp\left(-\frac{(\theta_r-\theta_{r,res}(\theta_\phi))^2}
    {\Delta_{\theta,res}^2}\right).
\end{equation}
$J_{\phi,res}(J_r)$ is chosen such that
$l\Omega_r(J_r,J_{\phi,res})+m\Omega_\phi(J_r,J_{\phi,res}) = \const$
for $J_z=0$, and $\theta_{r,res}(\theta_\phi)$ and is 
chosen such that
\begin{equation}\label{eq:res_theta}
l\theta_{r,res}+m\theta_\phi = C_\theta,
\end{equation}
where the constant $C_\theta=l\theta_{r,res}(\theta_\phi=0)$. 
 The values $\Delta_{J,res}$
and $\Delta_{\theta,res}$ give the width of the resonance peak around
the exact resonance lines in $J_\phi$ and $\theta_r$, respectively.

\begin{figure}
  \centerline{
    \includegraphics[width=.45\hsize]{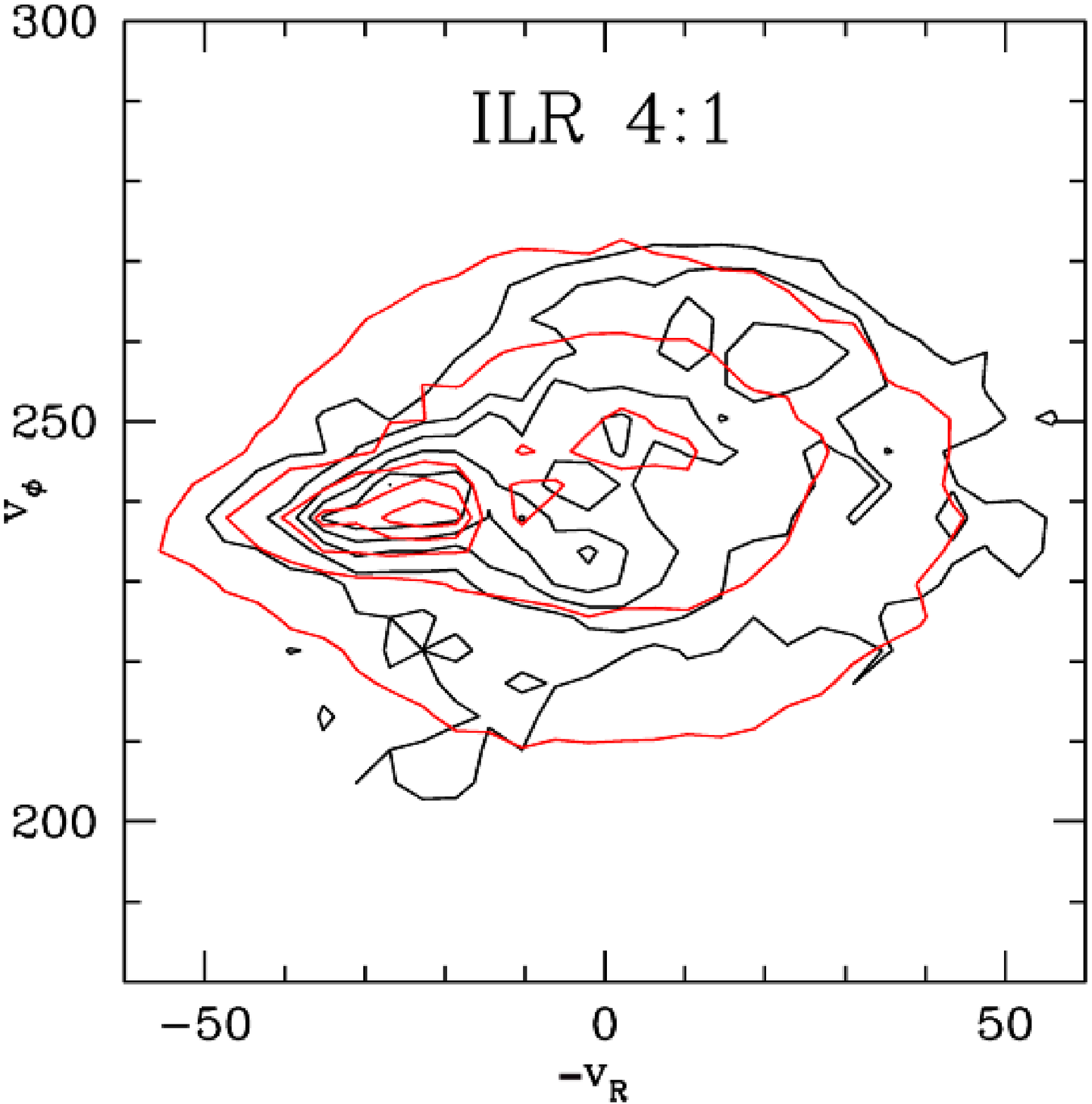}
    \includegraphics[width=.45\hsize]{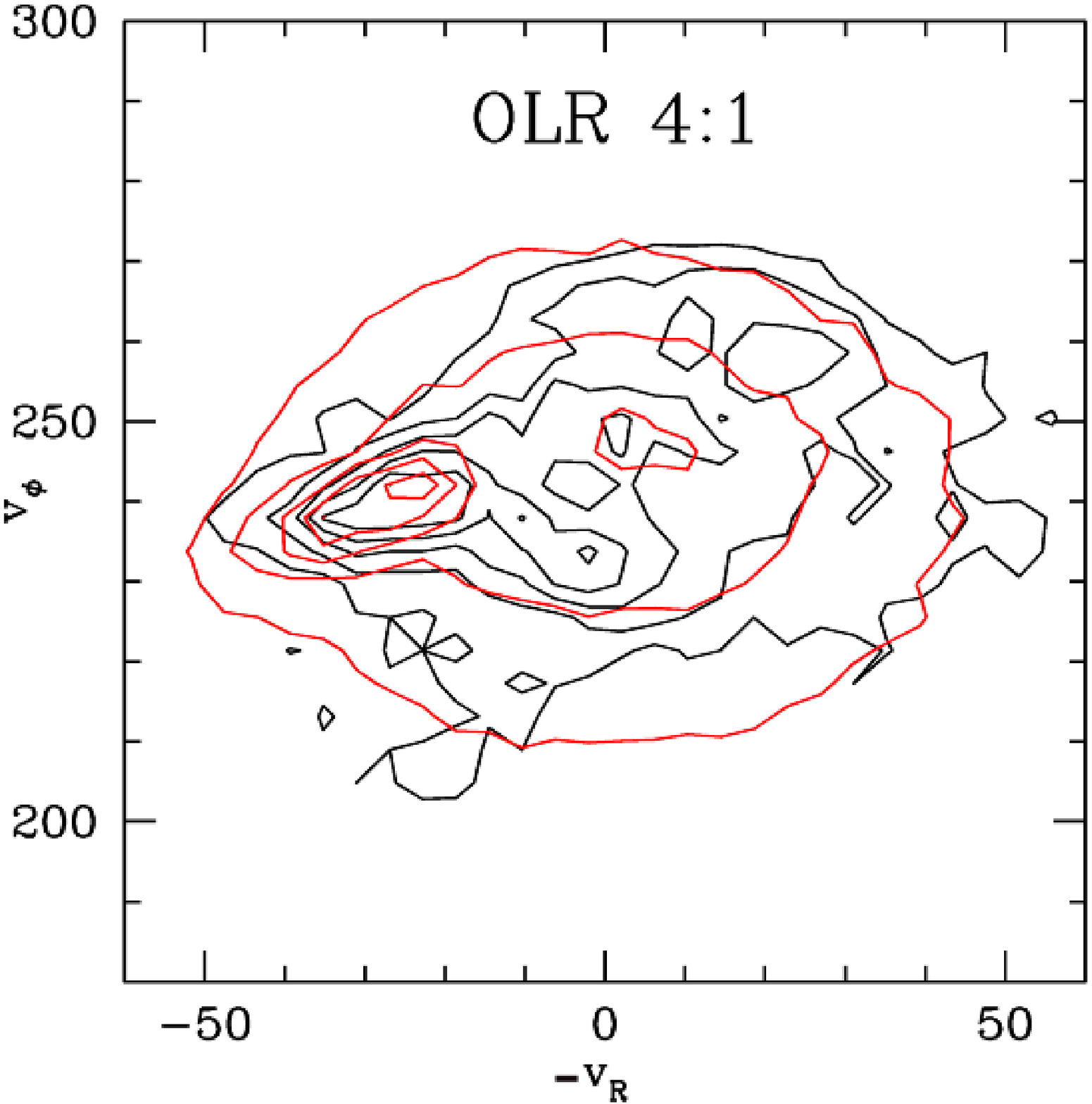}
  } \centerline{
    \includegraphics[width=.45\hsize]{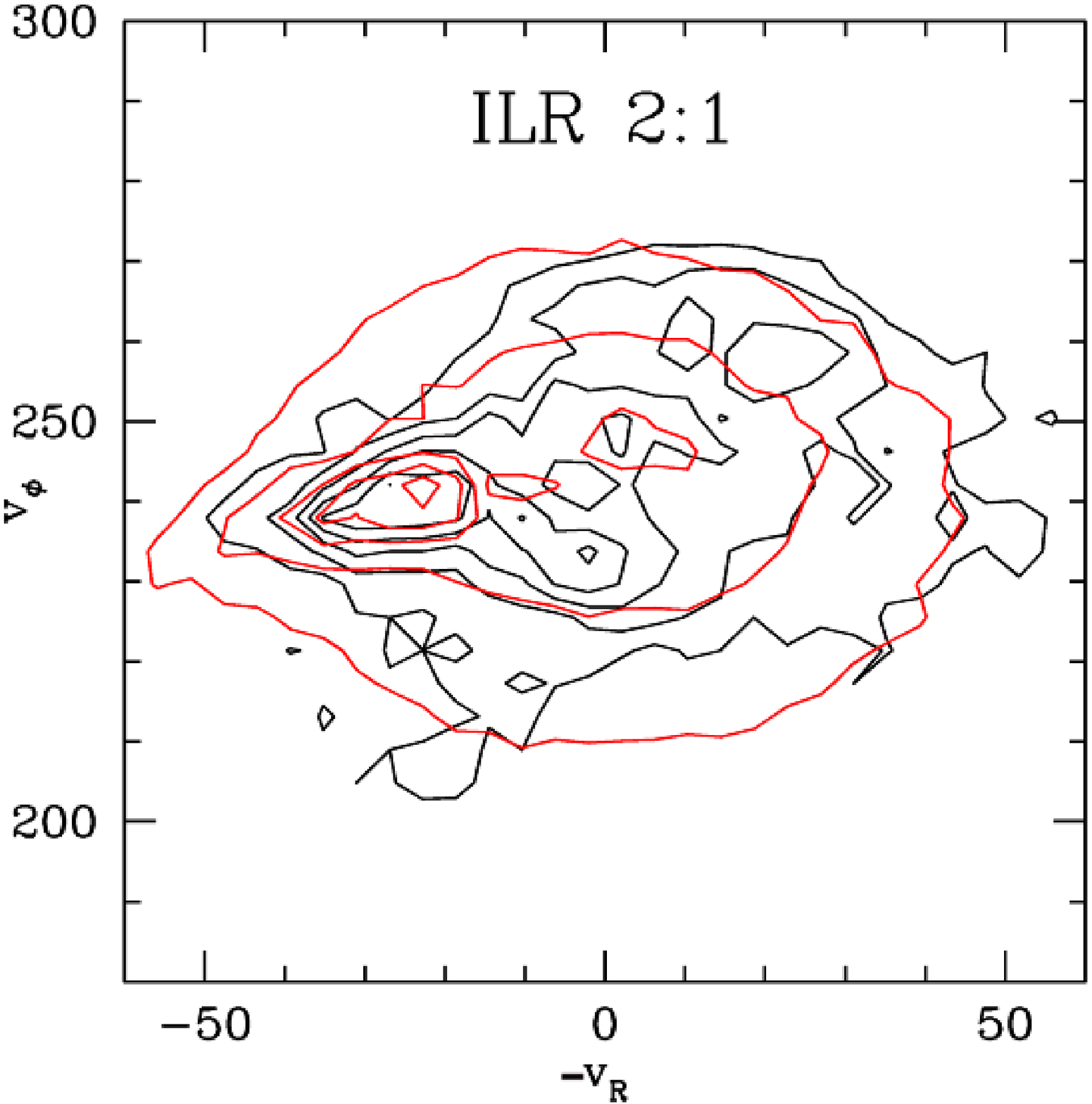}
    \includegraphics[width=.45\hsize]{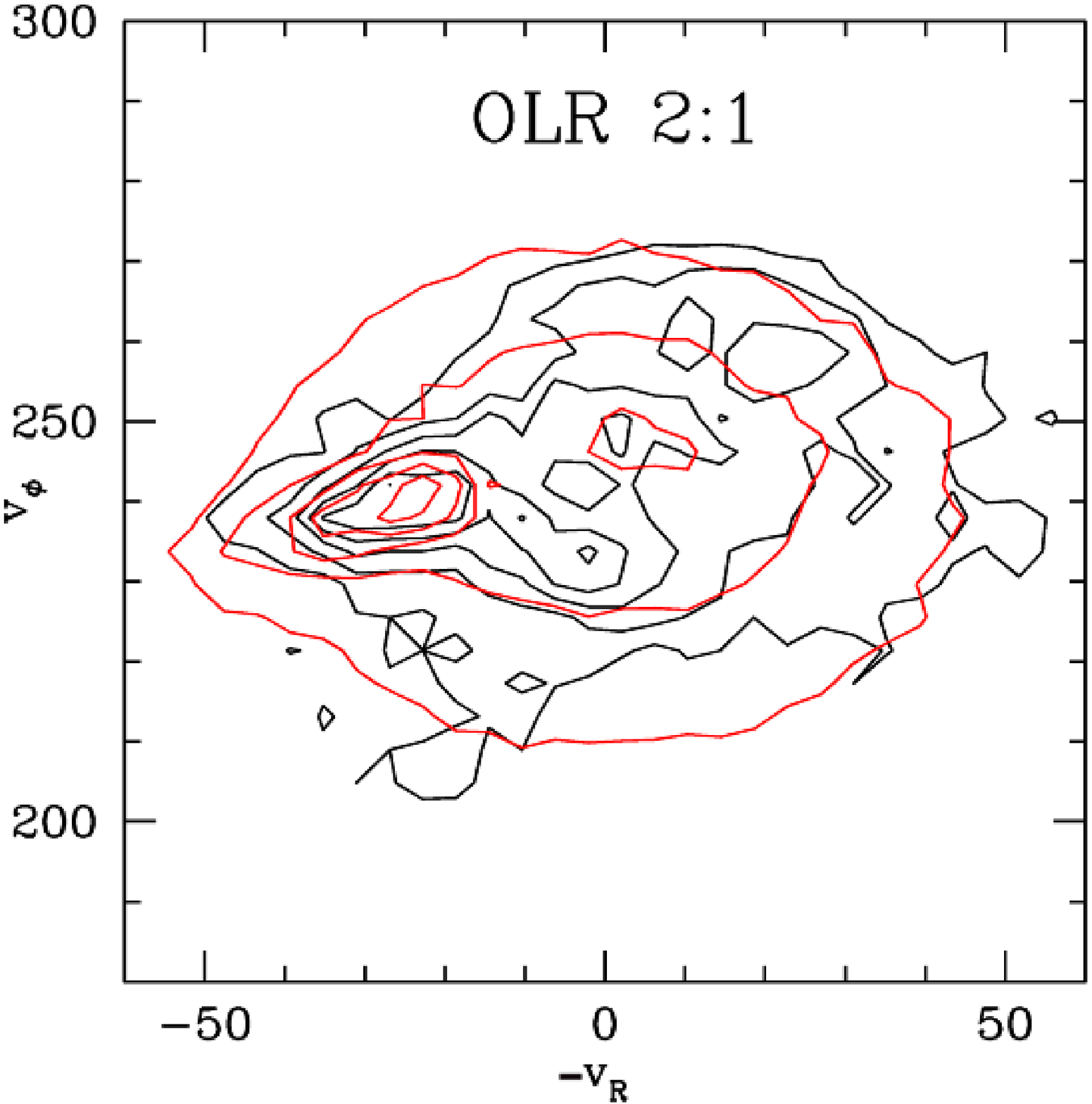}
  }
  \caption{
    Contour plots of the density in the $(-v_R)-v_\phi$ plane of the stars 
    observed by the GCS (black) and stars in the $200\pc$ around the sun 
    in each of the models described in Table~\ref{tab:dfres} (red). $-v_R$ is 
    used so that the plot reflects the familiar $U-V$ plane in the 
    Solar neighbourhood. The Hyades are found around $v_r=30\kms$,
    $v_\phi=240\kms$, and each of the models matches the observed
    overdensity well.
\label{fig:GCS_UV}
}
\end{figure}

The parameter values for the resonances considered in this study are
chosen to provide a reasonable match (by eye) 
to the Hyades as seen in the GCS,
and are shown in Table~\ref{tab:dfres}. This matching is a relatively 
simple process as the approximate positions of the resonance lines in 
action space are easy to determine from the density distribution of 
the GCS stars in action (Figure~\ref{fig:RL}), and the approximate positions
of the 
resonance lines in angle (and thus the value of $C_\theta$)
from the density distribution in 
$\theta_r-\theta_\phi$ plane (or marginalised versions thereof, e.g.
Fig. 8 or 5 of M11). 
Contour plots of the density
of stars in the $(-v_R)-v_\phi$ plane\footnote{It is more usual to
  show this plot in $U$ and $V$ but, since we also consider fields far
  from the Solar neighbourhood, we use $-v_R$ and $v_\phi$ which, at
  the Sun's position, are equivalent to $U$ and $V$ except offset by
  the Sun's velocity} in the Solar neighbourhood in the four models,
overlaid on the same figure plotted using the GCS data are shown in
Figure~\ref{fig:GCS_UV}.  This emphasises the point made by M11 that
the information provided by the distribution of stars in velocity
space in the Solar neighbourhood is insufficient to determine the type
of resonance which gives rise to the Hyades moving group. It is also
worth noting that very similar local velocity diagrams can be produced
with models that have significantly different values for various
parameters (e.g. either $\Delta_{J,res}$ or $\Delta_{\theta,res}$ can
vary by a factor of $\sim2$.

\section{Beyond the Solar neighbourhood}\label{sec:beyond}

\begin{figure}
  \centerline{
    \includegraphics[width=.45\hsize]{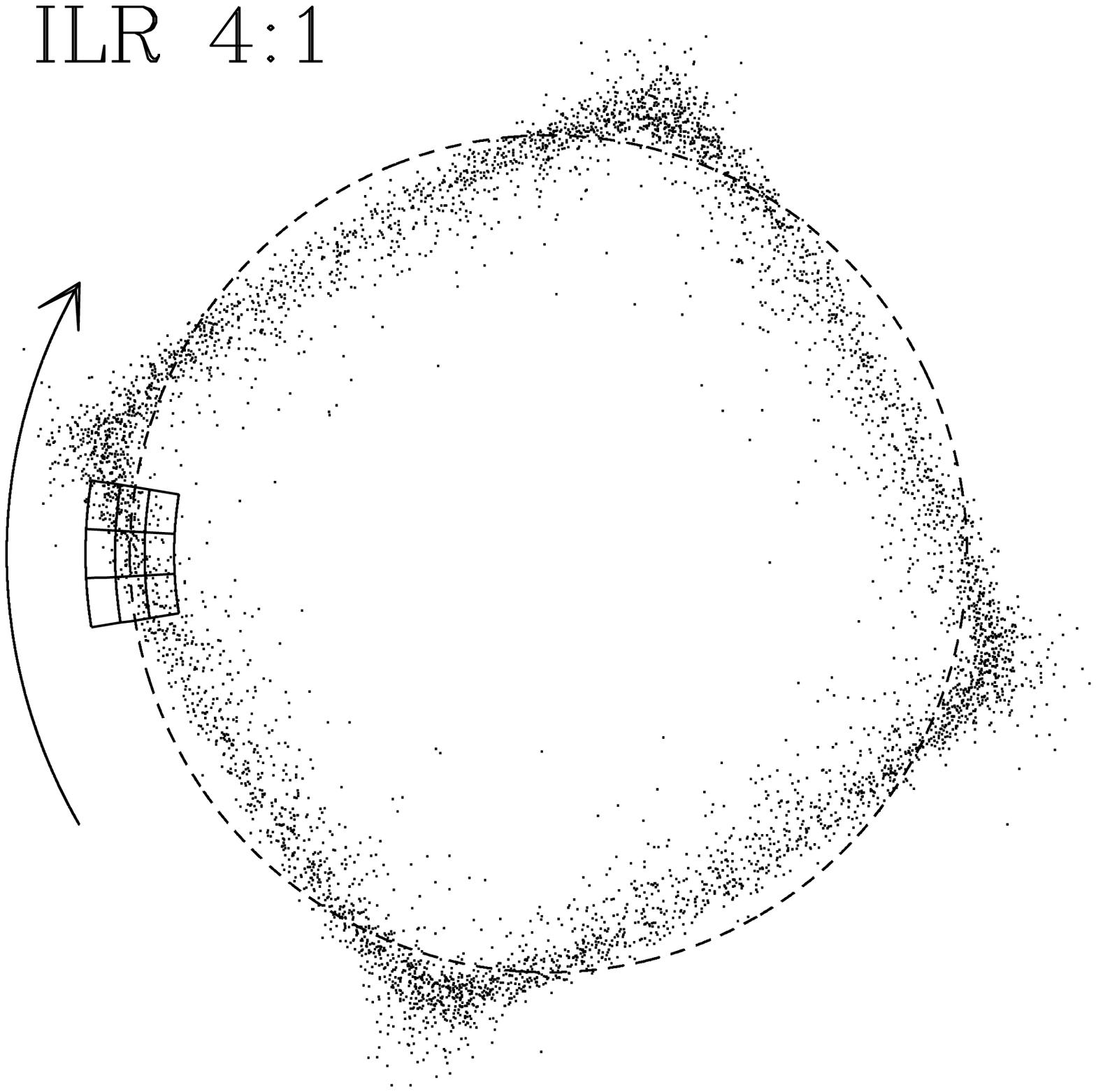}
    \includegraphics[width=.45\hsize]{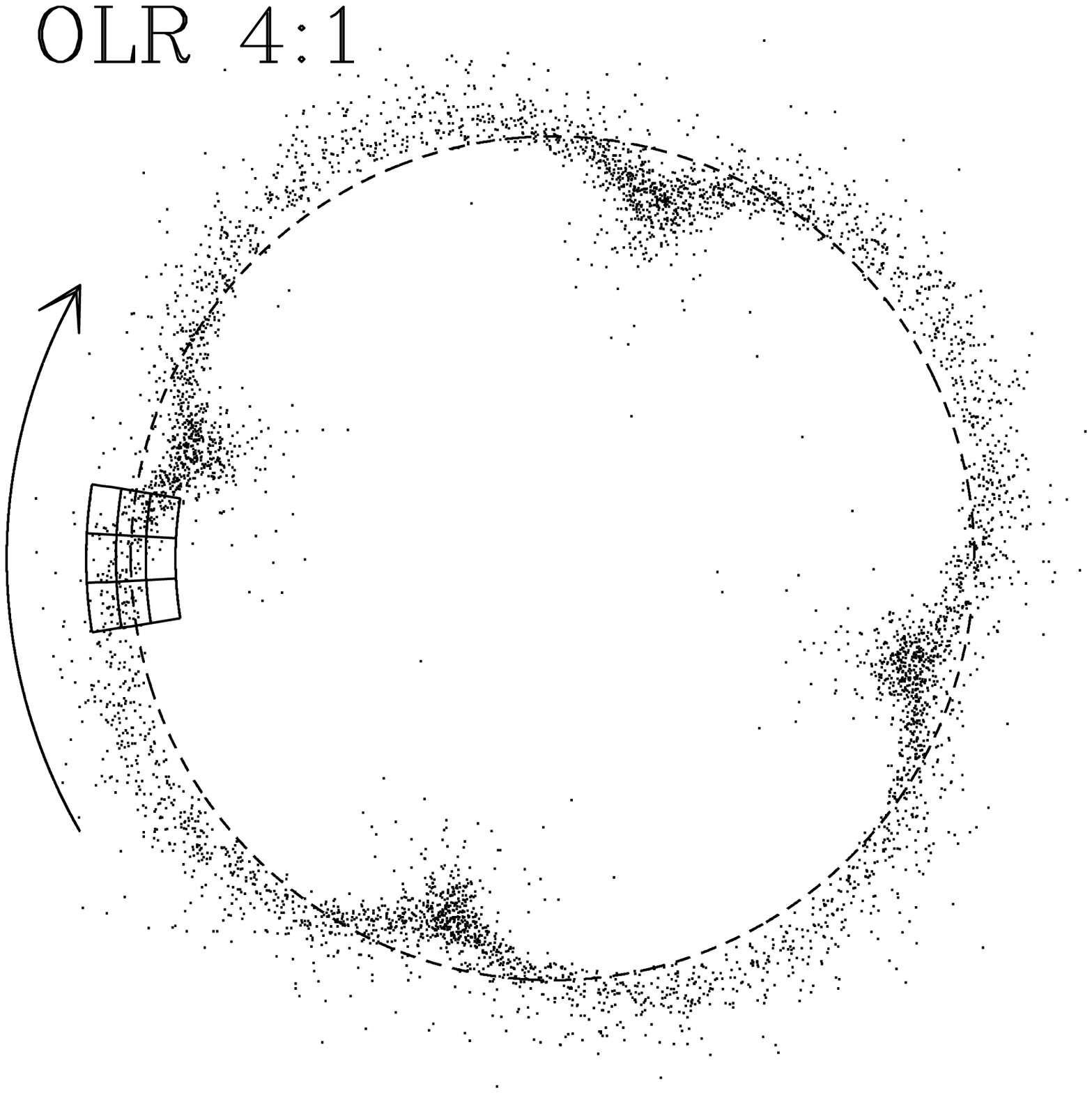}
  } \centerline{
    \includegraphics[width=.45\hsize]{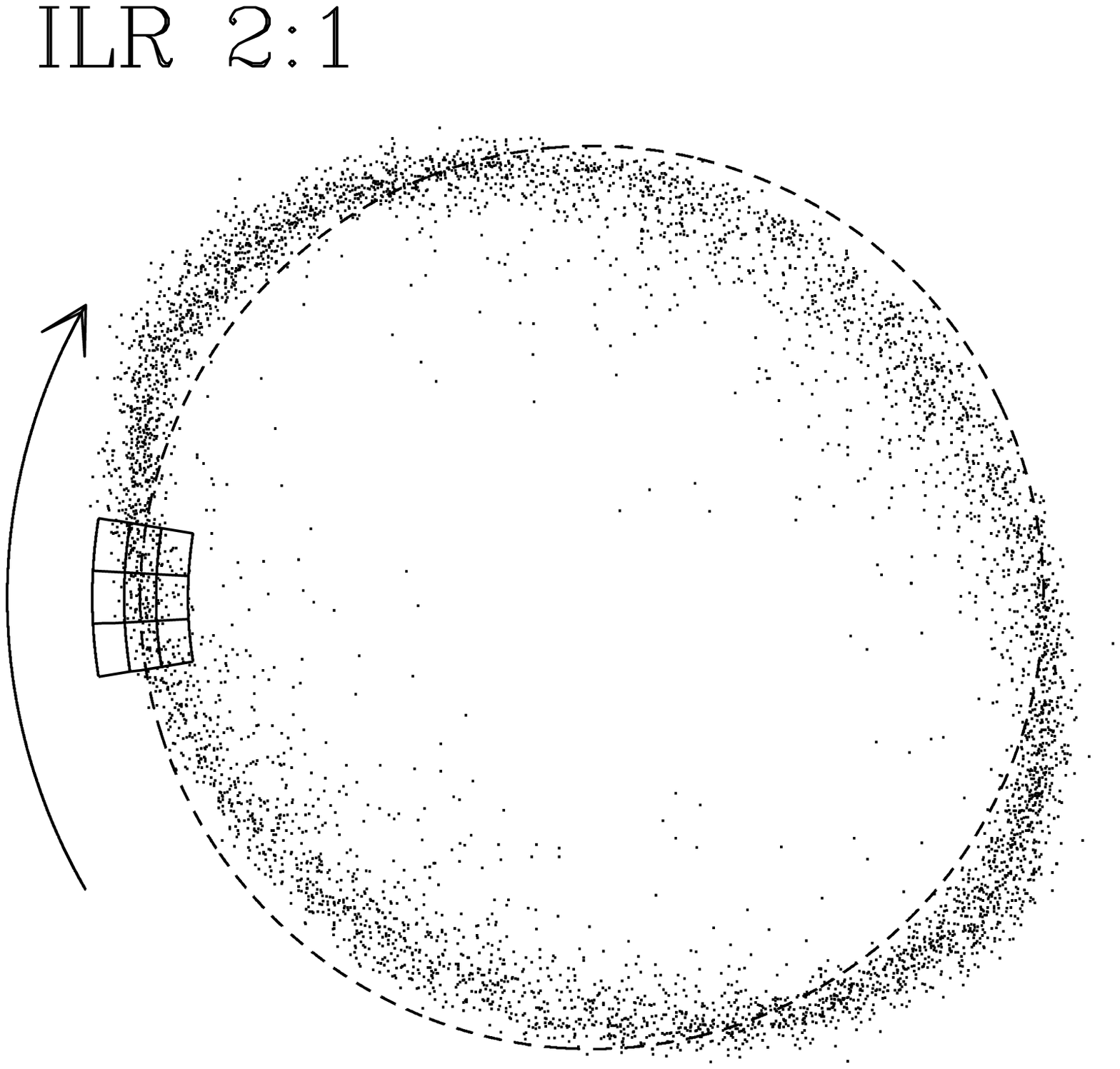}
    \includegraphics[width=.45\hsize]{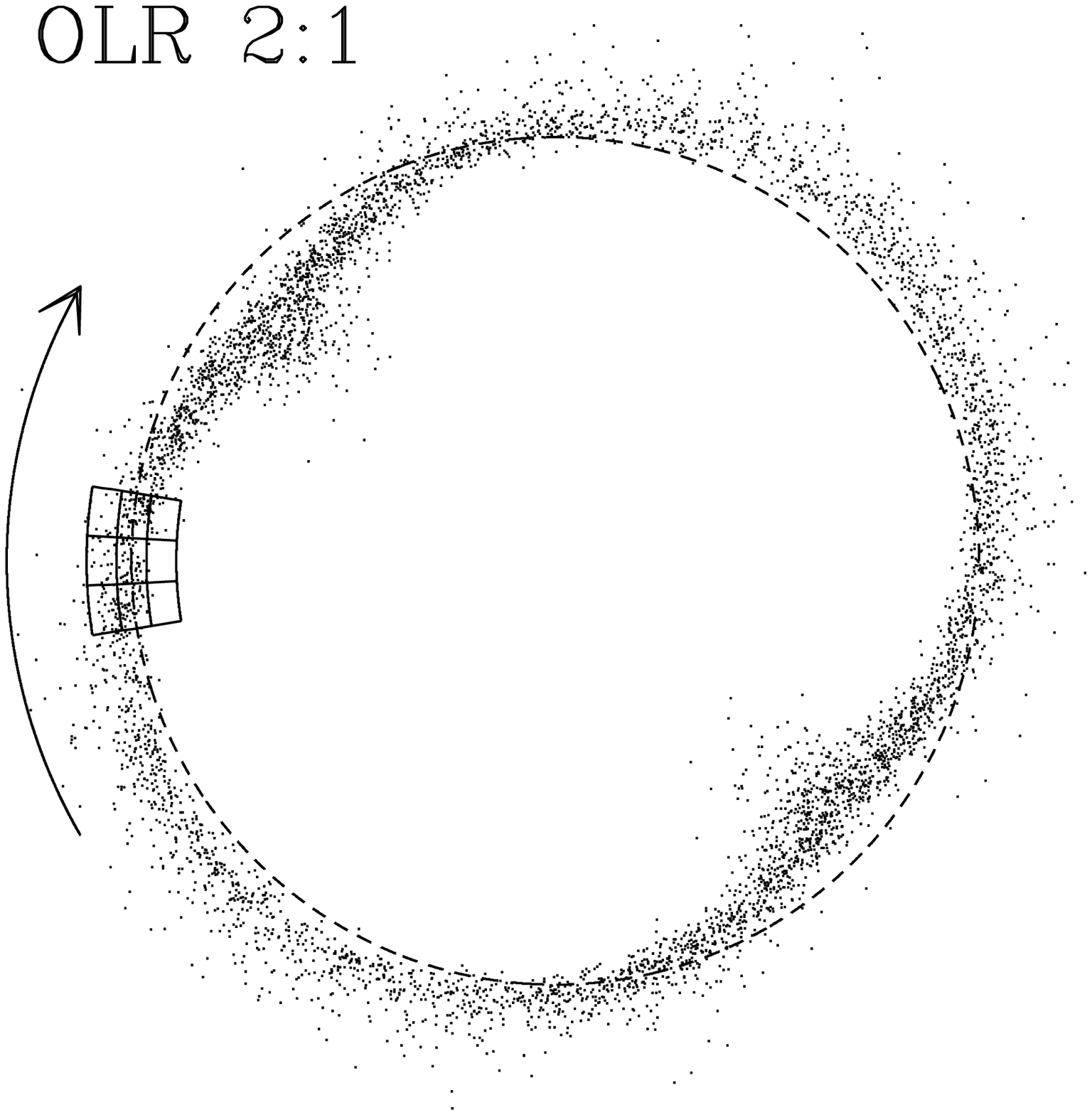}
  }
  \caption{
    Distribution of stars associated with the Hyades in the plane of
    the Galaxy for each of the four models (in each case these are
    $5000$ stars selected at random from the positive part of
    $f_\res$). 
    The arrow indicates the direction of Galactic rotation. 
    The \emph{dashed} line is drawn at the
    Solar radius. The nine bins on the left hand side of each diagram
    are the bins
    shown in Figure~\ref{fig:Rpbins}, so the Sun is at the middle of
    the central bin, and each bin spans $600\pc$ in radius and
    $6.1^\circ$ in azimuth.
    \label{fig:Galaxy}
  }
\end{figure}

\begin{figure}
  \centerline{
    \includegraphics[width=0.8\hsize]{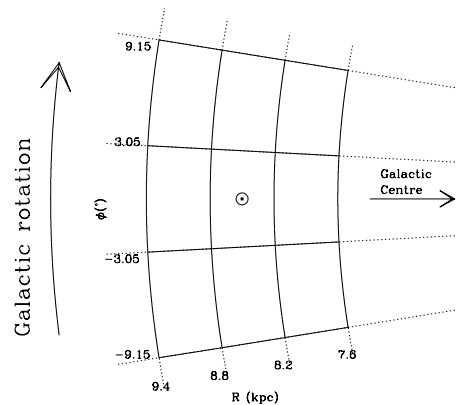}
  }
  \caption{
    Spatial bins in R \& $\phi$ used in the plots shown in
    Figures~\ref{fig:HYO_RP} and ~\ref{fig:ALL_RP} (\emph{solid}
    lines). \emph{Dotted} lines run in the $R$ and $\phi$ directions.
    The Sun is marked (as $\odot$) at $R=8.5$, $\phi=0$.  
\label{fig:Rpbins}
}
\end{figure}

It is clear that to determine the resonance responsible for the Hyades
moving group, we need to look beyond the Solar neighbourhood. In this
section we determine what each of the Hyades models predicts for the
distribution of stars in velocity space at different points in the
Galaxy.

In Figure~\ref{fig:Galaxy} we show the positions of stars associated
with the Hyades in the Galactic plane for all models. Much as one
would expect from the condition described by equation~\ref{eq:res_th},
the 4:1 resonances have groups of stars gathered at apocentre or
pericentre at four different azimuths ($\theta_r$ changing by $8\pi$
as $\theta_\phi$ changes by $2\pi$), producing a square or clover-leaf
pattern in position in the Galactic plane. Similarly the 2:1
resonances have groups at apo- or peri-centre at two different
azimuths ($\theta_r$ changing by $4\pi$ as $\theta_\phi$ changes by
$2\pi$), producing an oval pattern in position. 

In Figure~\ref{fig:Galaxy} we also show the position of the extended
Solar neighbourhood which we will be investigating in velocity space
to find the signature of the Hyades. The key thing to note is that the
``stream'' in position associated with the resonances runs in
different directions near to the Sun, depending on whether they are
inner or outer Lindblad resonances. ILRs produce streams that tend
to be at $R<R_0$ in the immediate anti-rotation direction, and towards
$R>R_0$ in the direction of rotation, whereas OLRs produce streams
that are the other way around. Note that each case the Hyades stars in
the Solar neighbourhood are all rotating in the same direction as the
Galaxy and moving radially outwards, so they are not ``following'' the
stream
in the Galactocentric frame (though they are in the frame rotating with the 
perturbation).

These streams are of non-negligible width, controlled by the parameter
$\Delta_{\theta,\res}$ in equation~\ref{eq:beta}. It should be
possible to determine the value of $\Delta_{\theta,res}$ (or something
similar, for a different model), as well the type of resonance, from
observations by determining the range in $R$ (at a given $\phi$) over
which the Hyades can be found (in velocity space). \cite{Se10}
attempted to estimate the time since the resonance occurred by
comparing the width of the resonance in frequency to the width of the
resonance in angle. That result is flawed because of the unrecognised
impact of selection effects on the observed spread in angle, but
determining the true value of $\Delta_{\theta,res}$ would enable us to
make an improved estimate. It should be noted that this can only be an
upper estimate, without knowing the width in $\bolth$ intrinsic in the
libration of the resonantly trapped stars.
 

\begin{figure*}
  \centerline{
    \includegraphics[width=0.75\hsize]{
      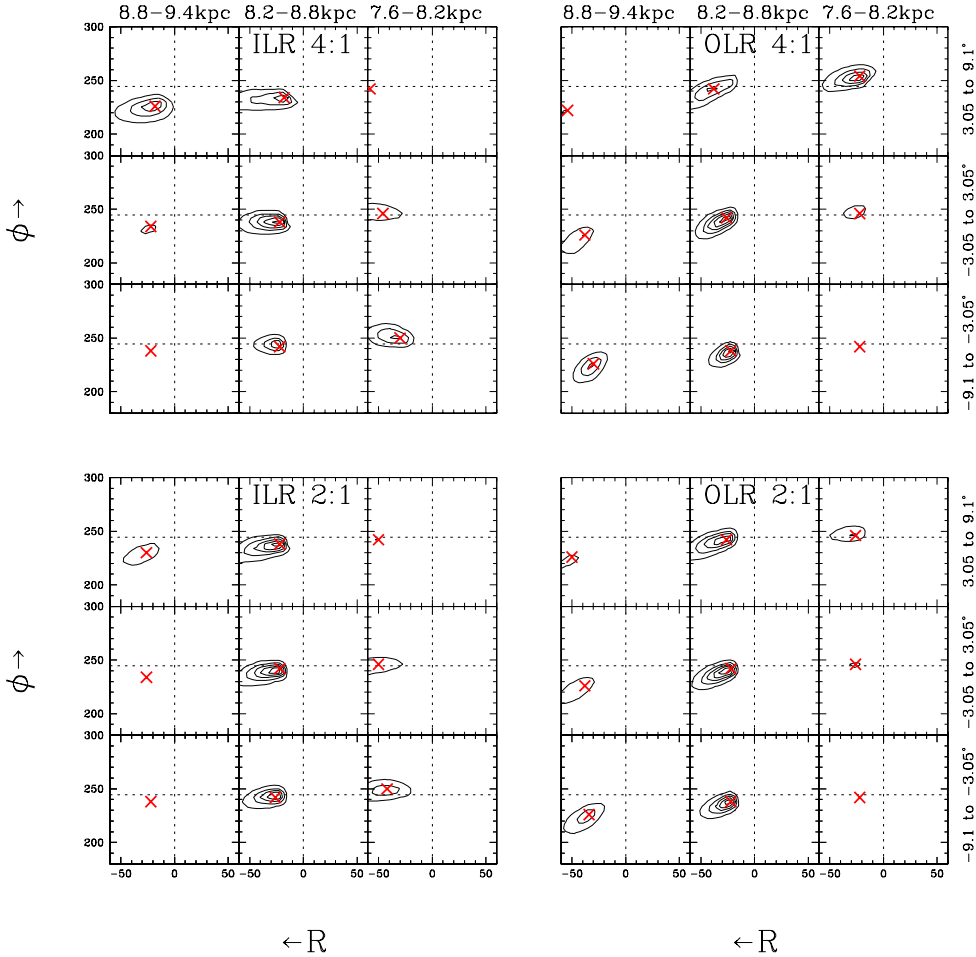}
  }
  \caption{ The four different panels correspond to the four Hyades
    models (as labelled). Each panel is divided into nine plots,
    corresponding to the nine spatial bins shown in
    Figure~\ref{fig:Rpbins}, and shown in the same orientation
    (i.e. plots in the left of each panel all correspond to bins between
    $8.8$ and $9.4\kpc$ from the Galactic centre and those at the
    bottom of each panel all correspond to the range $-9.15 < \phi <
    -3.05$). Each plot is a 
    contour plot of the distribution of stars associated with the 
    Hyades (i.e. $f_{\res}$ and not $f_{mix}$) in the $(-v_R)-v_\phi$ plane. 
    In each individual plot the range in $-v_R$ is 
    $-60$ to $60\kms$ and that in $v_\phi$ is $180$ to $300\kms$ (i.e. the 
    same range as Figure~\ref{fig:GCS_UV}), with dotted lines marking 
    $v_R=0\kms$ and $v_\phi = 244.5\kms = v_{0}$, the circular speed at the 
    Sun. Contours are placed in each plot 
    where the density is $10,30,50,70$ or $90$ per
    cent of the highest density seen in the central plot of the panel (the bin 
    which contains the Solar neighbourhood). Red crosses mark the 
    highest density point in each plot.
\label{fig:HYO_RP}
}
\end{figure*}
 
\begin{figure*}
  \centerline{
    \includegraphics[width=0.75\hsize]{
      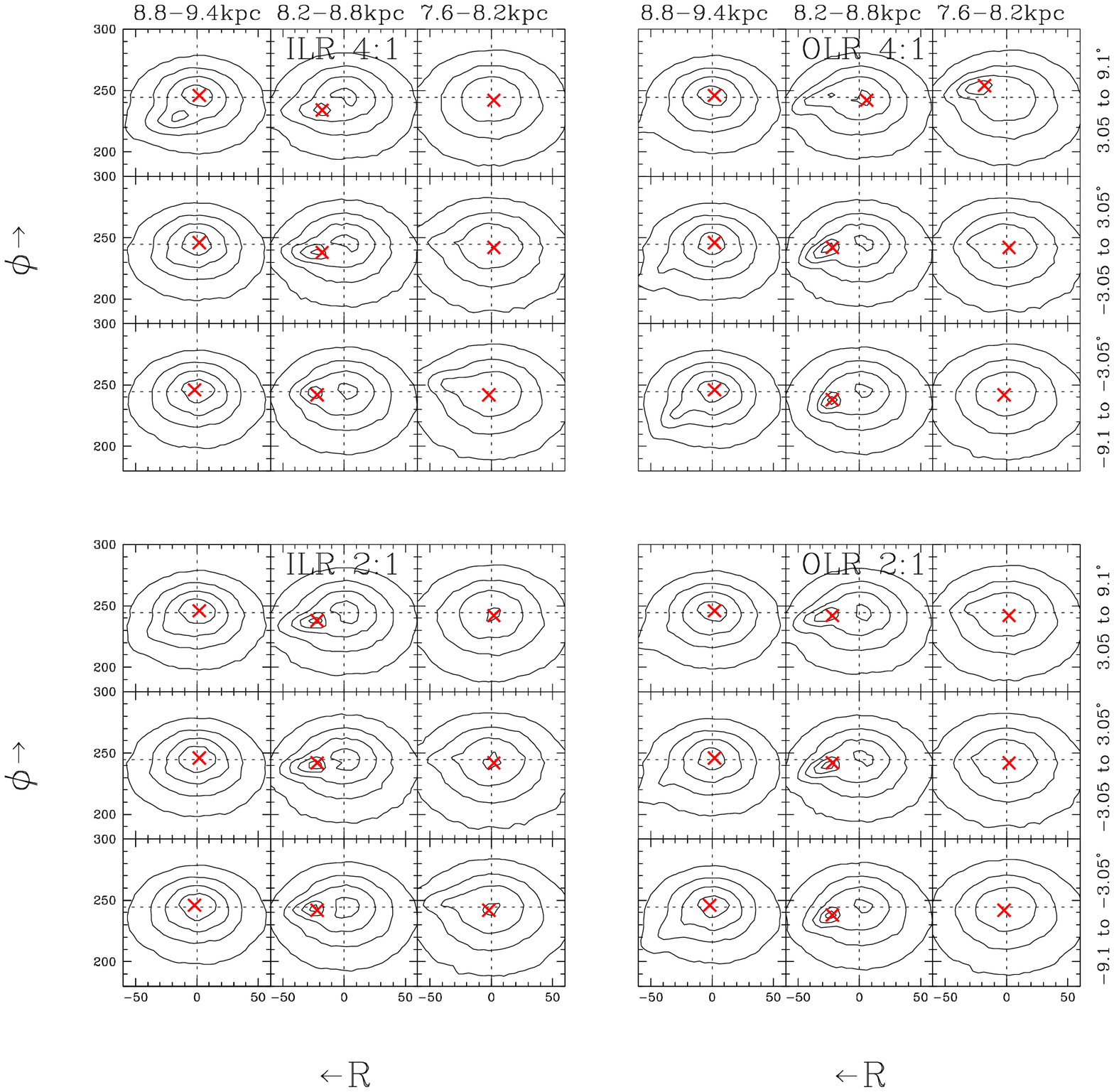}
  }
  \caption{The same plots as Figure~\ref{fig:HYO_RP} except that the 
    full \df \ is used (i.e. both $f_{res}$ \emph{and} $f_{mix}$),
    giving a sense of the difficulty in picking out 
    the Hyades from a smooth background. 
\label{fig:ALL_RP}
}
\end{figure*}

In Figure~\ref{fig:HYO_RP} we show the distribution in velocity space
of stars associated with the Hyades in each model, separated into bins
in the Galactic plane each of which covers $600\pc$ in Galactocentric
radius $R$, $6.1^\circ$ in Galactocentric angle $\phi$ (corresponding
to $\sim900\pc$ at the Solar radius), and only contains stars which
are within $300\pc$ of the Galactic plane. These bins are illustrated
in Figure~\ref{fig:Rpbins}.

In Figure~\ref{fig:HYO_RP} the Hyades can clearly be seen as a strong
feature in all three azimuthal bins at the Solar radius for all
models. In each case the Hyades appear at a similar point in the
$v_r-v_\phi$ plane as they do in the Solar neighbourhood. This is in
keeping with the overdensity observed by A12. In bins at
lower $R$, the Hyades are found at higher $v_\phi$, and in bins at
higher $R$, they are found at lower $v_\phi$. This is because in all
cases the Hyades are associated with a narrow range in angular
momentum, which naturally produces a correlation between $R$ and
$v_\phi$. The peak of the distribution associated with the Hyades
moves from $v_\phi\sim250\kms$ in the inner bin to $v_\phi\sim230\kms$
(varying slightly between the models). This is a somewhat smaller
change than one would naively expect from angular momentum
conservation given that the bin centres are $600\pc$ apart, but this
is because stars in the outer bins tend to be close to the central bin
rather than spread evenly across the bin.

There are two notable differences between the ILR models and the OLR
models. Probably the most useful difference, in terms of telling one
model from another, is the variation in the number of stars associated
with the Hyades in each spatial bin.  The two ILR models tend to have
many Hyades stars in the outer Galaxy, and few in the inner, in the
direction of Galactic rotation and many in the inner Galaxy, and few
in the outer, in the anti-rotation direction. The opposite is true in
the case of the OLR models. This effect is most pronounced in the 4:1
resonances, with the 2:1 resonances placing few Hyades stars far from
the Solar radius in this range of $\phi$. This is because of the
differences in the paths of the ``streams'' illustrated in
Figure~\ref{fig:Galaxy}.  The second difference is in the $v_r$ value
of the peak in the velocity distribution in the inner or outer
bins. The ILR models have a peak at $v_r\sim50\kms$ in the bins
$7.6\kpc<R<8.2\kpc$ and $v_r\sim22\kms$ in the bins
$8.8\kpc<R<9.4\kpc$, whereas in the OLR models this is reversed, with
$v_r\sim22\kms$ for $7.6\kpc<R<8.2\kpc$ and $v_r\sim50\kms$ for
$8.8\kpc<R<9.4\kpc$.

In Figure~\ref{fig:ALL_RP} we show the same figures for the model
\emph{including} the smooth background \df. This shows that in the
inner and outer bins with relatively strong signatures of the Hyades,
we can expect to observe the Hyades above the background, but the weak
signatures are lost in the background. It should be noted that this
is, in some ways, a best case scenario, as we have assumed negligible
uncertainty on velocity, a large number of observed stars (so shot
noise doesn't hide the Hyades) and ignore other substructure in the
velocity diagram that might cause confusion (for example, from the
dynamical structures associated with the Pleiades and Sirius moving
groups).

\begin{figure*}
  \centerline{
    \includegraphics[width=.125\hsize]{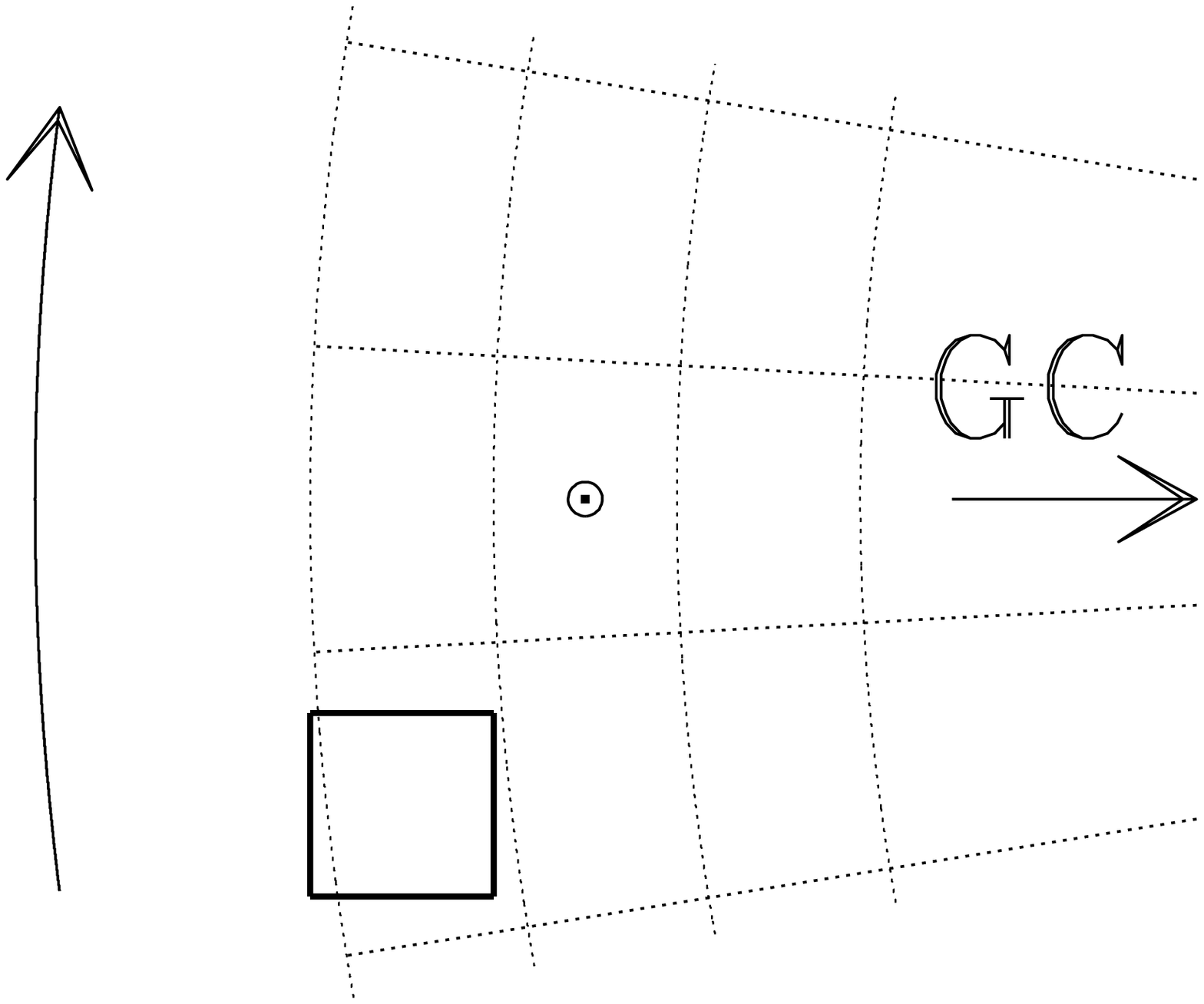}
    \includegraphics[width=.125\hsize]{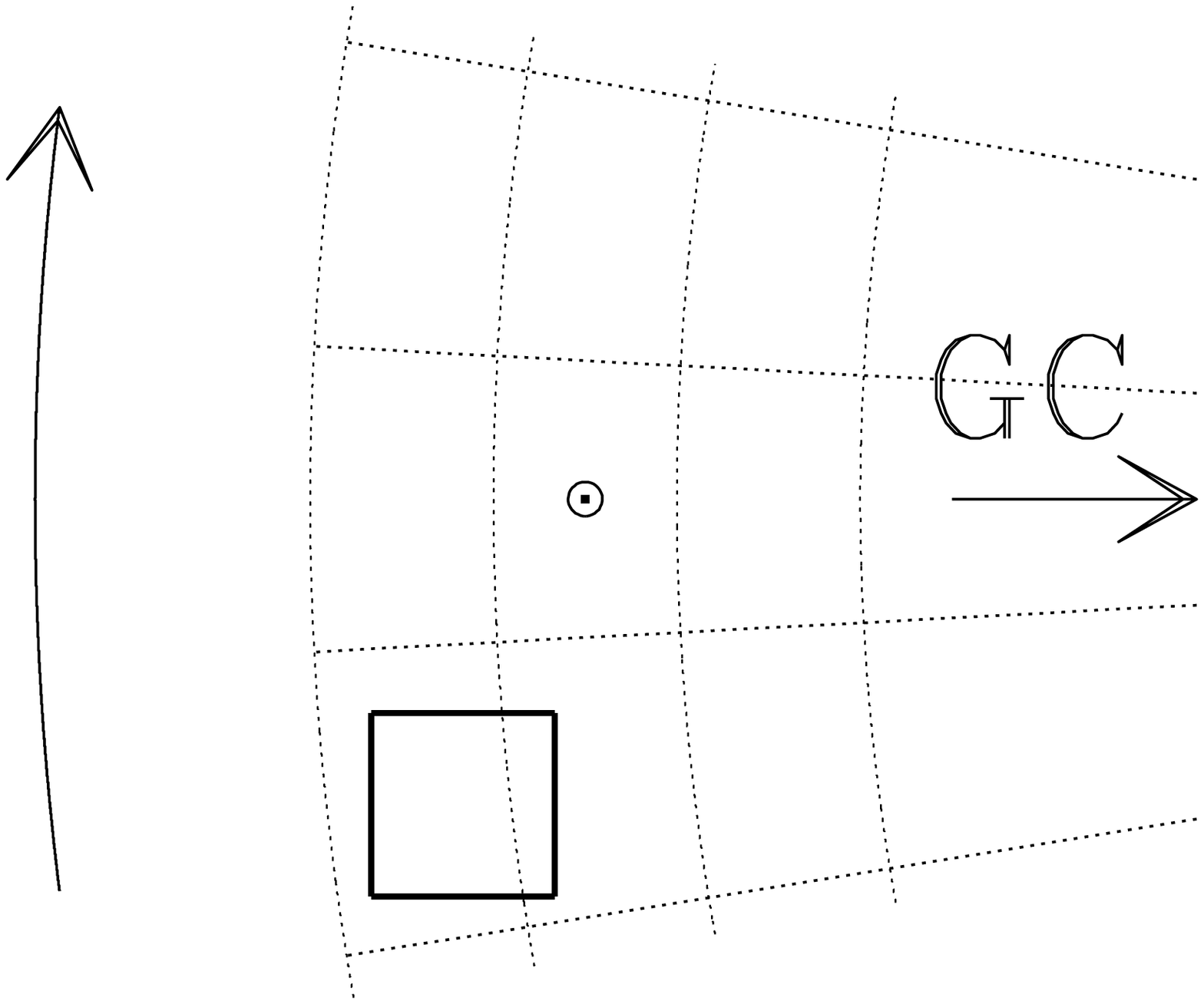}
    \includegraphics[width=.125\hsize]{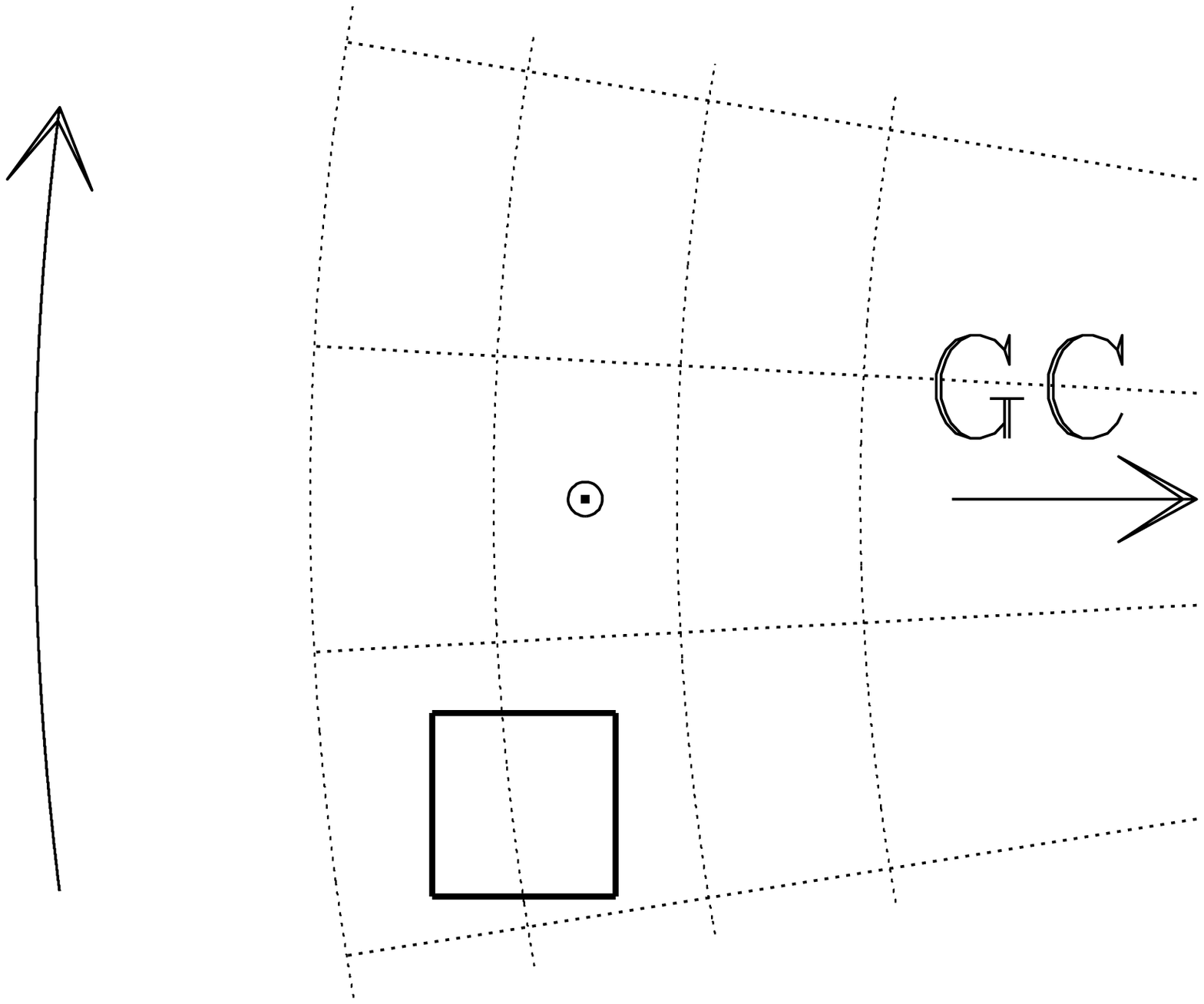}
    \includegraphics[width=.125\hsize]{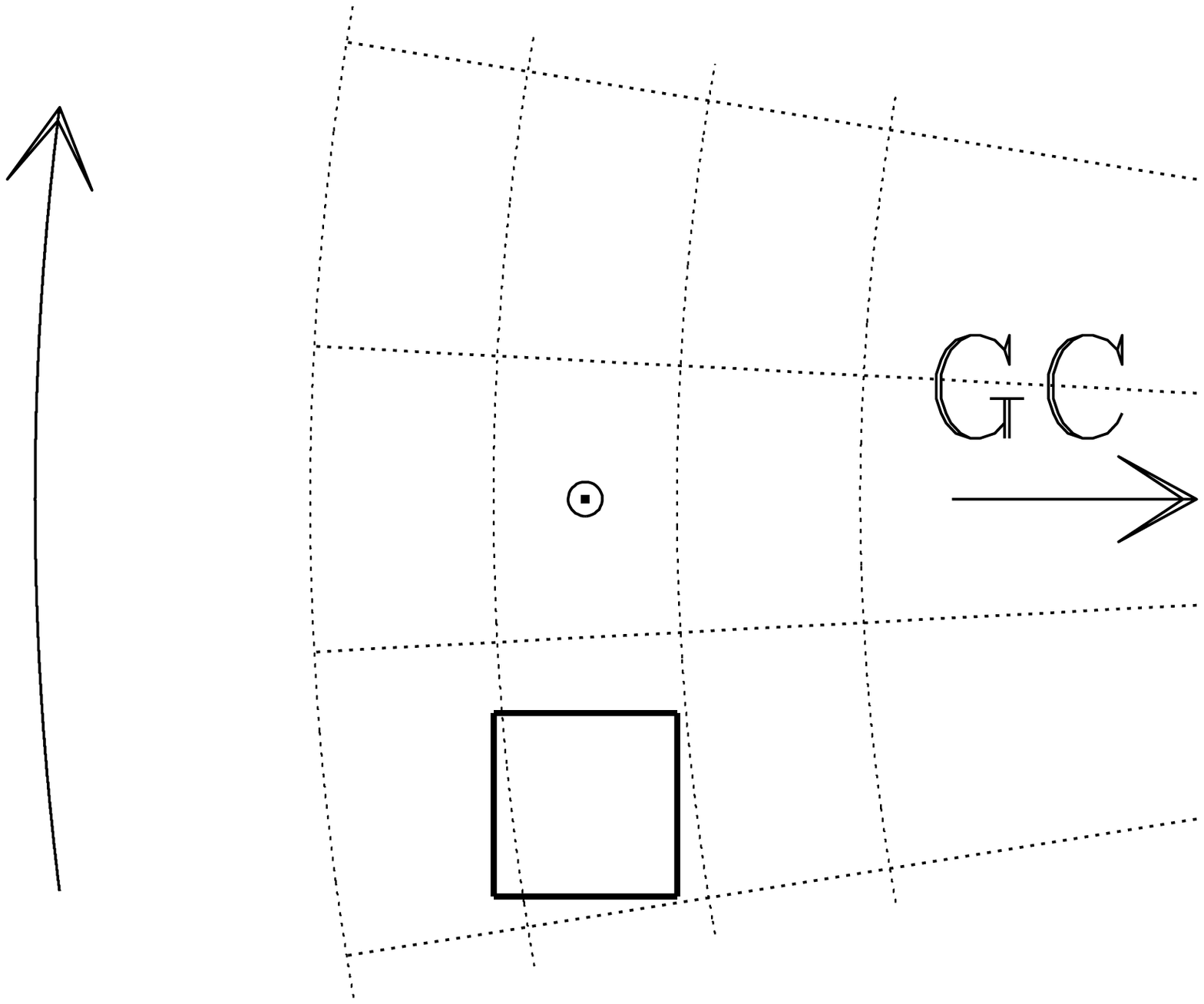}
    \includegraphics[width=.125\hsize]{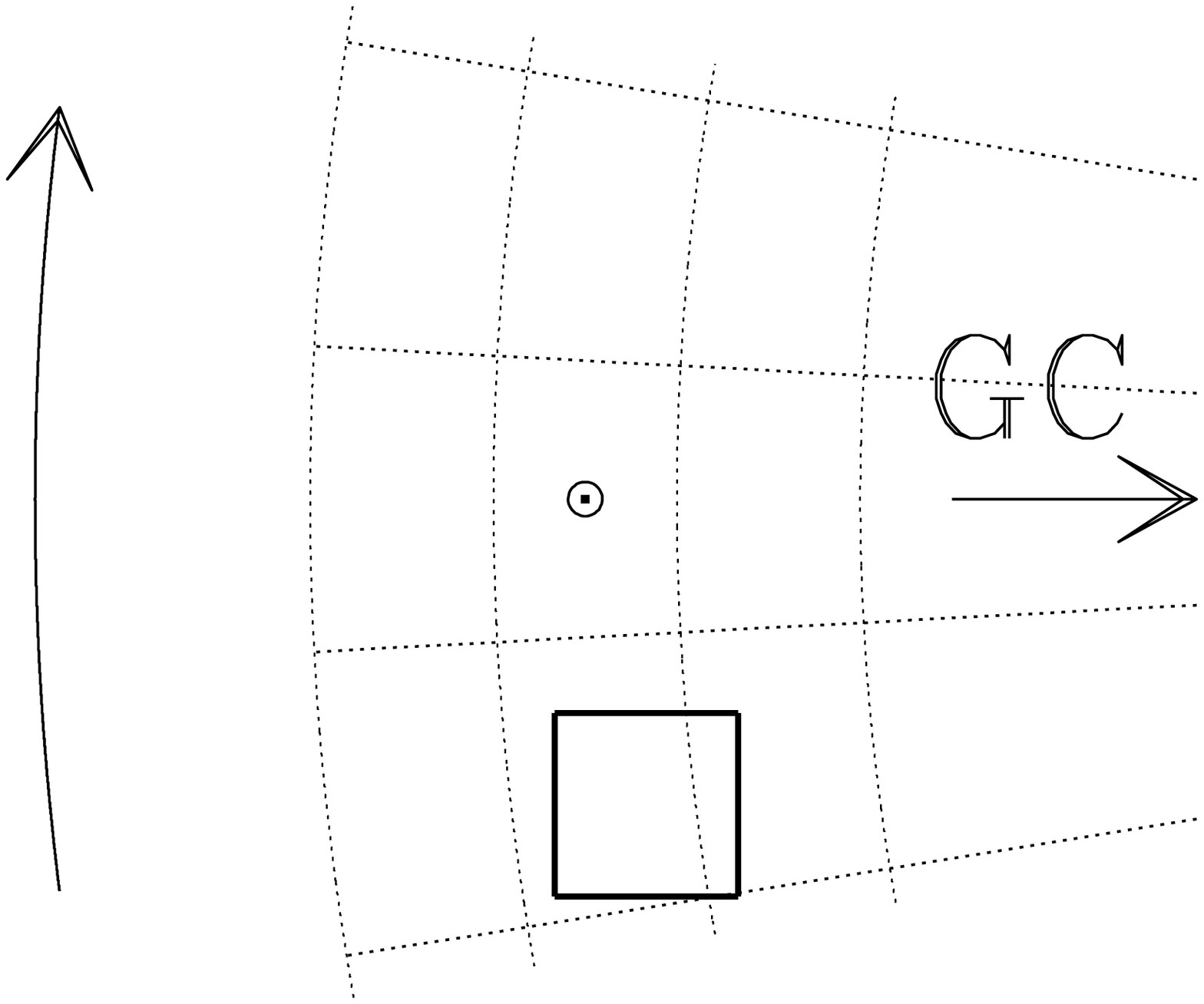}
    \includegraphics[width=.125\hsize]{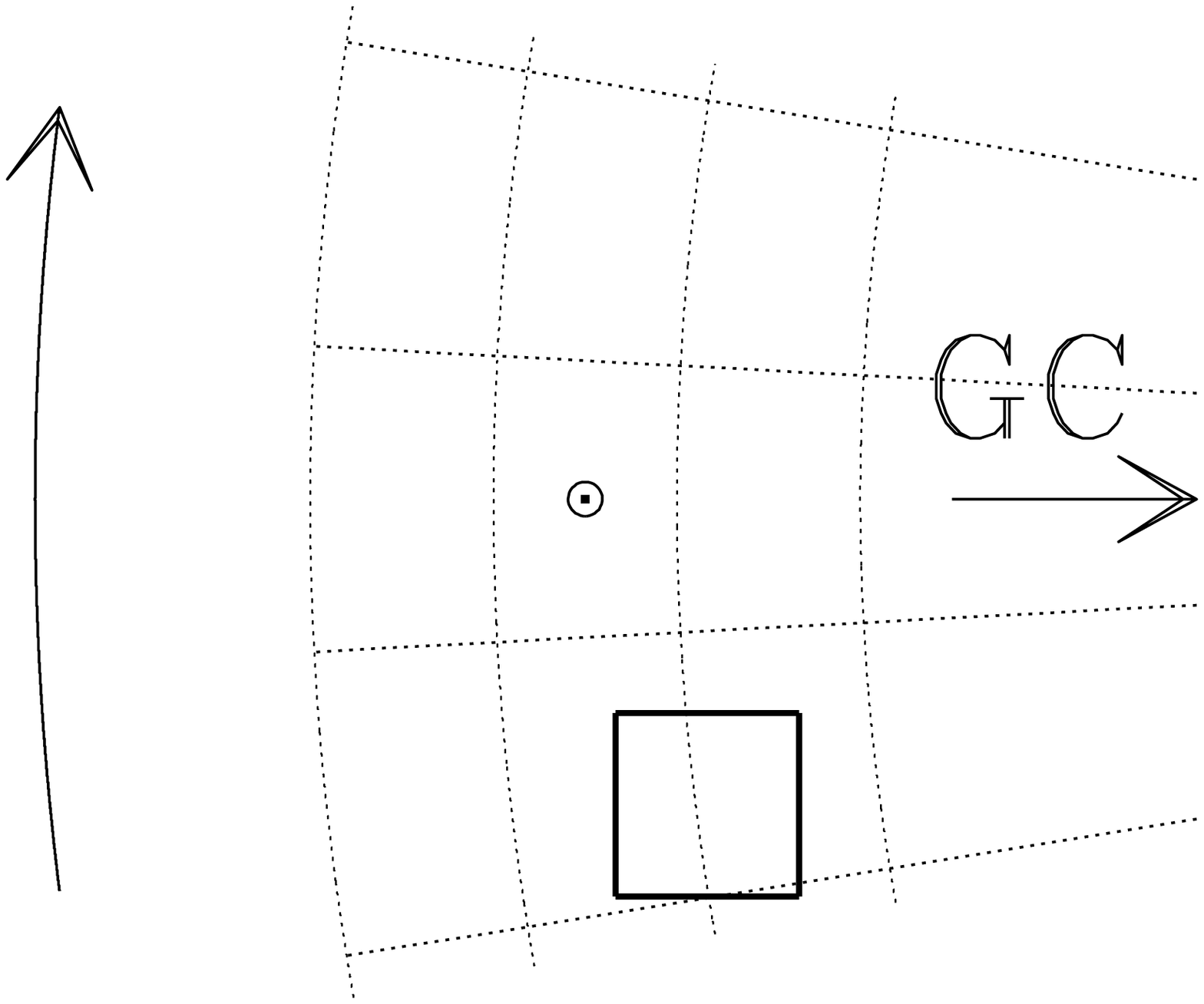}
    \includegraphics[width=.125\hsize]{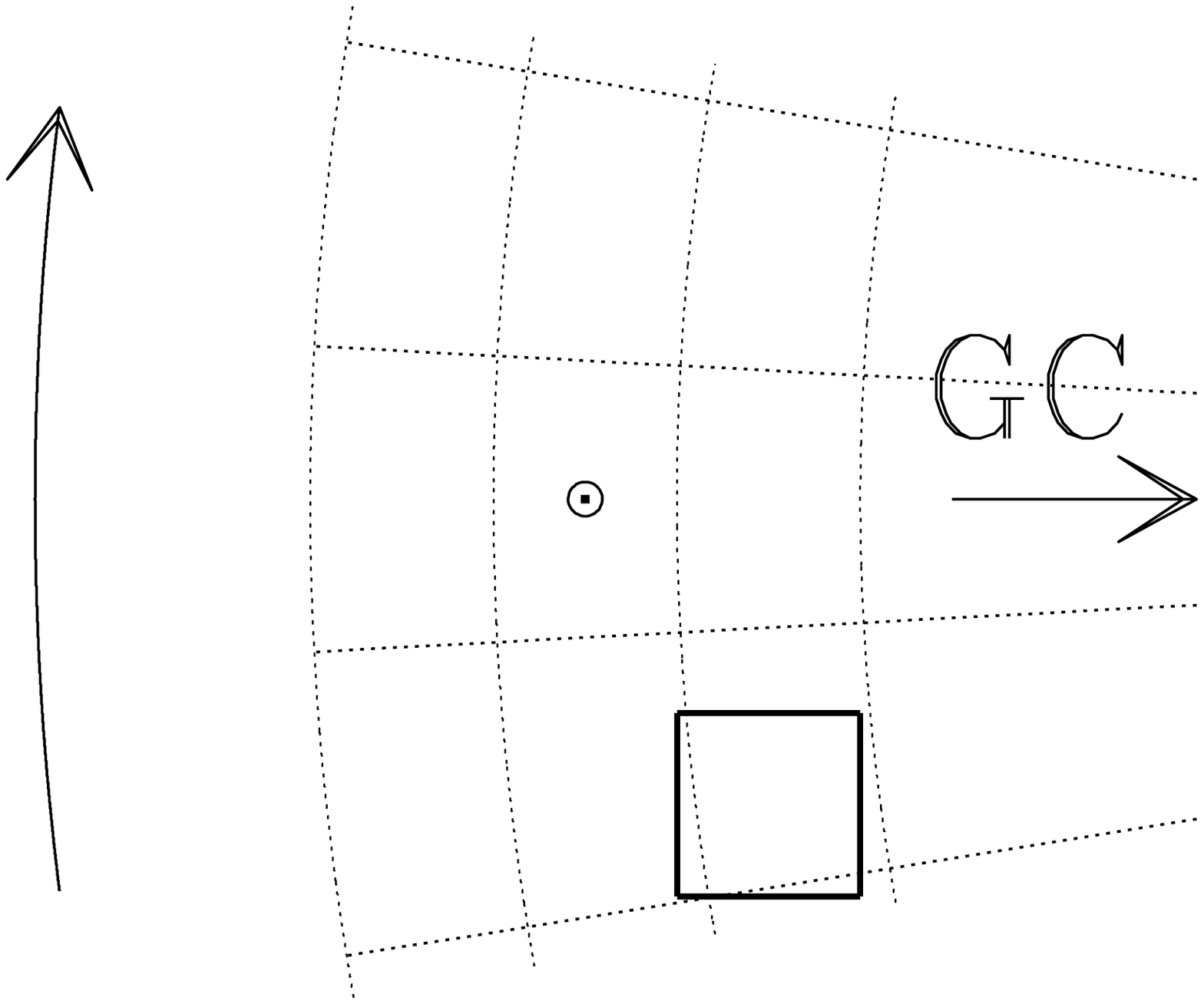}
  }
  \caption{
    Spatial bins used in the plots show in Figures~\ref{fig:RAVE} and
    \ref{fig:RAVEe} (\emph{solid lines}). The \emph{dotted} lines
    run in the $R$ and $\phi$ directions at the same places as in
    Figure~\ref{fig:Rpbins} to indicate scale. 
\label{fig:RAVEbins}
}
\end{figure*}

We can look more closely at the distribution of stars in spatial bins
similar to those used by A12. If we take a Galactocentric
Cartesian coordinate system $(X,Y,Z)$, with the Sun at
$X=R_0=8.5\kpc$, $Y=0$, and with $Y$ increasing in the direction of
Galactic rotation at the Sun, then we can take stars with $-1.3\kpc
<Y<-0.7\kpc$, and in windows in the range $7.6\kpc <X<9.4\kpc$ to
correspond to the areas probed by A12's work
(though they chose to take $R_0=7.8\kpc$). To probe the area in
detail, while still using large spatial bins to maximise the numbers
of stars per bin in a given survey, we consider overlapping bins. Each
bin is $0.6\kpc$ wide in $X$, and we consider bins with limits that
are shifted by $0.2\kpc$ in $X$ from each other. These bins are
illustrated in Figure~\ref{fig:RAVEbins}.

In Figure~\ref{fig:RAVE} we show the distribution of stars in the
$(-v_R) - v_\phi$ space in each of these spatial bins. Again it is
clear that the major difference between the various models is
whereabouts in the Galaxy the Hyades remains a strong feature. Since
these bins are in the anti-rotation direction from the Sun, the OLR
models show a strong Hyades feature in the outer bins, and the ILR
models show it in the inner bins.

Finally we look at the consequences of observational uncertainties on
the structures observed in the $(-v_R) - v_\phi$ plane for the various
models. If we assume an uncertainty of $2\kms$ in line-of-sight
velocity, $1\masyr$ in proper motion, and $20$ per cent in distance
(approximately the uncertainty on RAVE observations), and apply this
to our models, we find the ``observed'' distribution in the
$(-v_R)-v_\phi$ plane shown in Figure~\ref{fig:RAVEe}. The most
obvious consequence is that the entire velocity distribution gets
stretched in the direction associated with the uncertainty in
transverse velocity. This is because the dominant velocity uncertainty
is that associated with the distance uncertainty. This uncertainty
also weakens the signature of the Hyades in each case. The differences
between the two models is discernible, but it is seriously faded. It
is worth noting that the signatures of the ILR models are visible in
the innermost spatial bin, but those of the OLR models are not in the
outermost.

\subsection{Comparison to observations}

\begin{figure*}
  \centerline{
    \includegraphics[height=.125\hsize,angle=270]{HyadesPlots/blank.eps}
    \includegraphics[height=.125\hsize,angle=270]{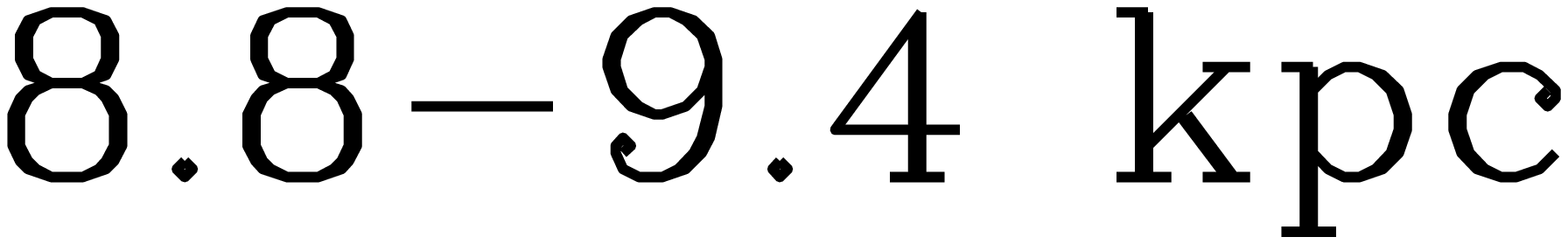}
    \includegraphics[height=.125\hsize,angle=270]{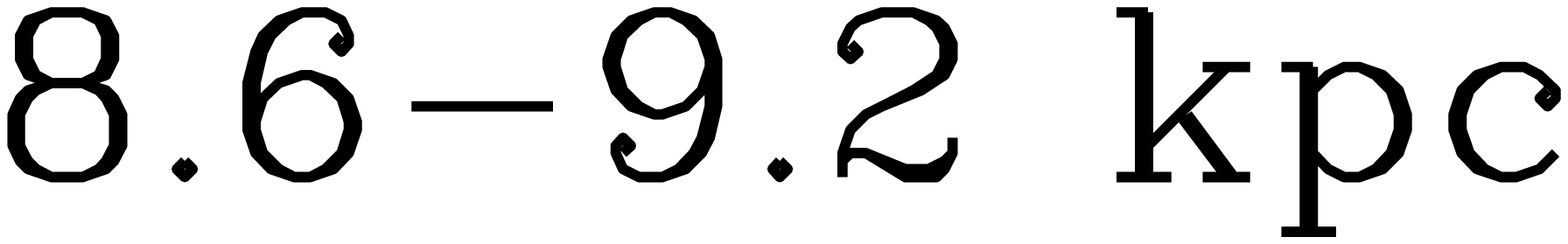}
    \includegraphics[height=.125\hsize,angle=270]{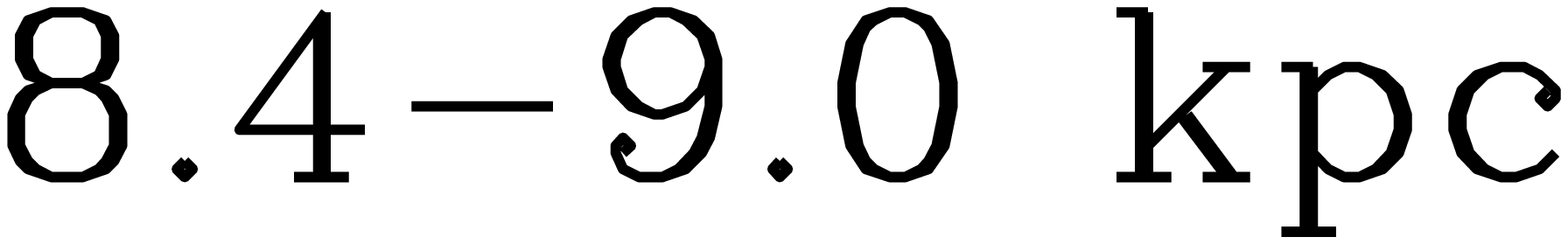}
    \includegraphics[height=.125\hsize,angle=270]{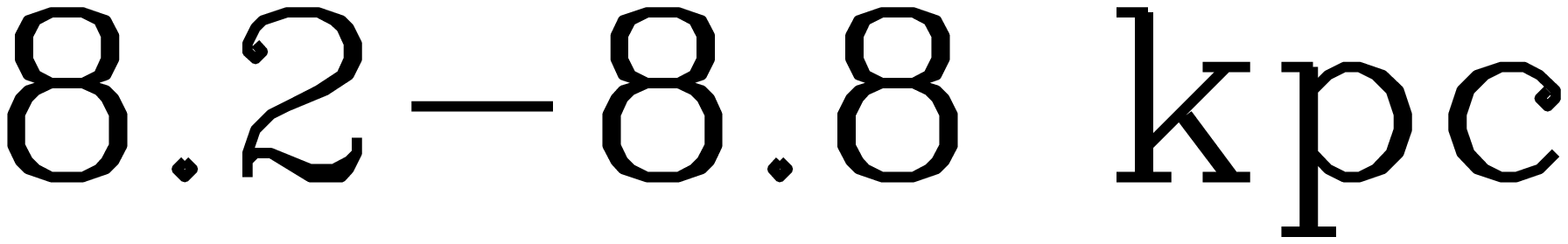}
    \includegraphics[height=.125\hsize,angle=270]{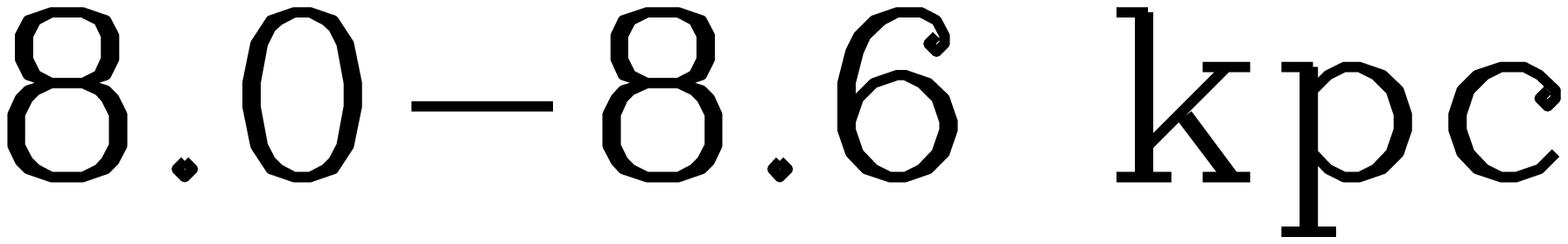}
    \includegraphics[height=.125\hsize,angle=270]{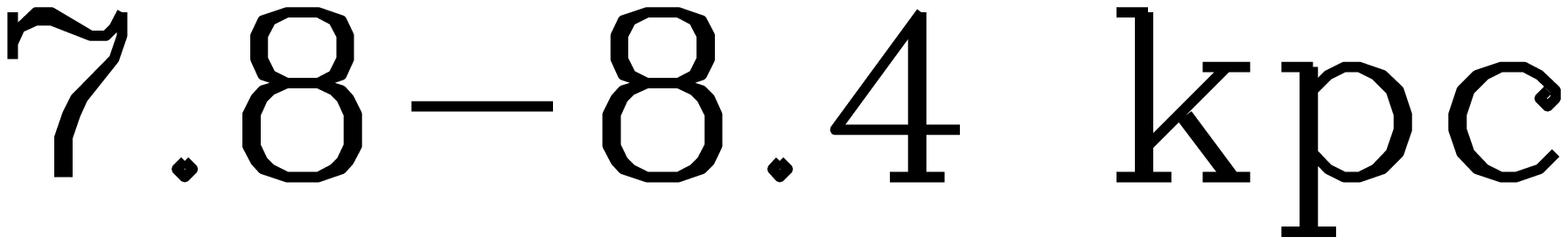}
    \includegraphics[height=.125\hsize,angle=270]{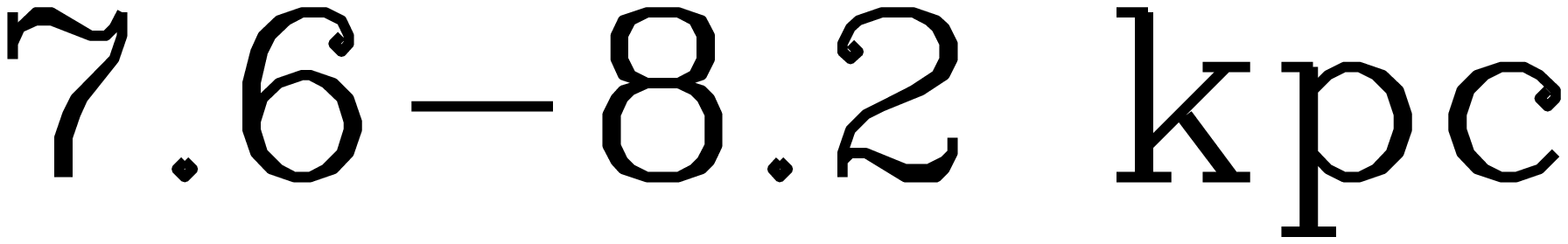}
  } \centerline{
    \includegraphics[width=.125\hsize]{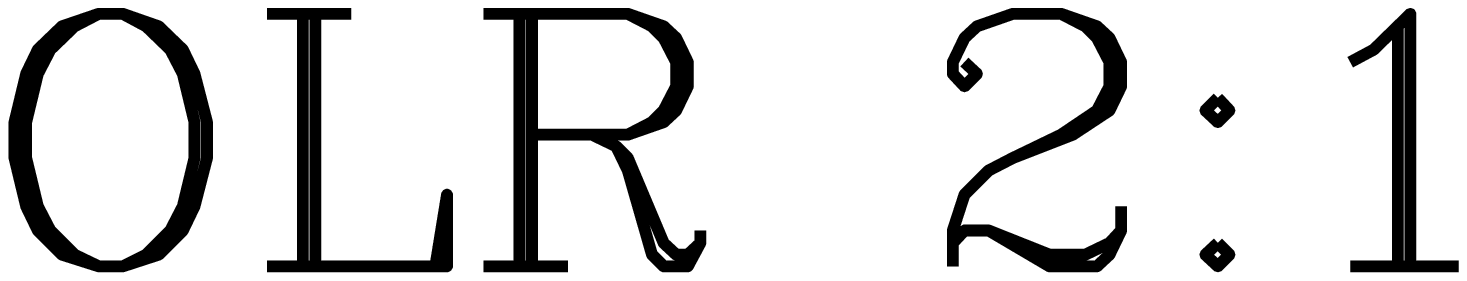}
    \includegraphics[width=.125\hsize]{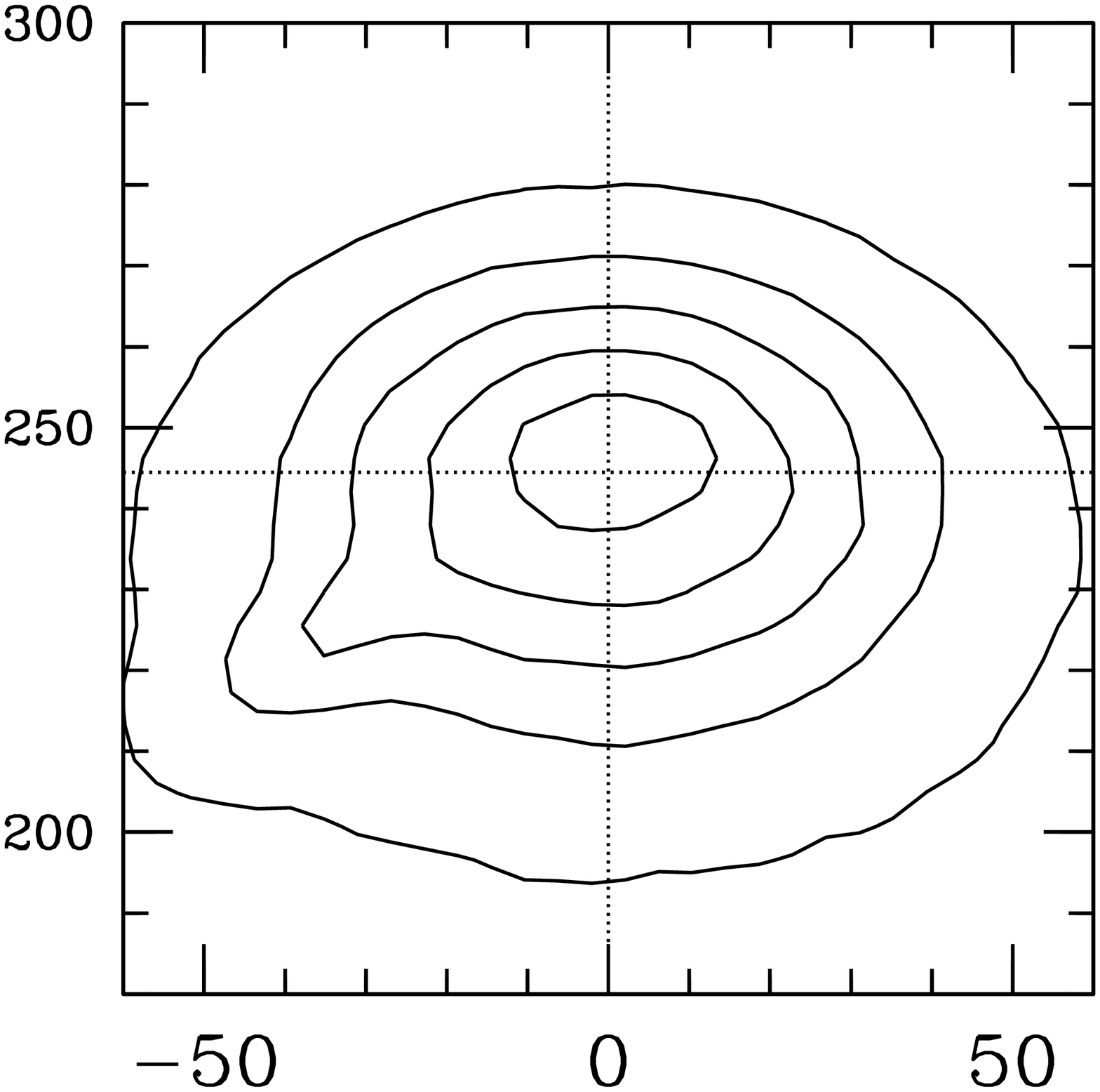}
    \includegraphics[width=.125\hsize]{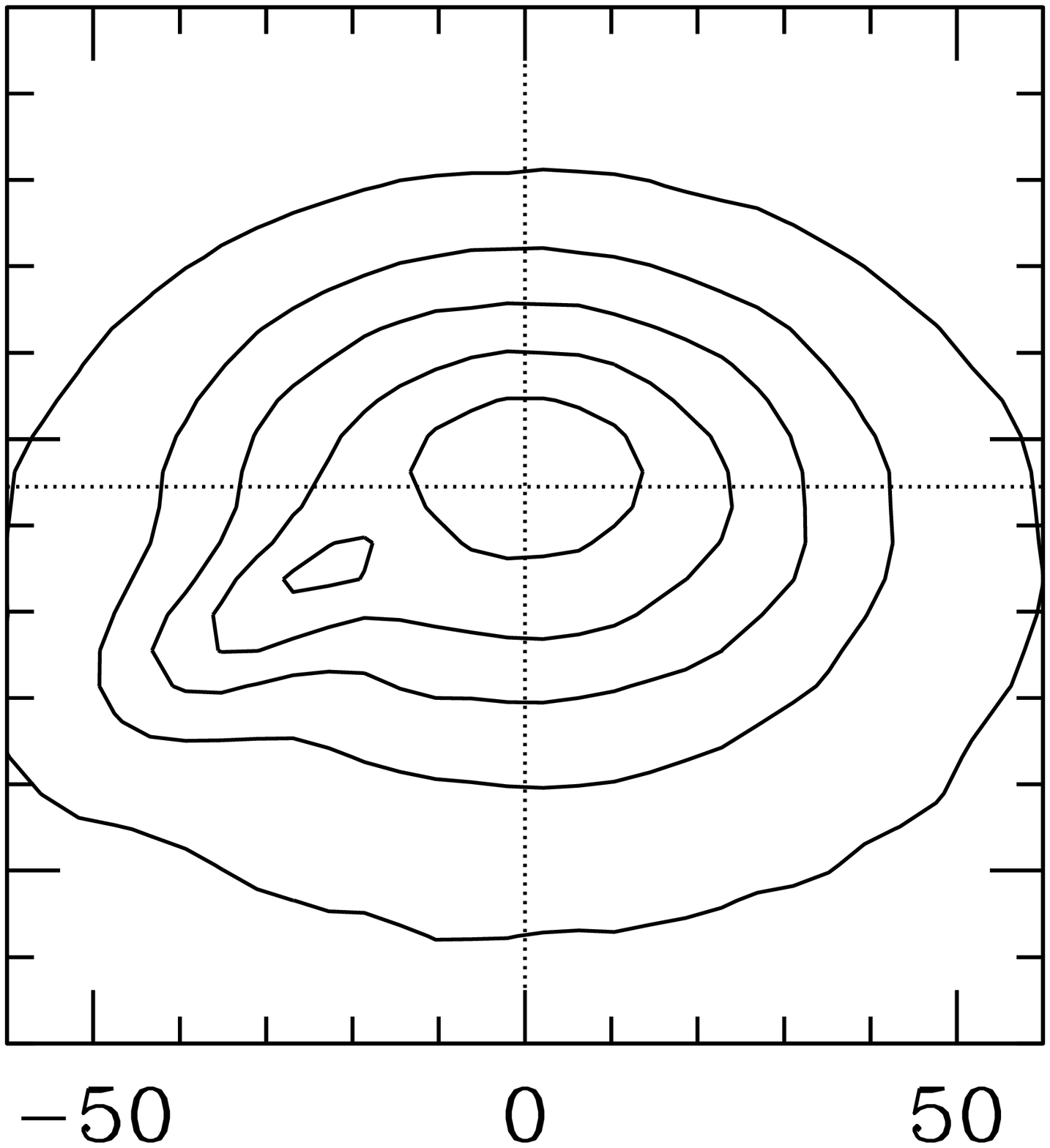}
    \includegraphics[width=.125\hsize]{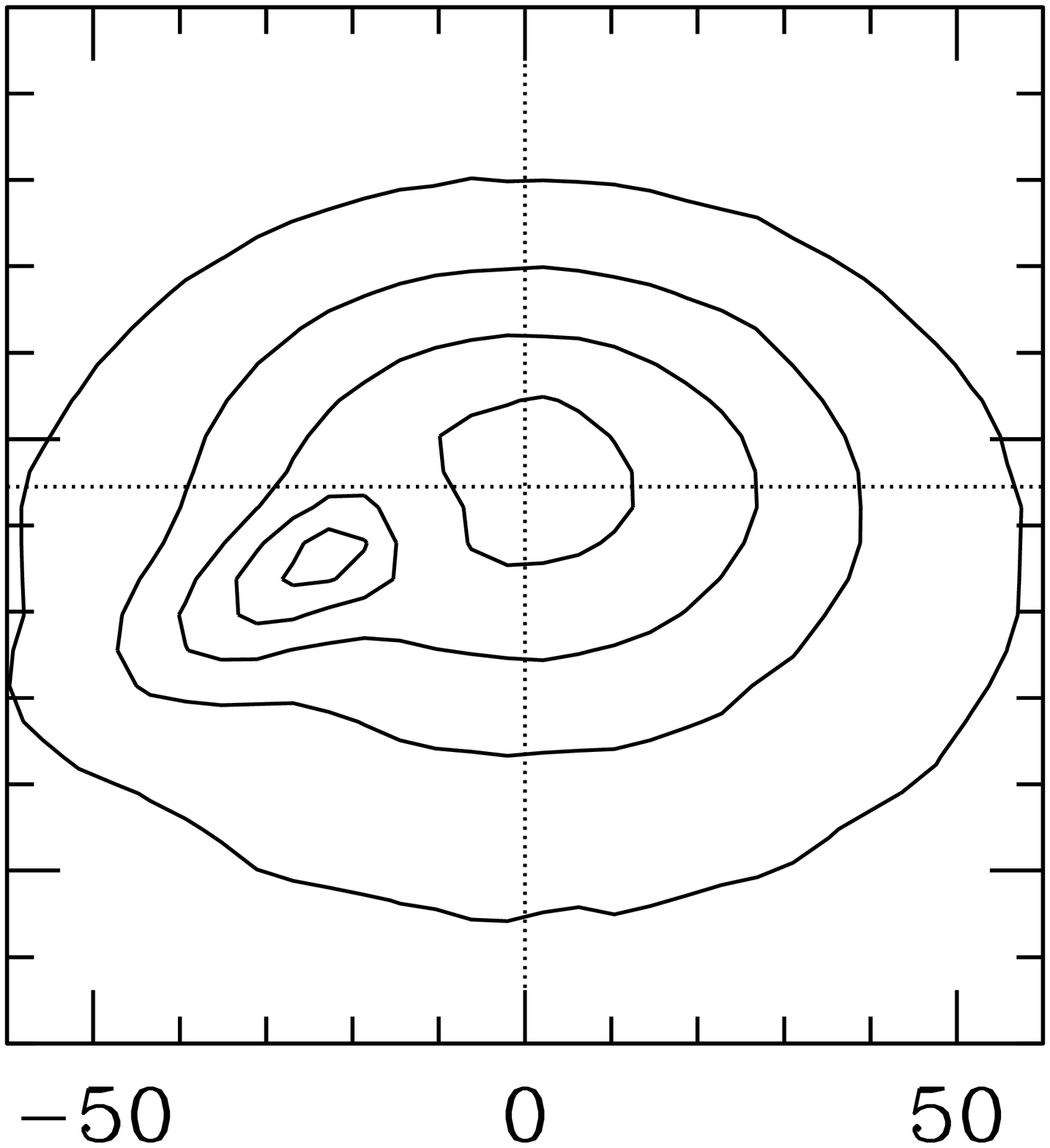}
    \includegraphics[width=.125\hsize]{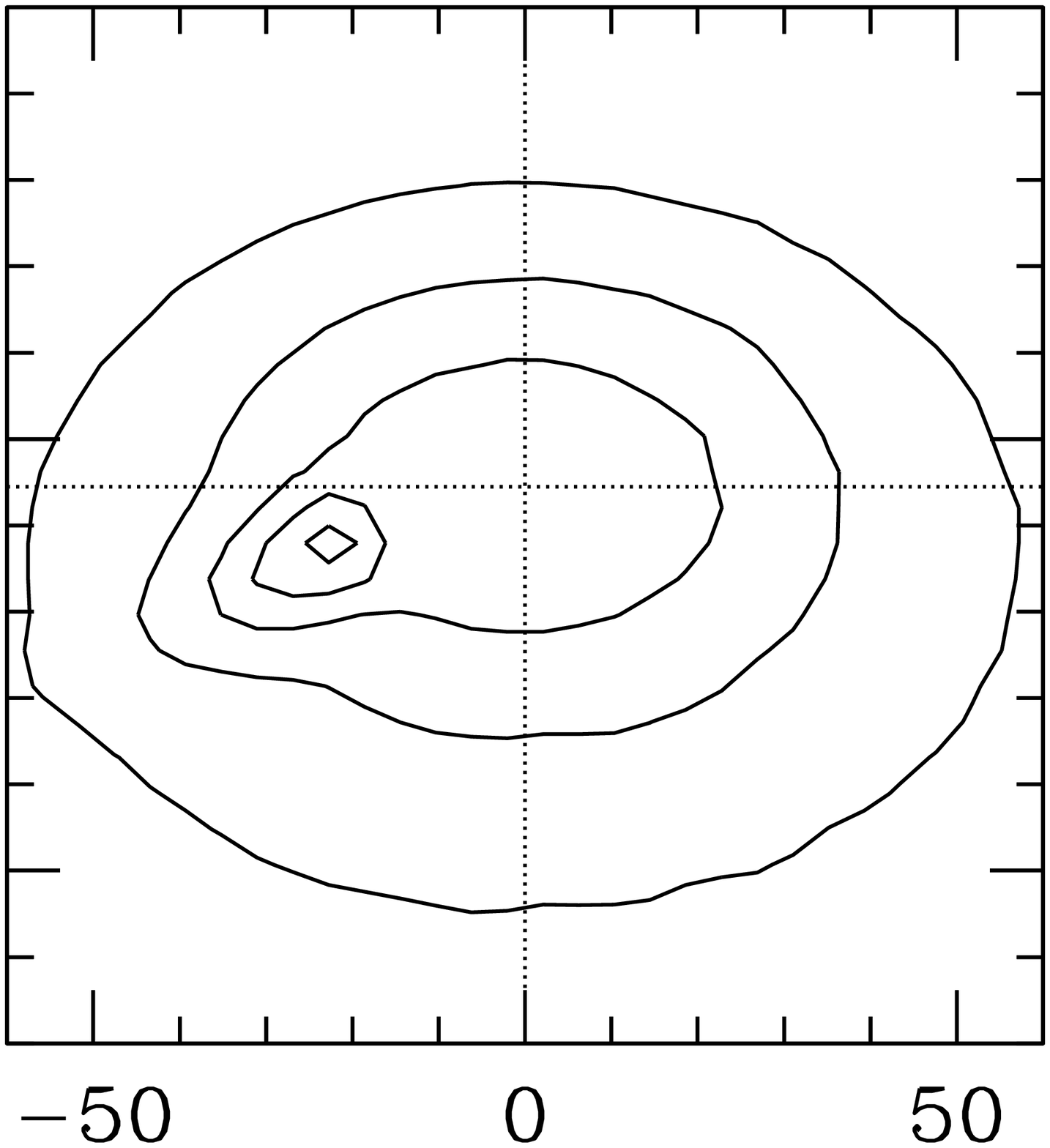}
    \includegraphics[width=.125\hsize]{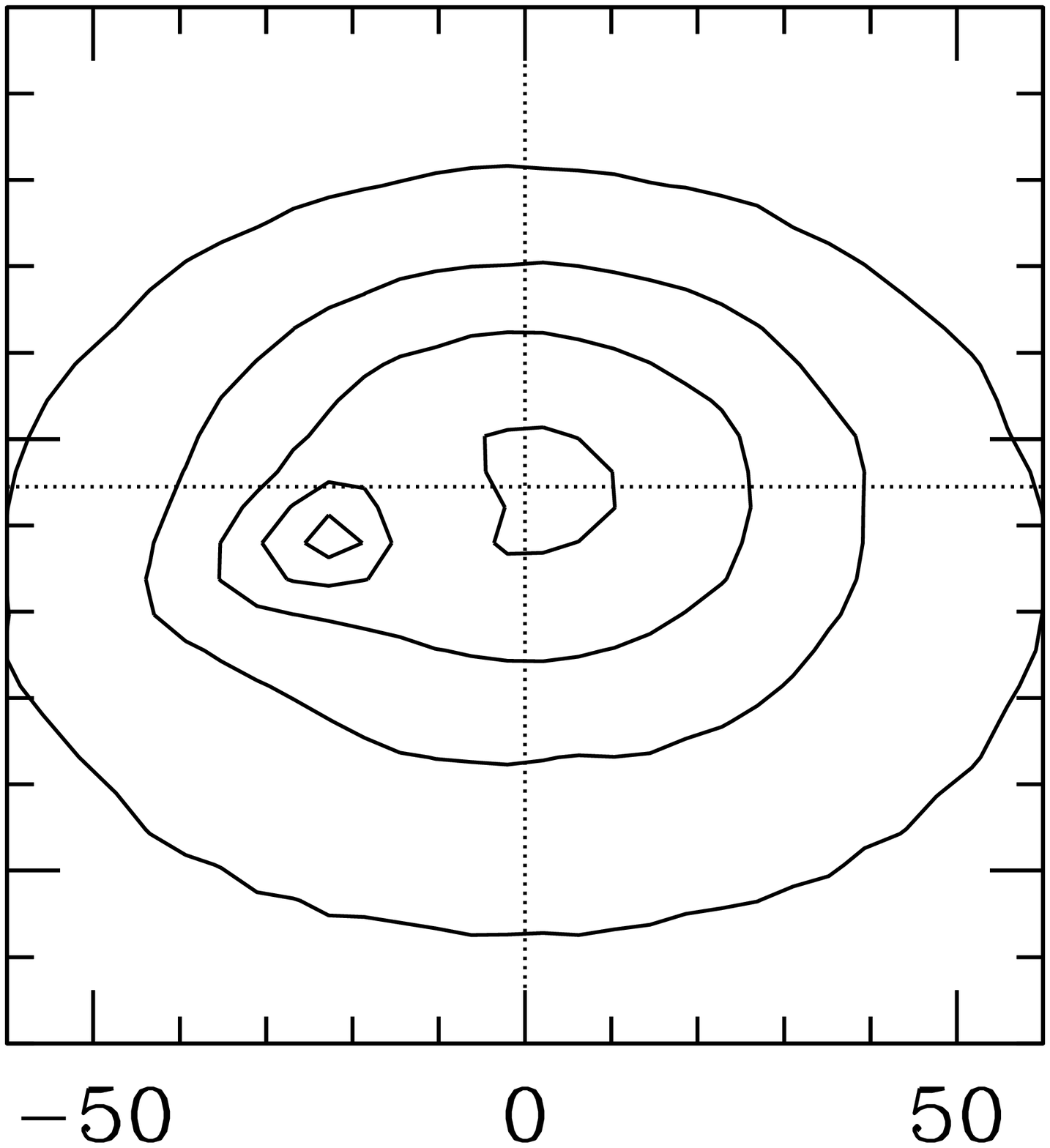}
    \includegraphics[width=.125\hsize]{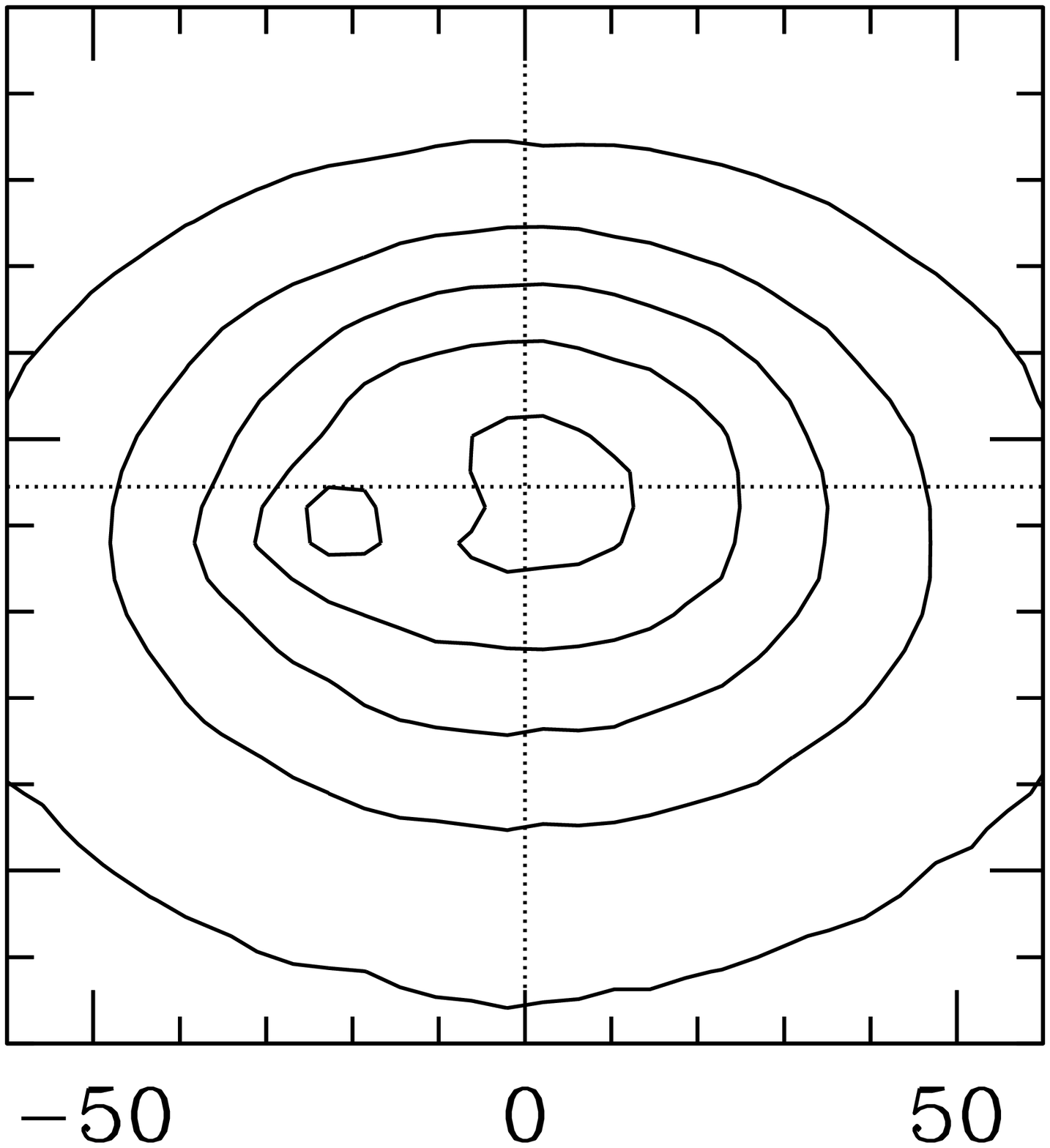}
    \includegraphics[width=.125\hsize]{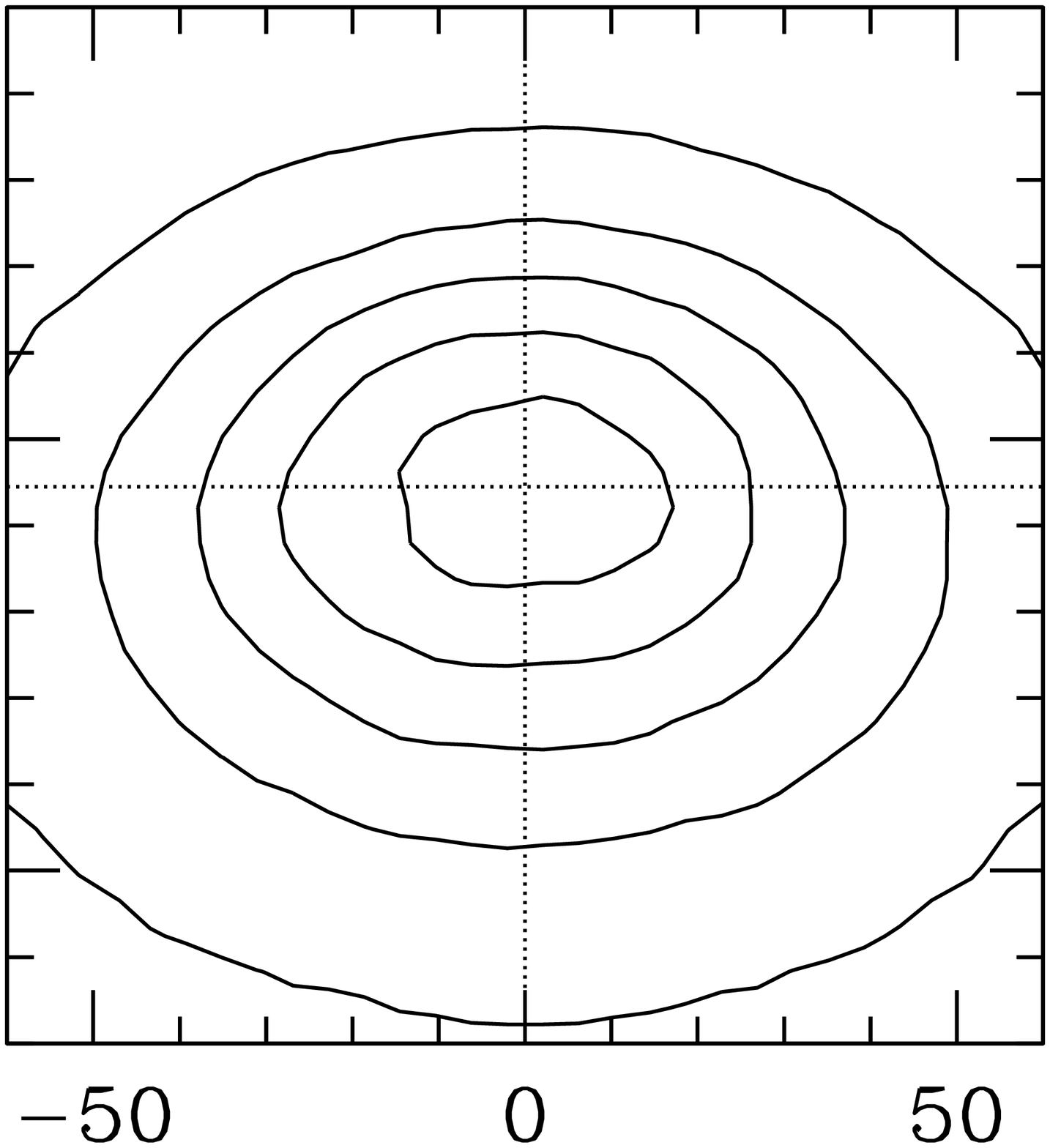}
  } \centerline{
    \includegraphics[width=.125\hsize]{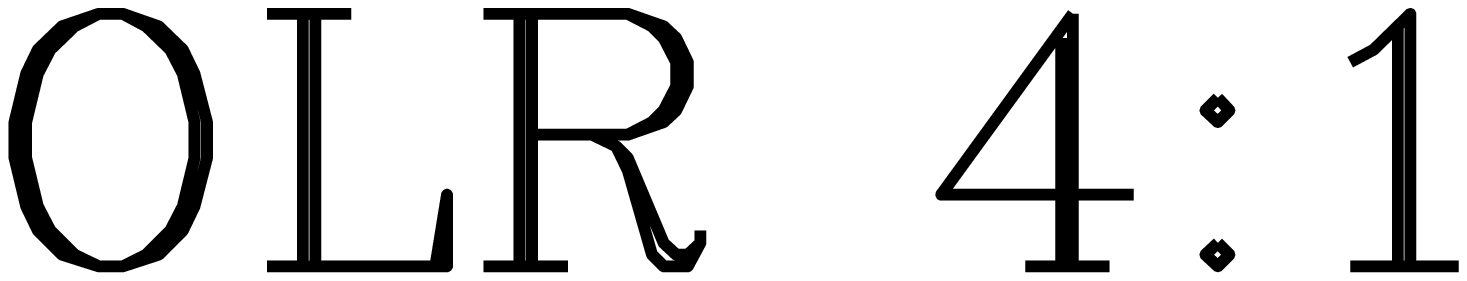}
    \includegraphics[width=.125\hsize]{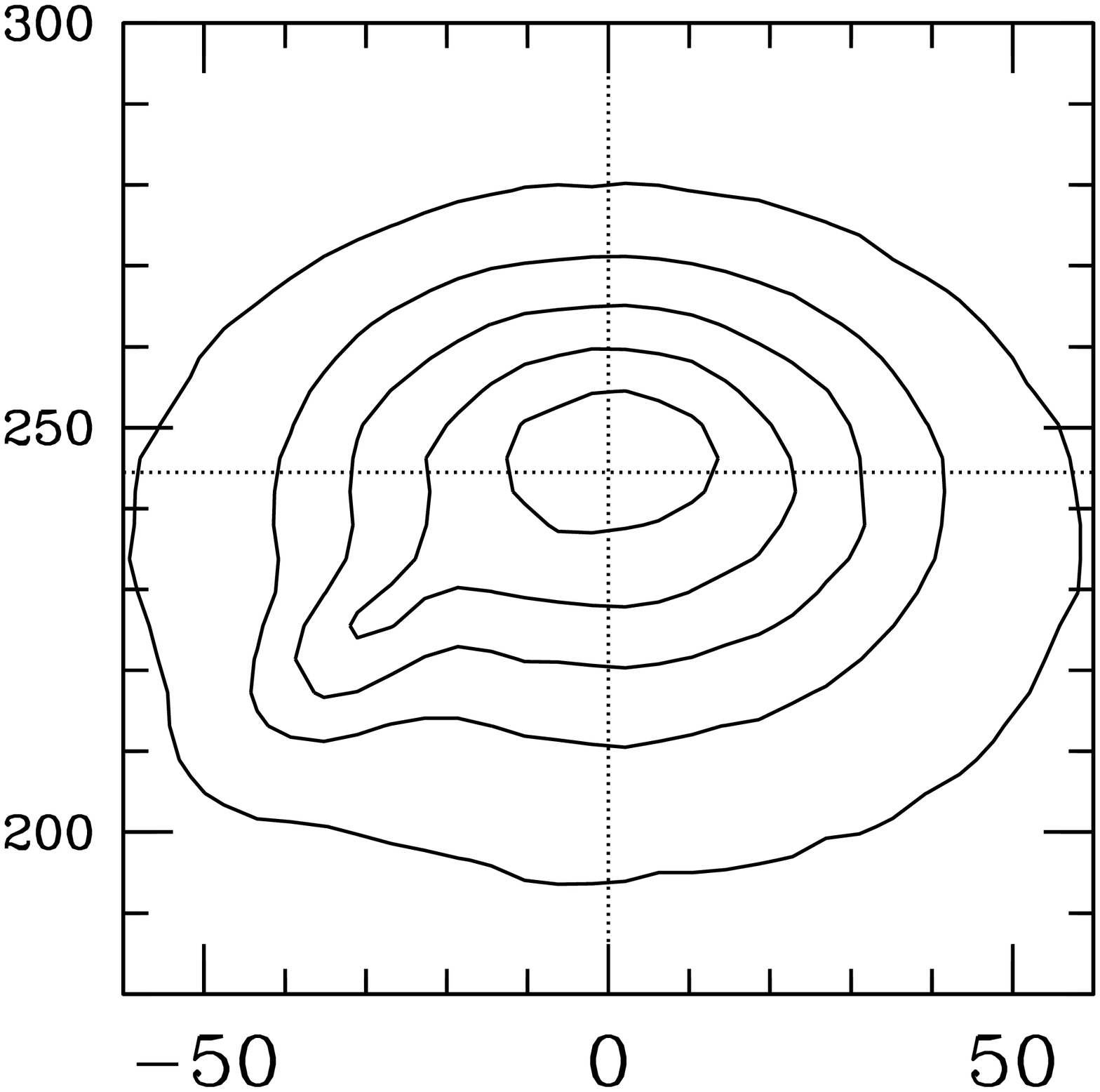}
    \includegraphics[width=.125\hsize]{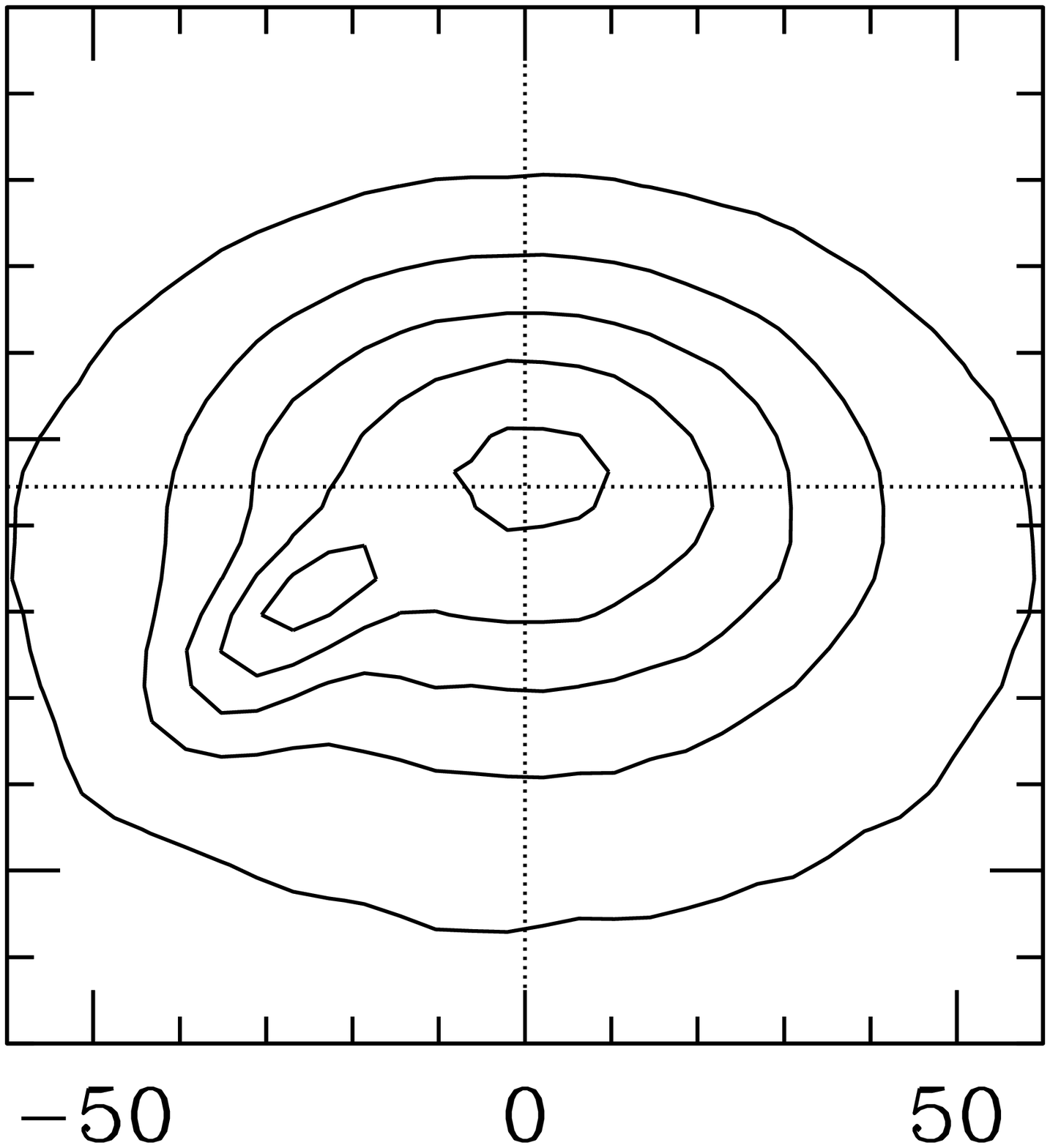}
    \includegraphics[width=.125\hsize]{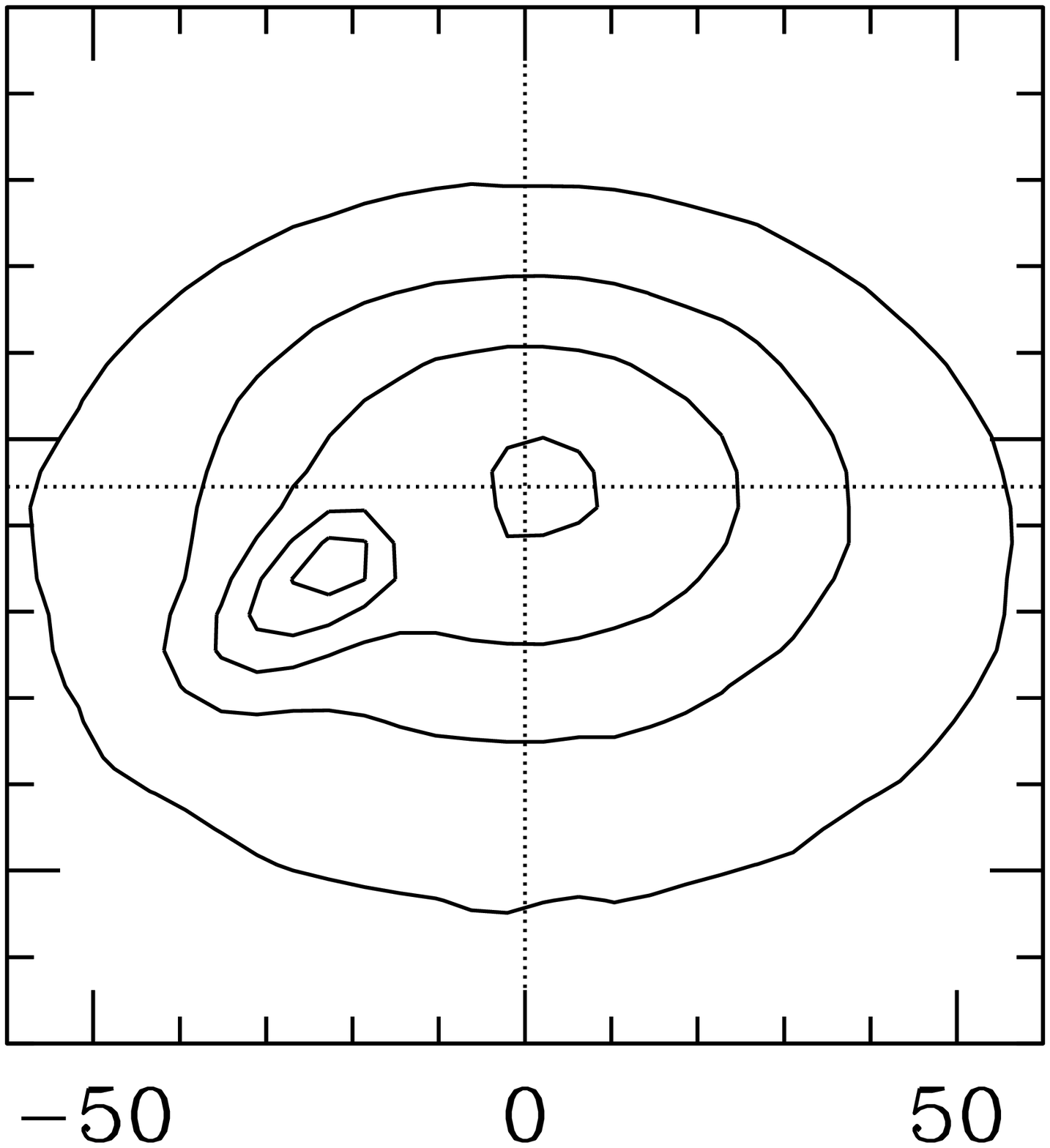}
    \includegraphics[width=.125\hsize]{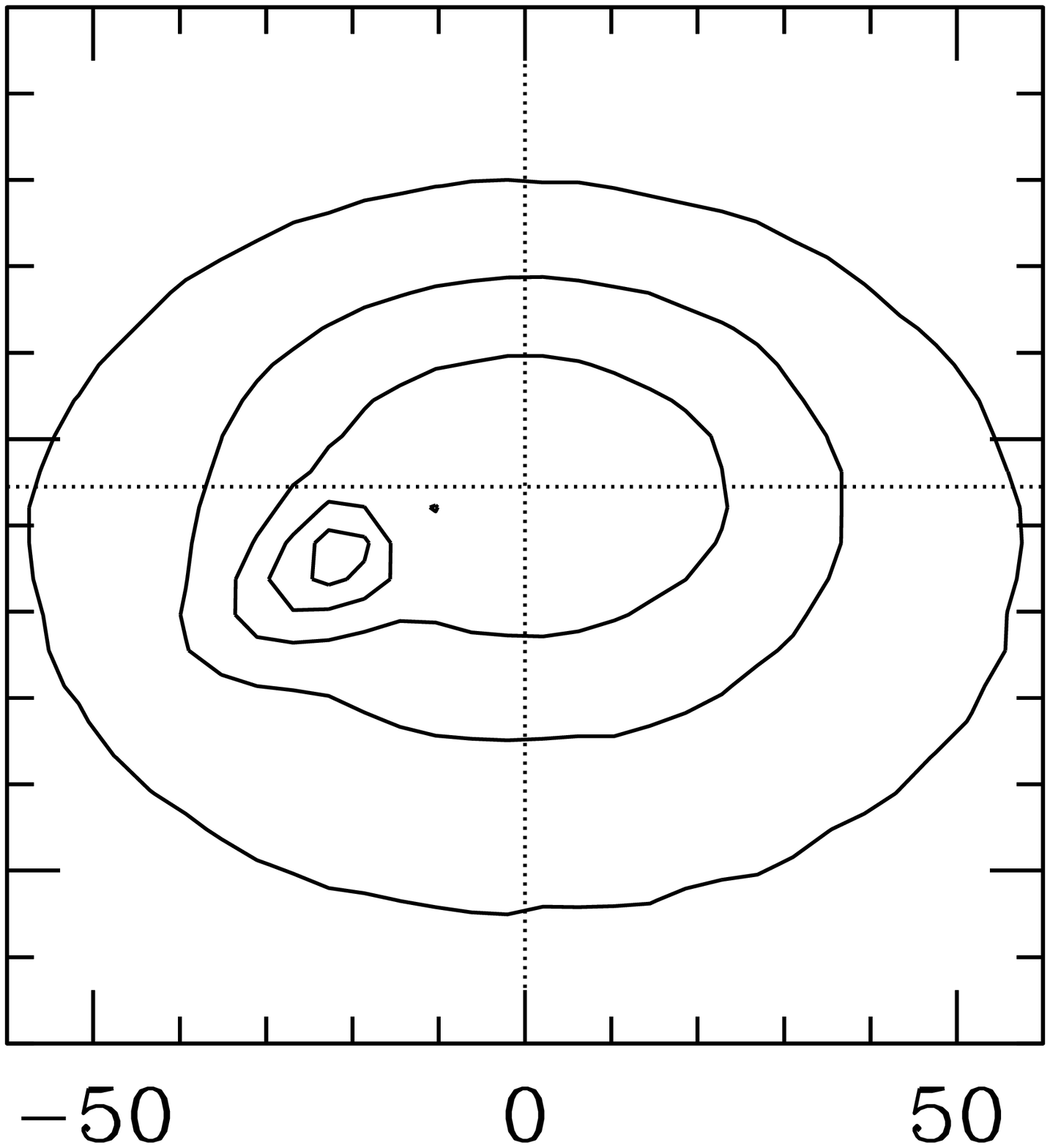}
    \includegraphics[width=.125\hsize]{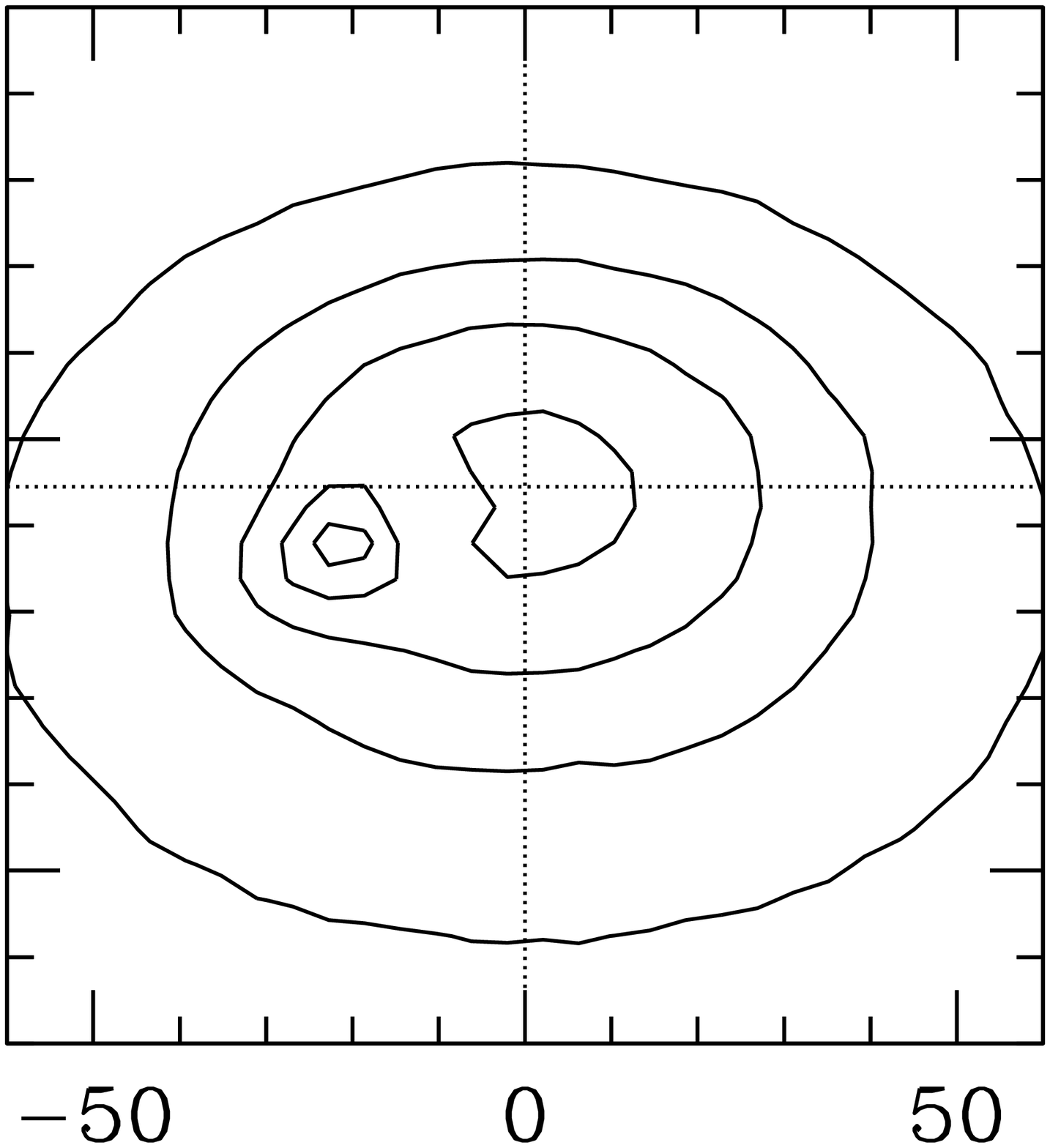}
    \includegraphics[width=.125\hsize]{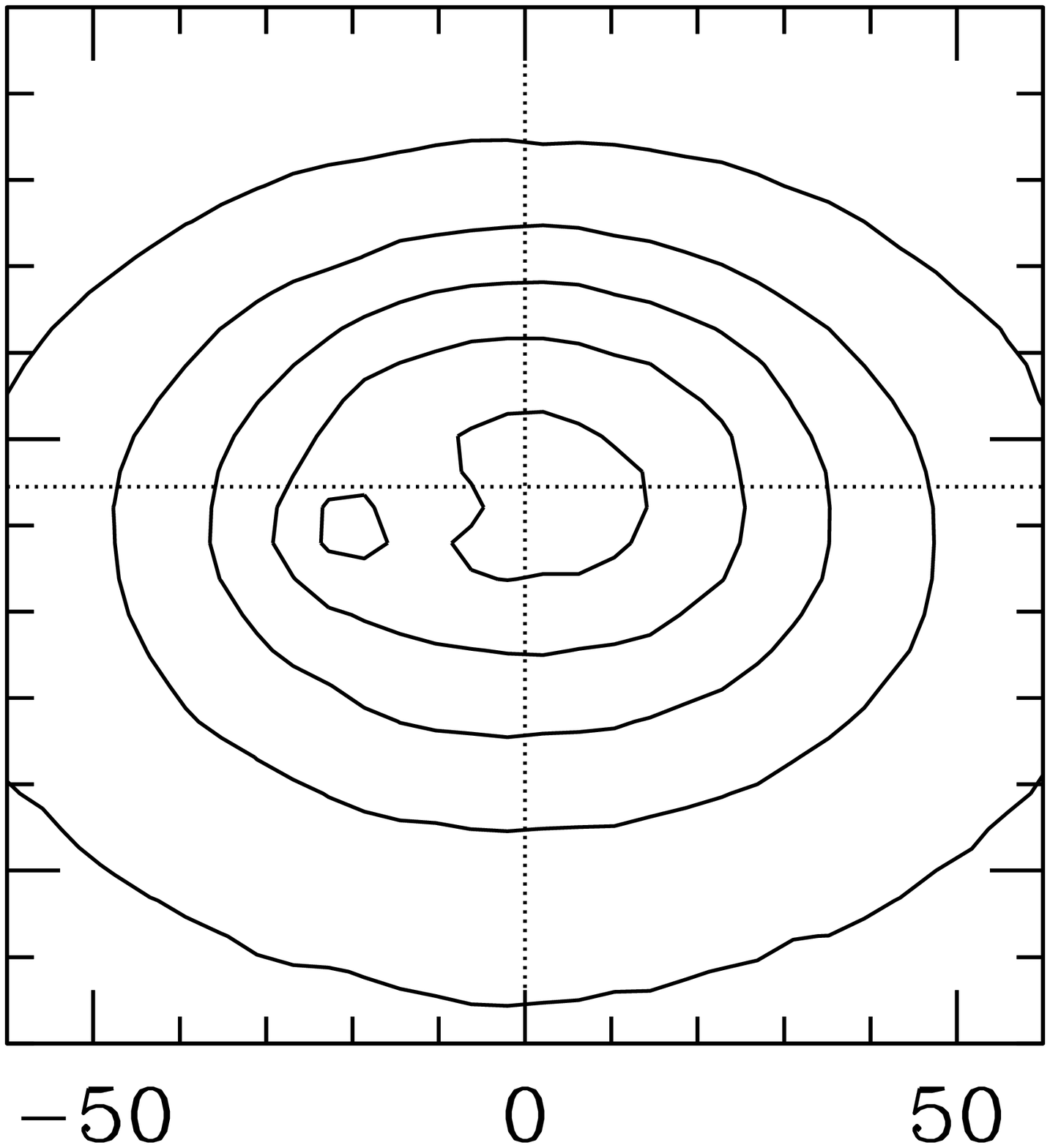}
    \includegraphics[width=.125\hsize]{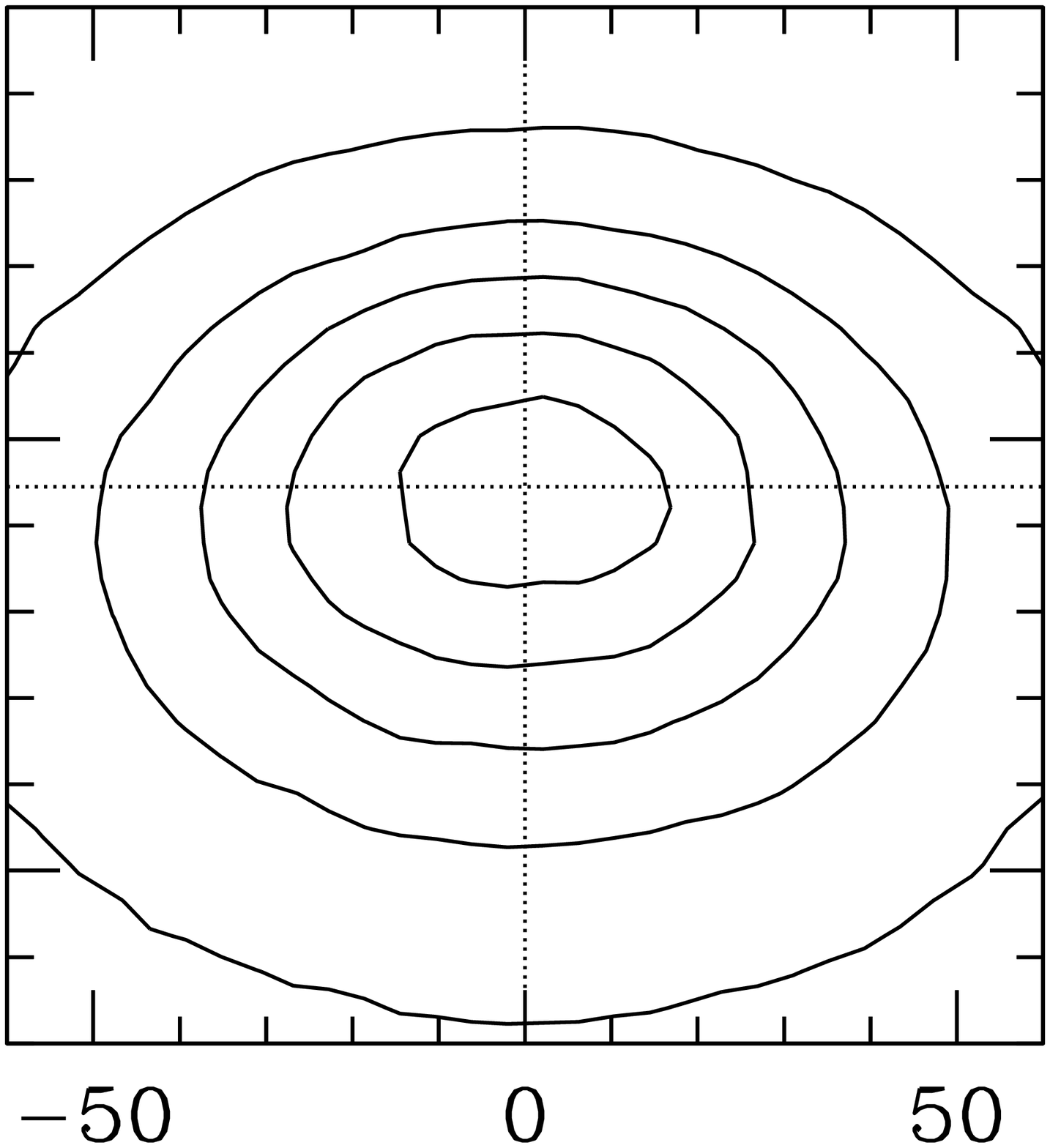}
  } \centerline{
    \includegraphics[width=.125\hsize]{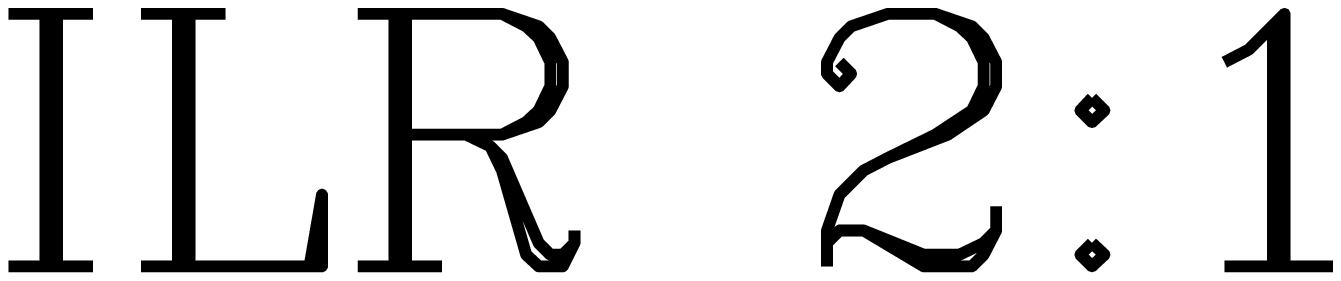}
    \includegraphics[width=.125\hsize]{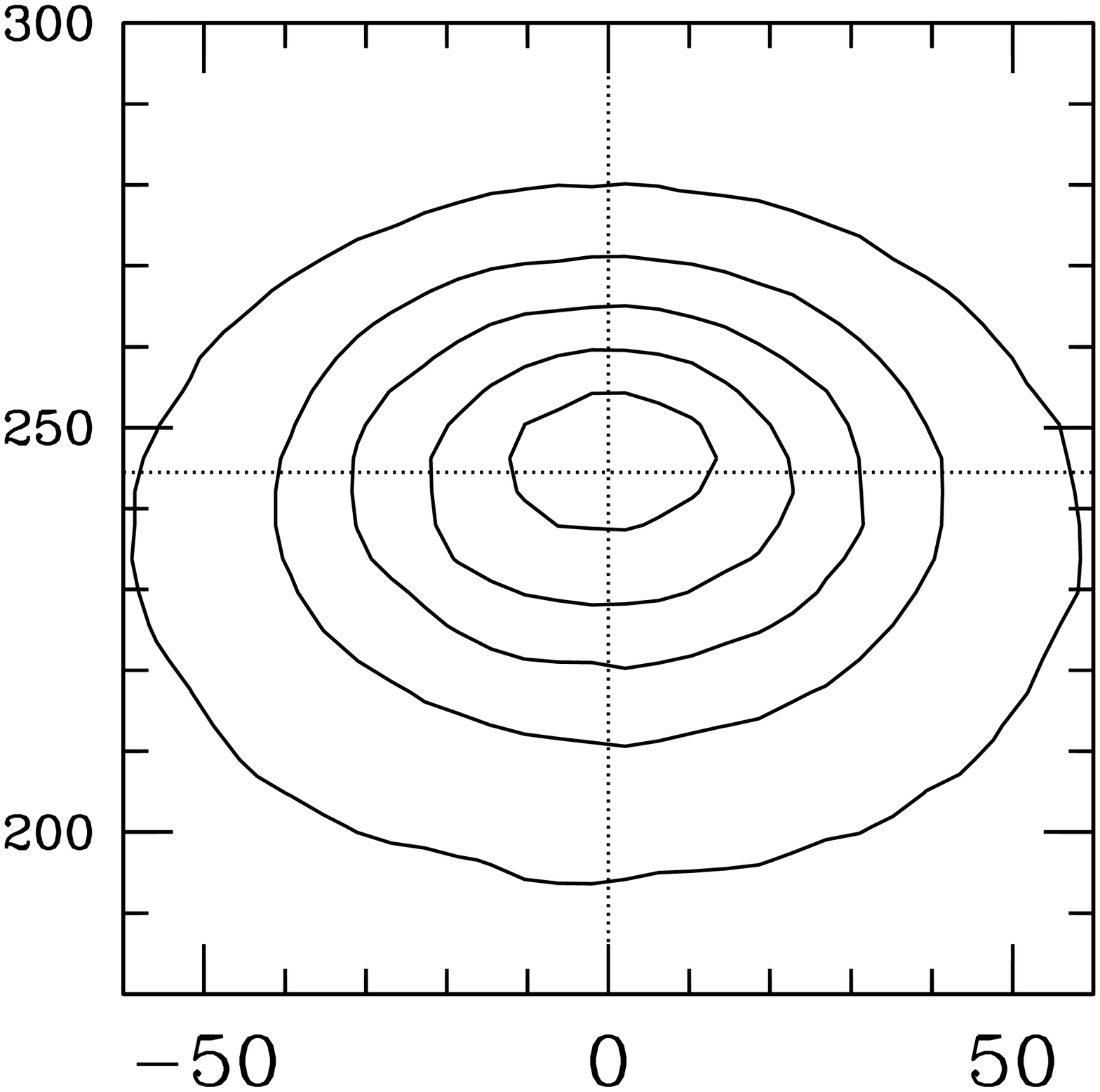}
    \includegraphics[width=.125\hsize]{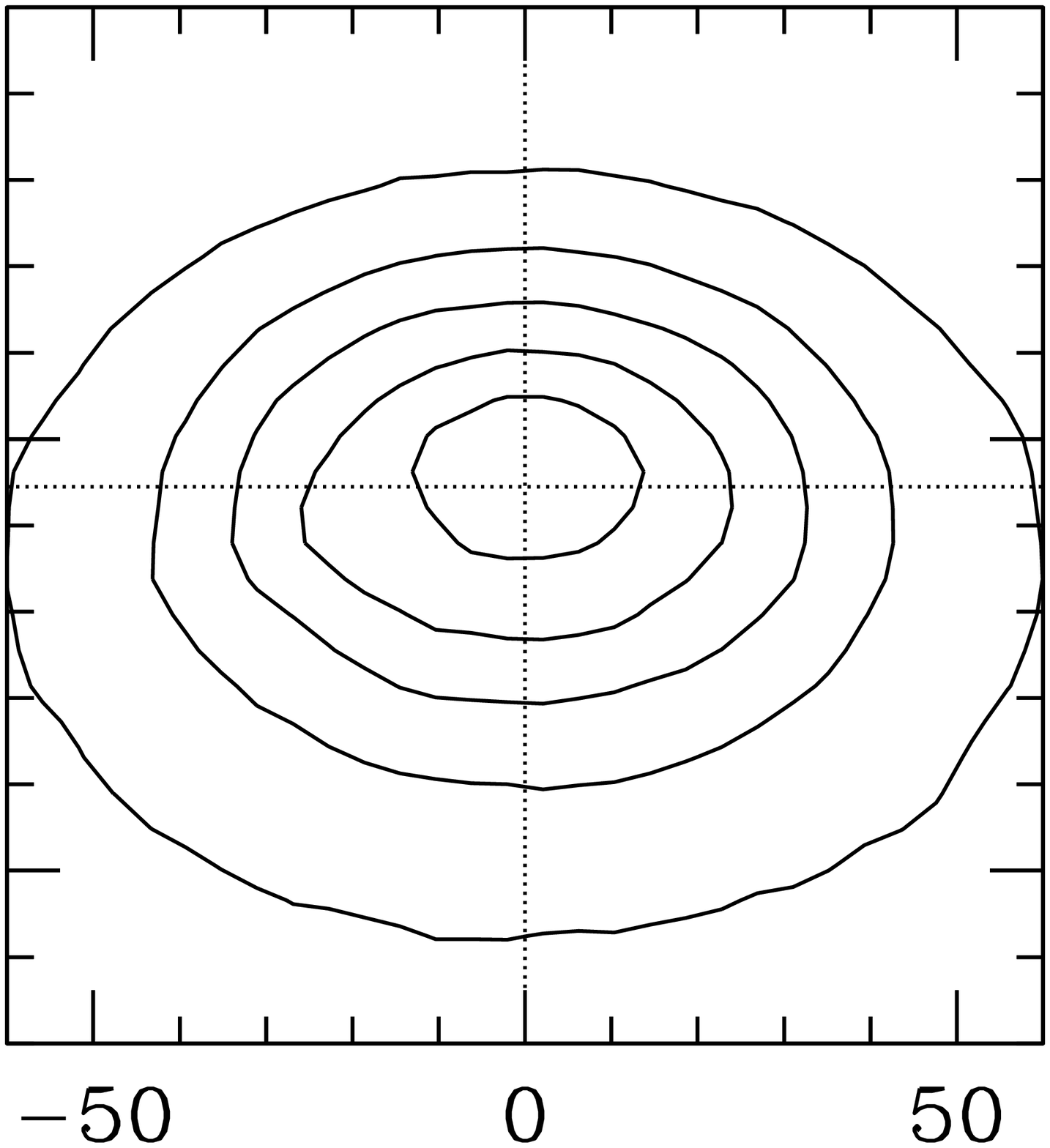}
    \includegraphics[width=.125\hsize]{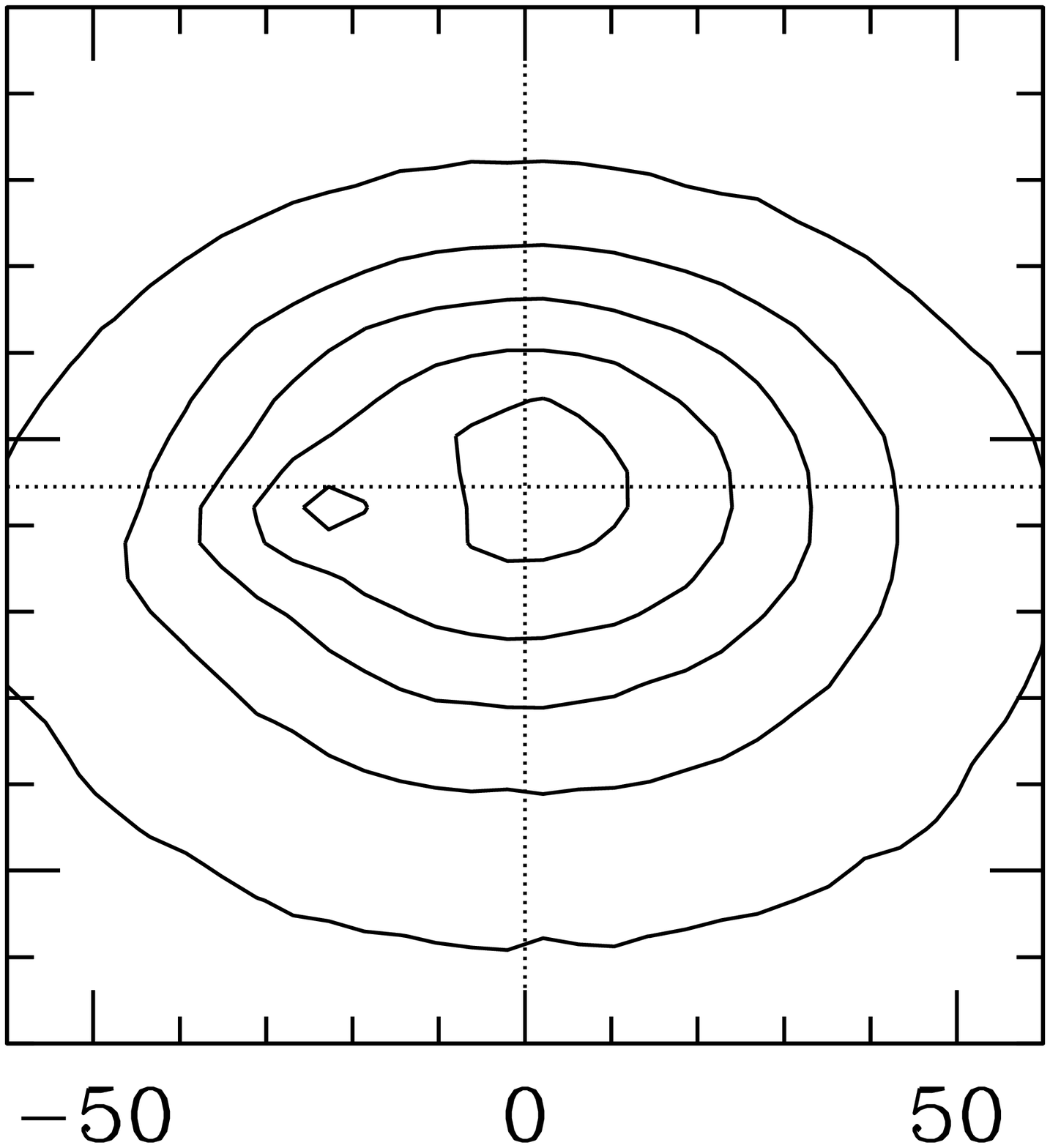}
    \includegraphics[width=.125\hsize]{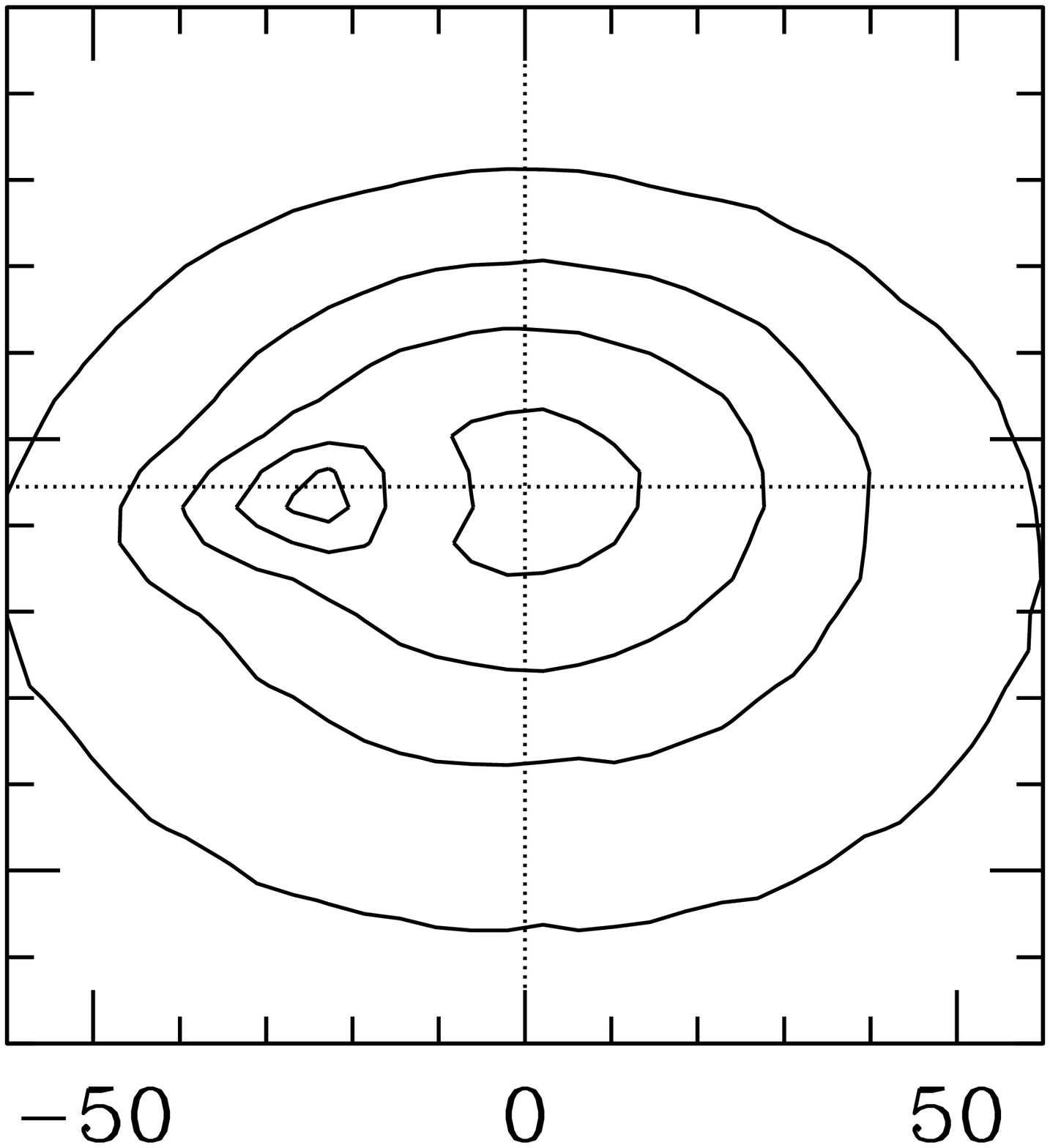}
    \includegraphics[width=.125\hsize]{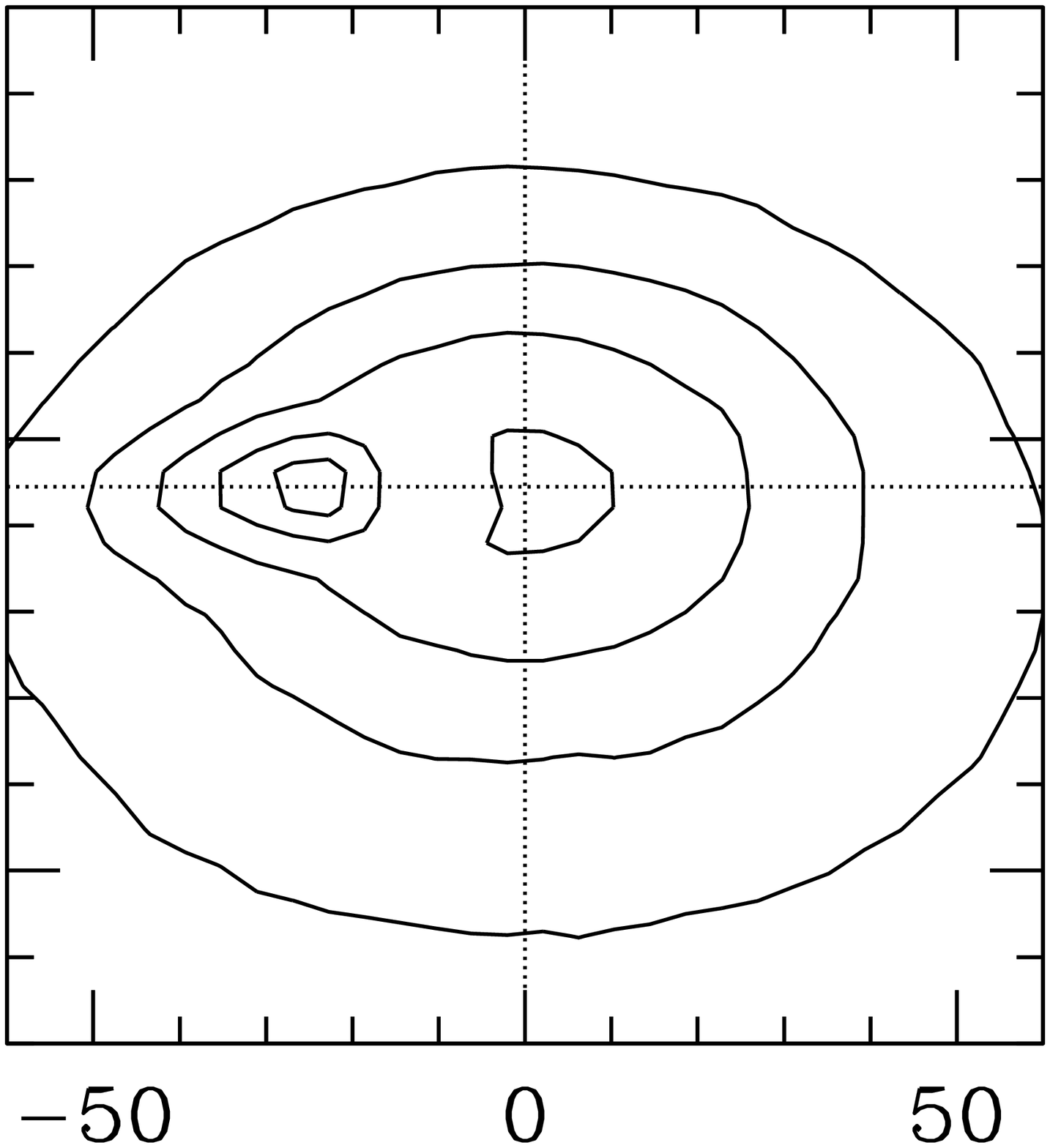}
    \includegraphics[width=.125\hsize]{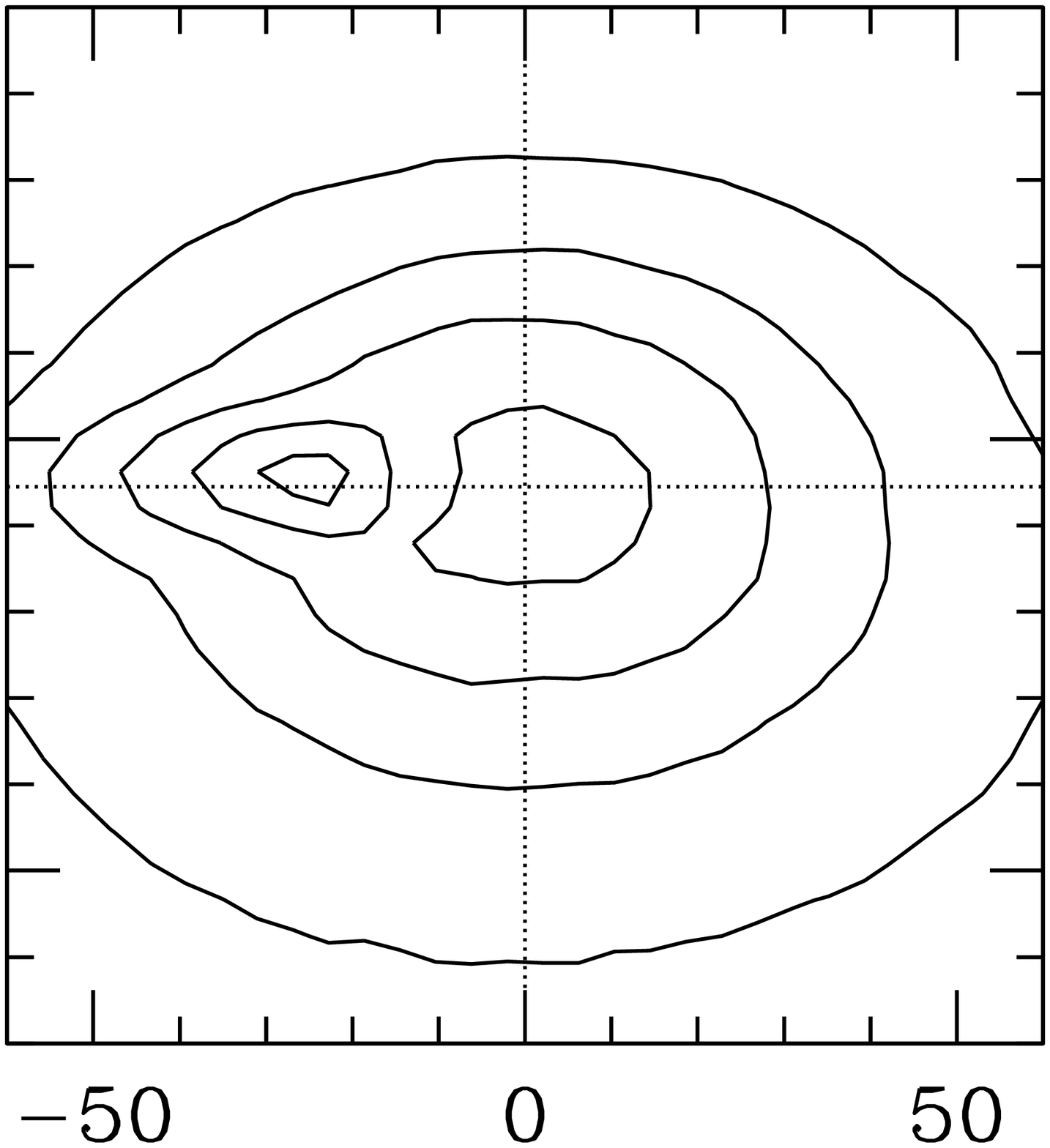}
    \includegraphics[width=.125\hsize]{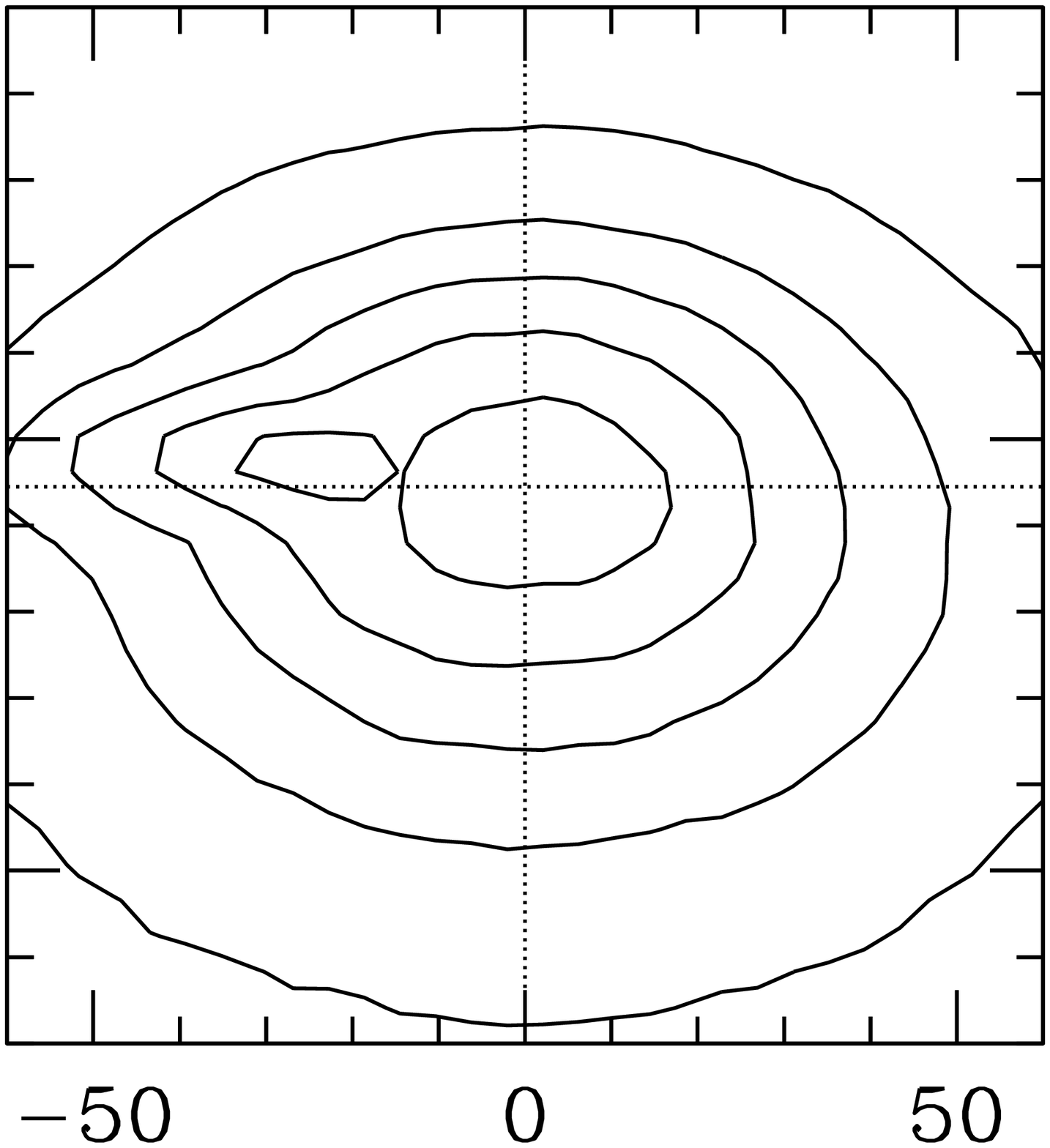}
  } \centerline{
    \includegraphics[width=.125\hsize]{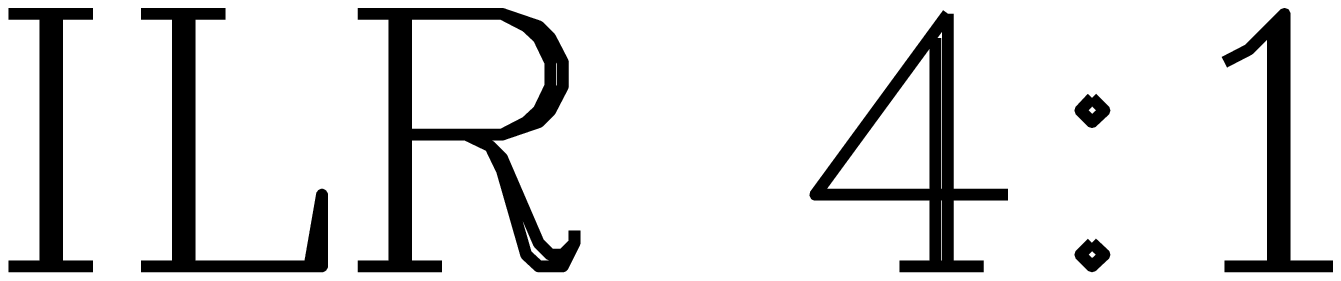}
    \includegraphics[width=.125\hsize]{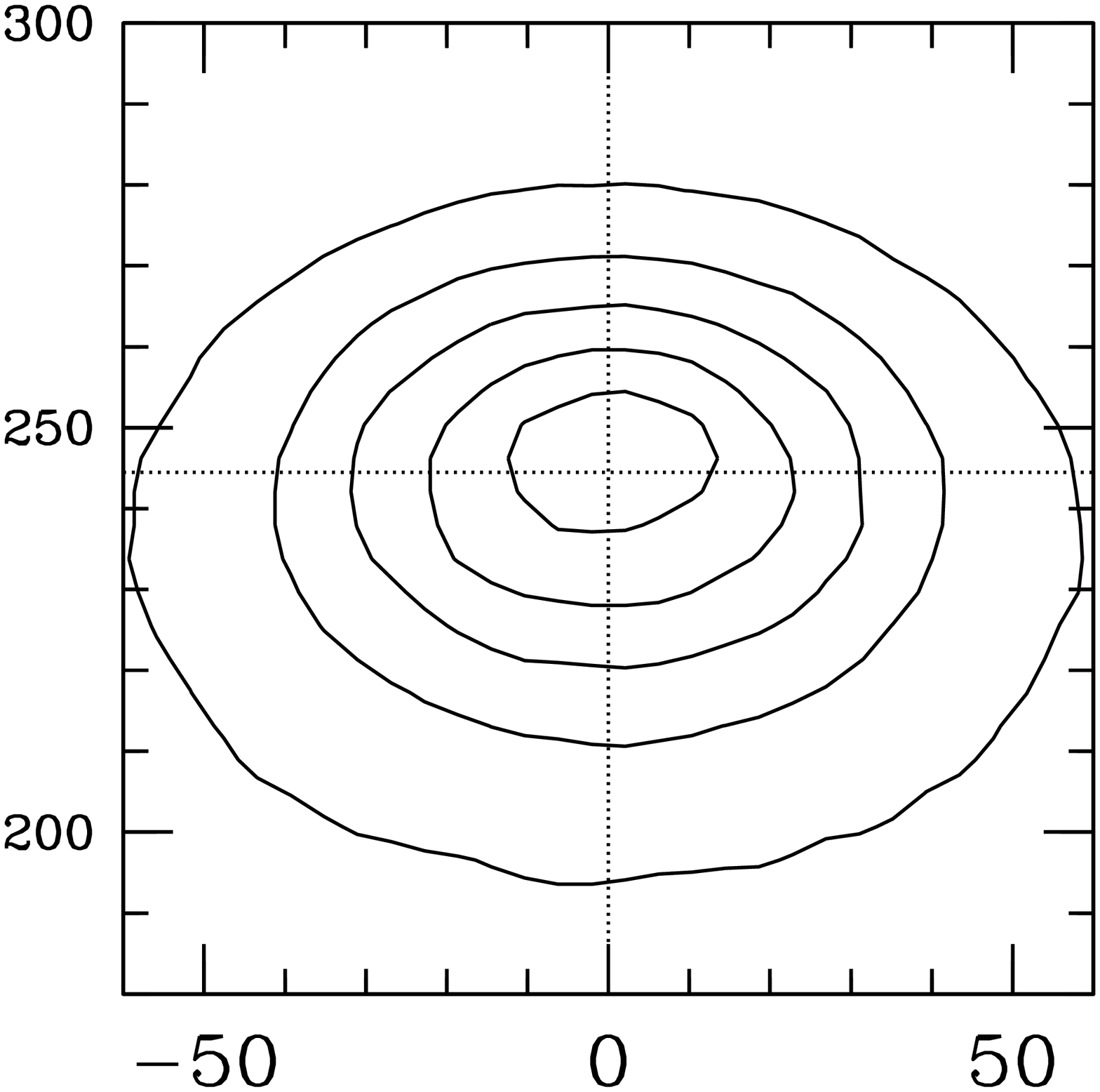}
    \includegraphics[width=.125\hsize]{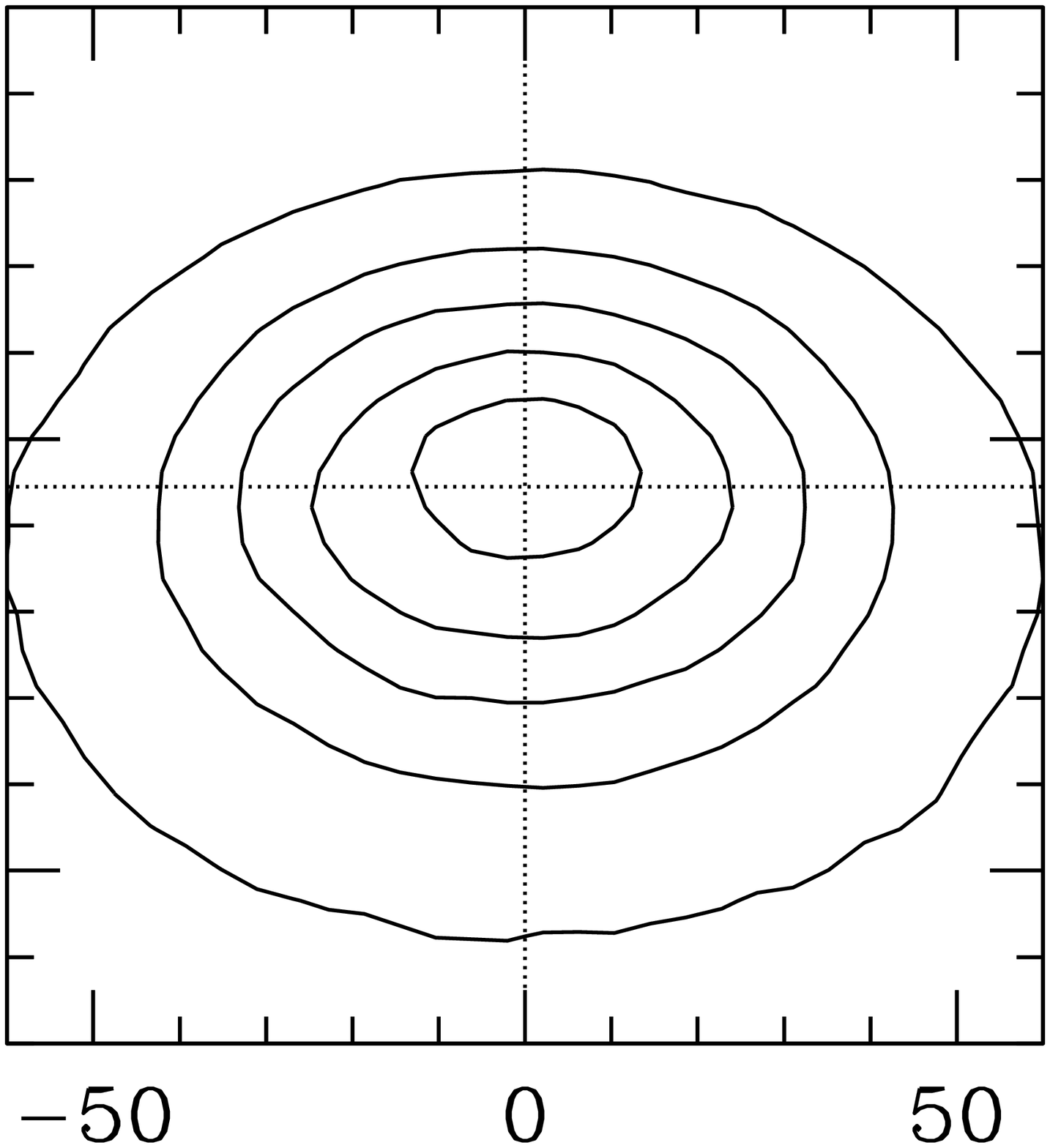}
    \includegraphics[width=.125\hsize]{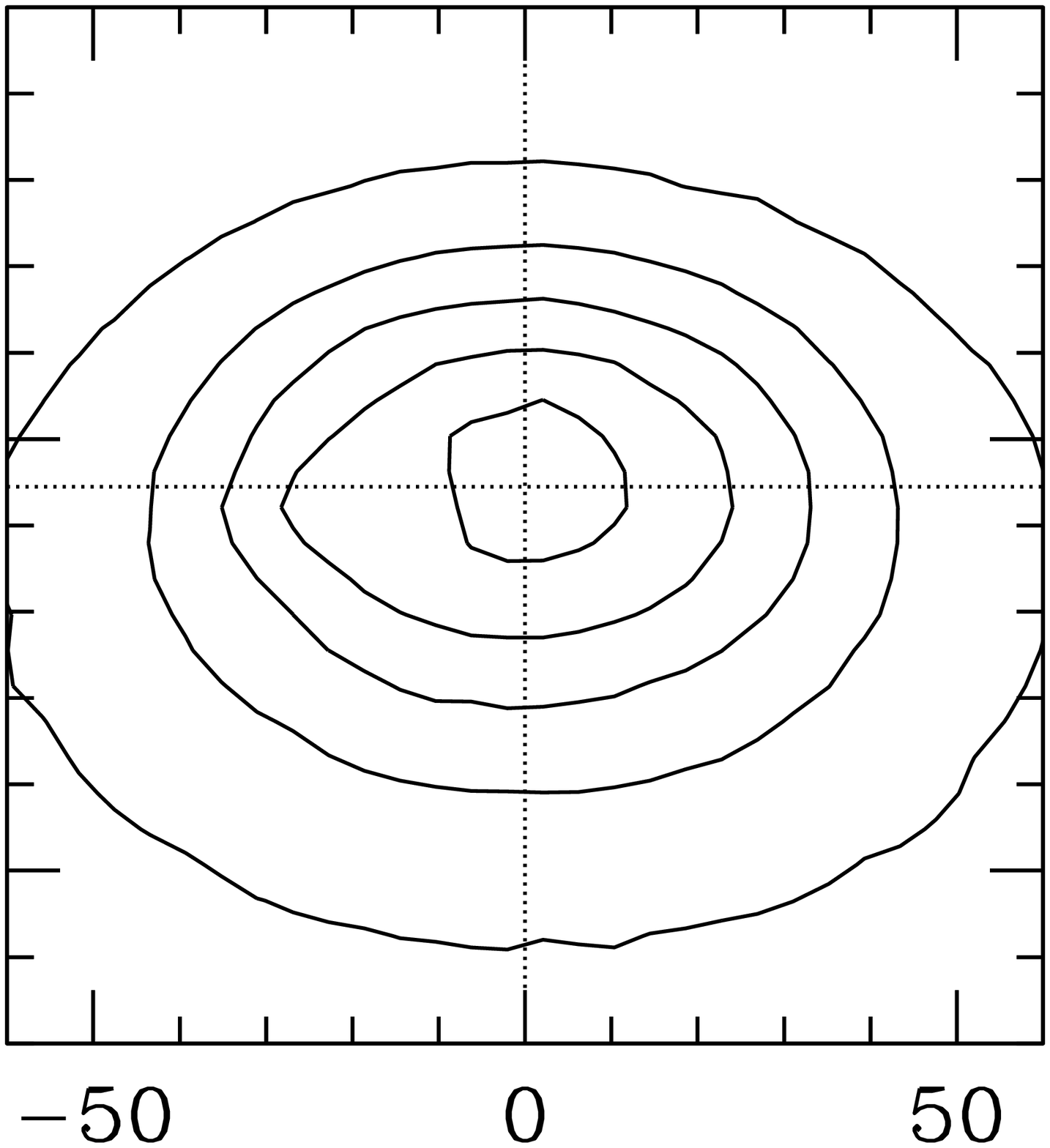}
    \includegraphics[width=.125\hsize]{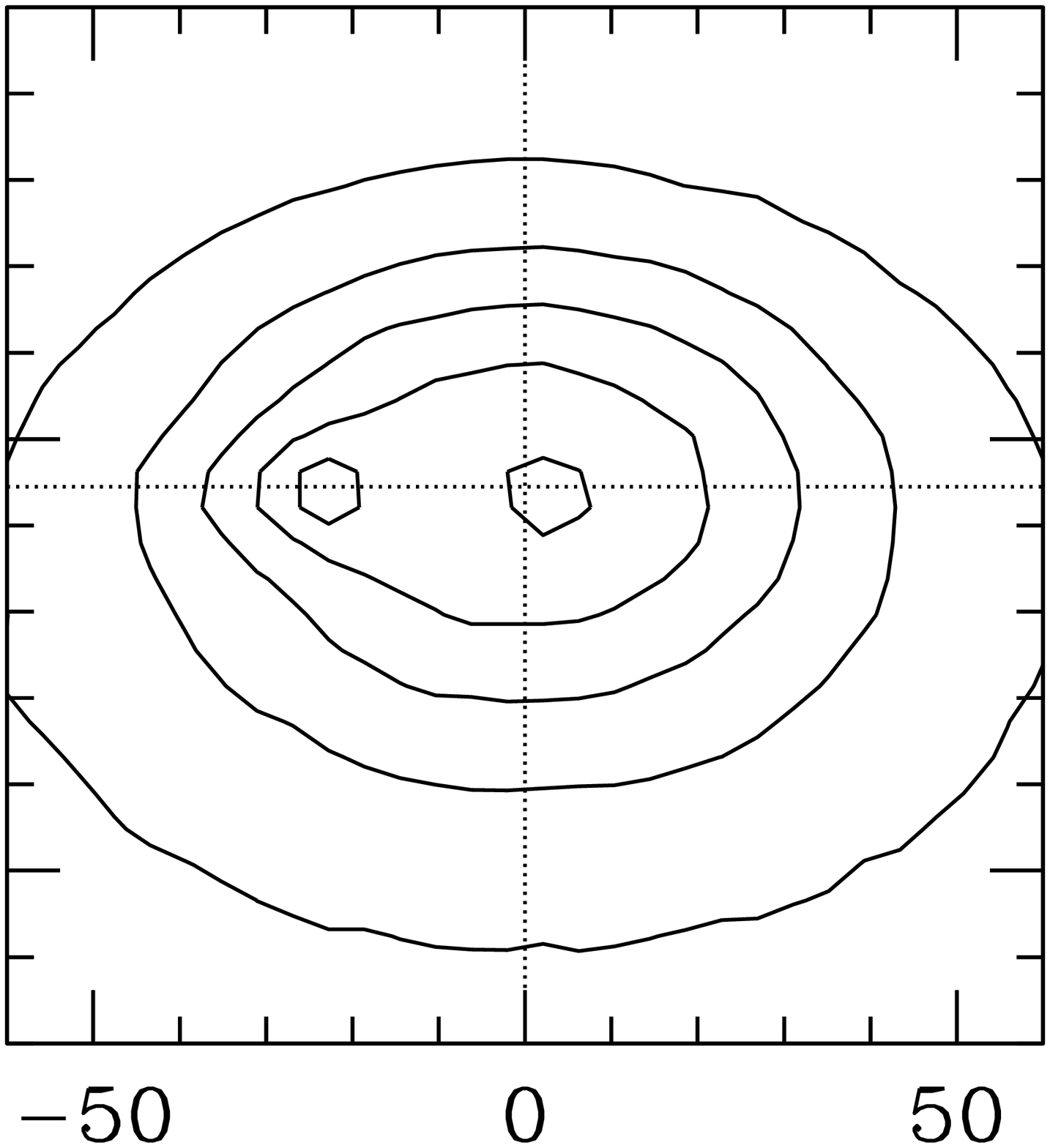}
    \includegraphics[width=.125\hsize]{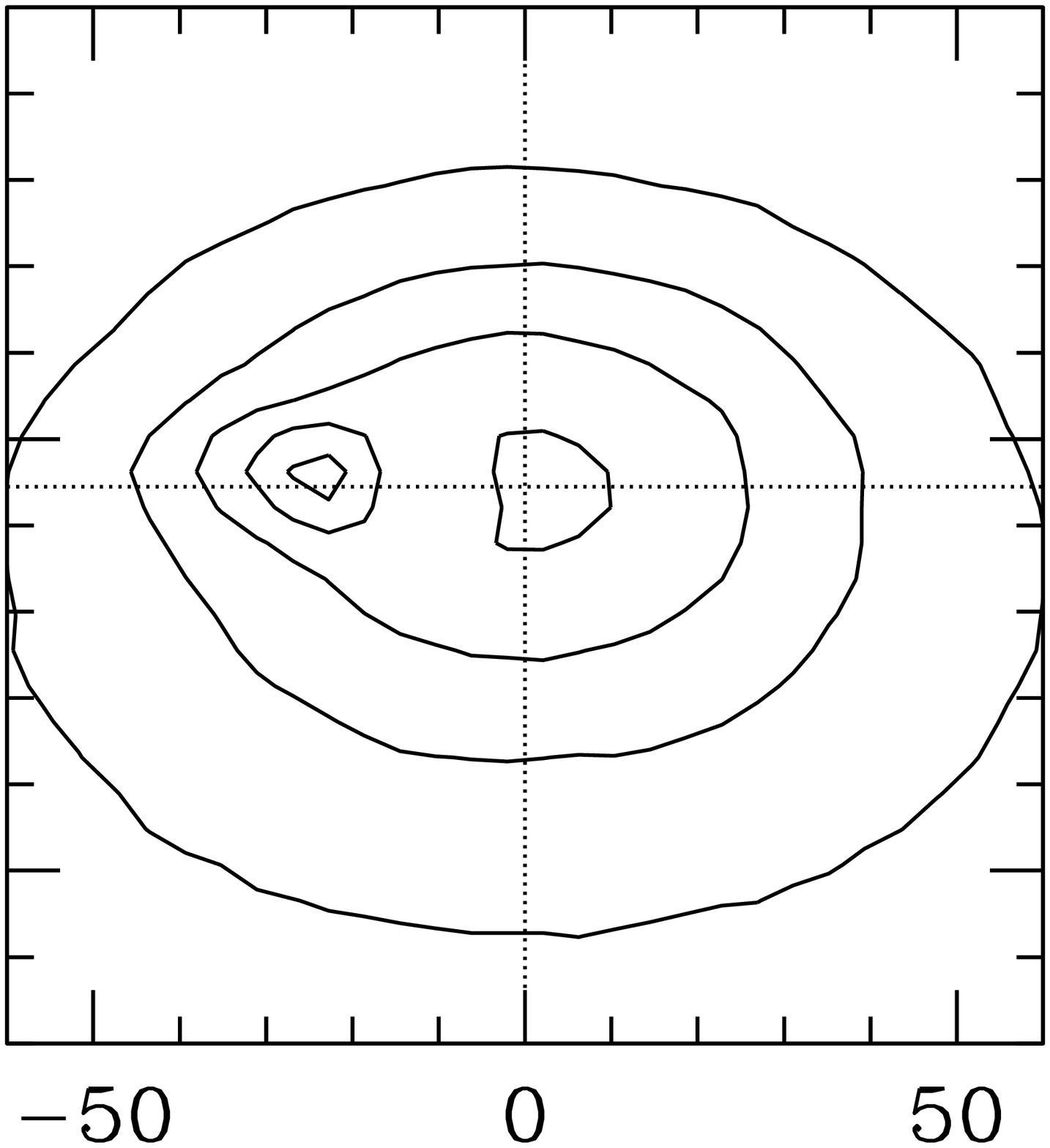}
    \includegraphics[width=.125\hsize]{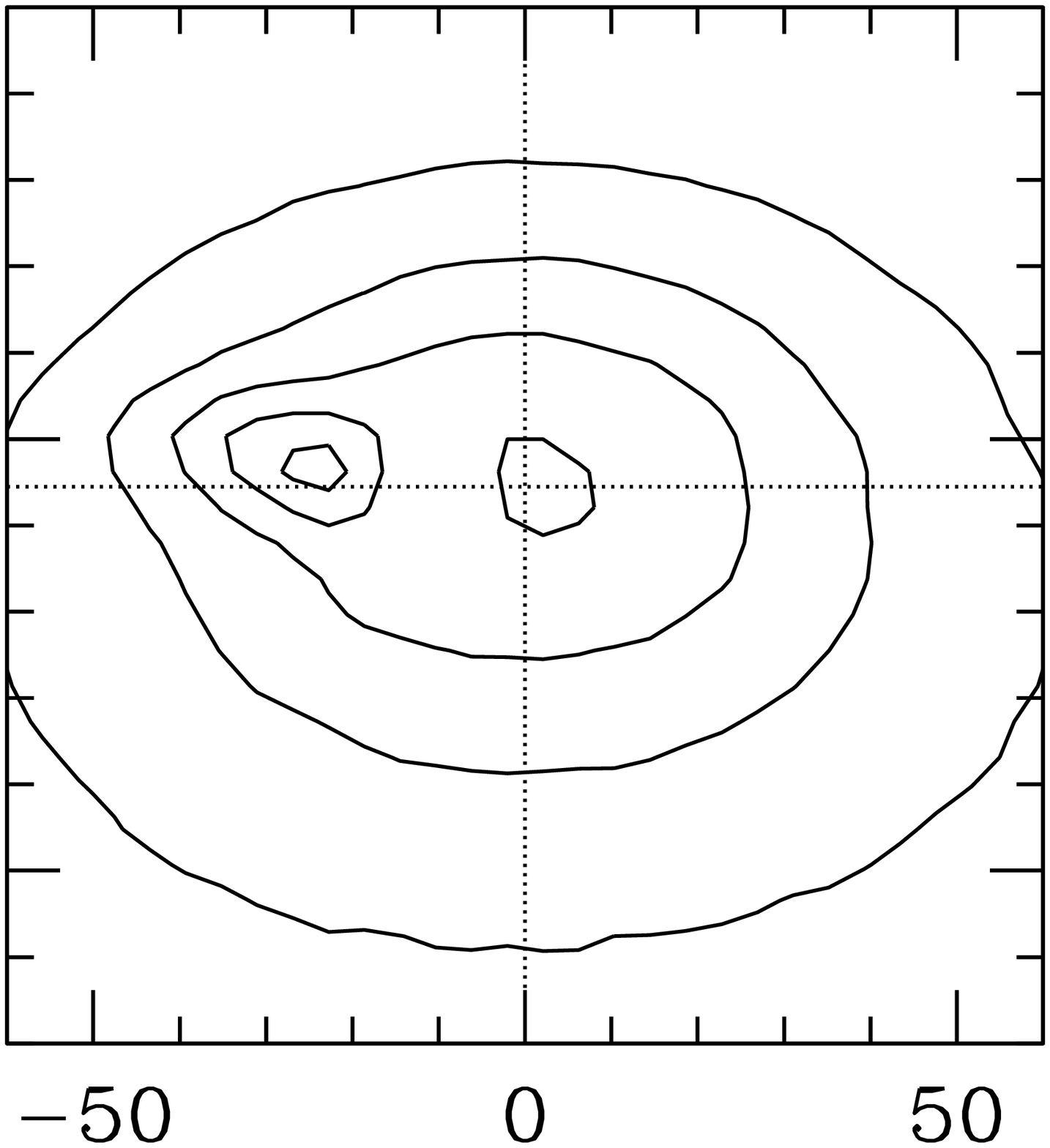}
    \includegraphics[width=.125\hsize]{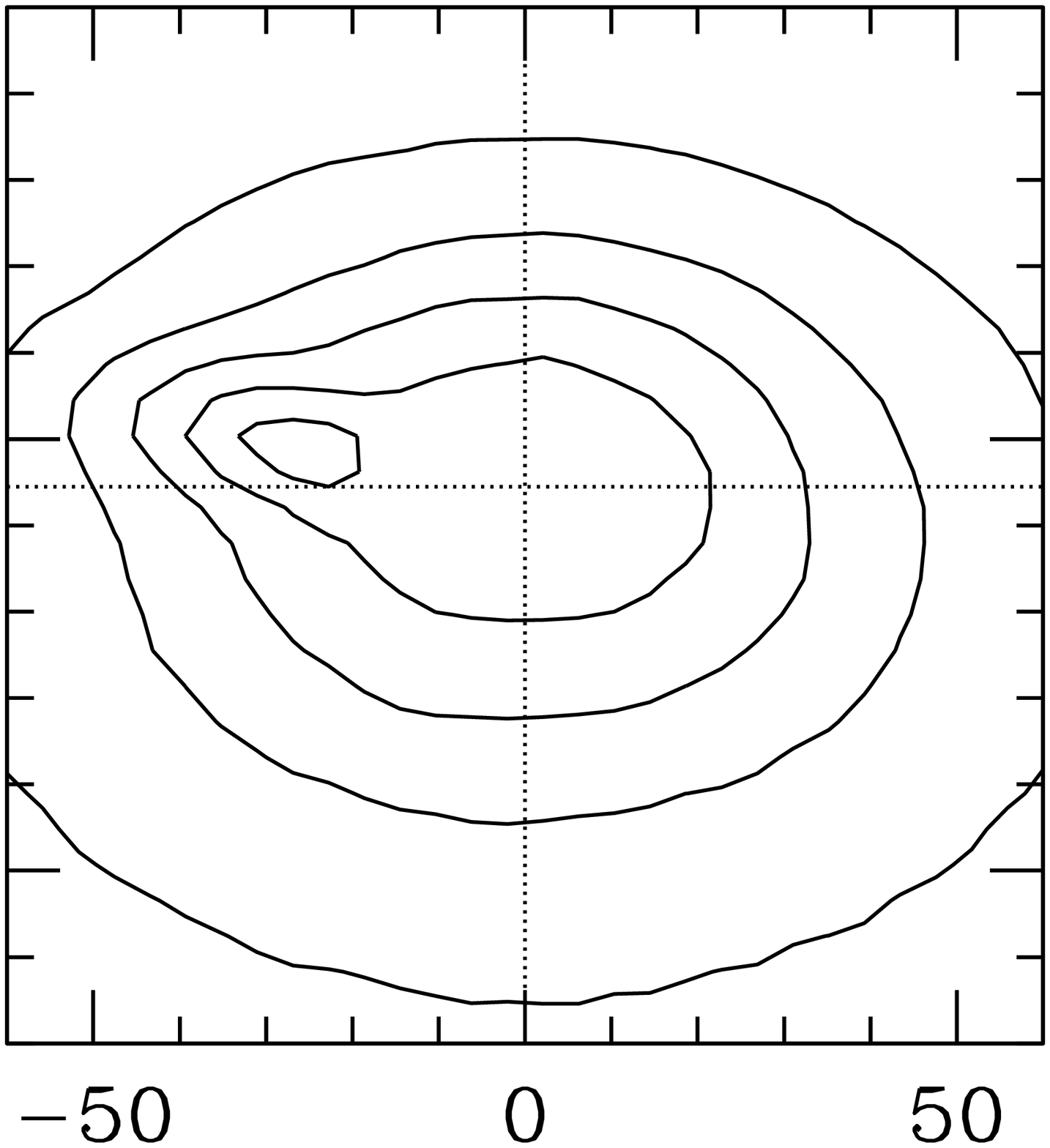}
  }
  \caption{
    Contour plots of the density in the $(-v_R)-v_\phi$ plane of the stars in 
    the bins shown in Figure~\ref{fig:RAVEbins} for each of the four models. 
    Each plot covers the range $-60\kms < (-v_R) < 60\kms$, 
    $180\kms < v_\phi < 300\kms$.
\label{fig:RAVE}
}
\end{figure*}

\begin{figure*}
  \centerline{
    \includegraphics[height=.125\hsize,angle=270]{HyadesPlots/blank.eps}
    \includegraphics[height=.125\hsize,angle=270]{HyadesPlots/88kpc.eps}
    \includegraphics[height=.125\hsize,angle=270]{HyadesPlots/86kpc.eps}
    \includegraphics[height=.125\hsize,angle=270]{HyadesPlots/84kpc.eps}
    \includegraphics[height=.125\hsize,angle=270]{HyadesPlots/82kpc.eps}
    \includegraphics[height=.125\hsize,angle=270]{HyadesPlots/80kpc.eps}
    \includegraphics[height=.125\hsize,angle=270]{HyadesPlots/78kpc.eps}
    \includegraphics[height=.125\hsize,angle=270]{HyadesPlots/76kpc.eps}
  } \centerline{
    \includegraphics[width=.125\hsize]{HyadesPlots/OLR21_lab.eps}
    \includegraphics[width=.125\hsize]{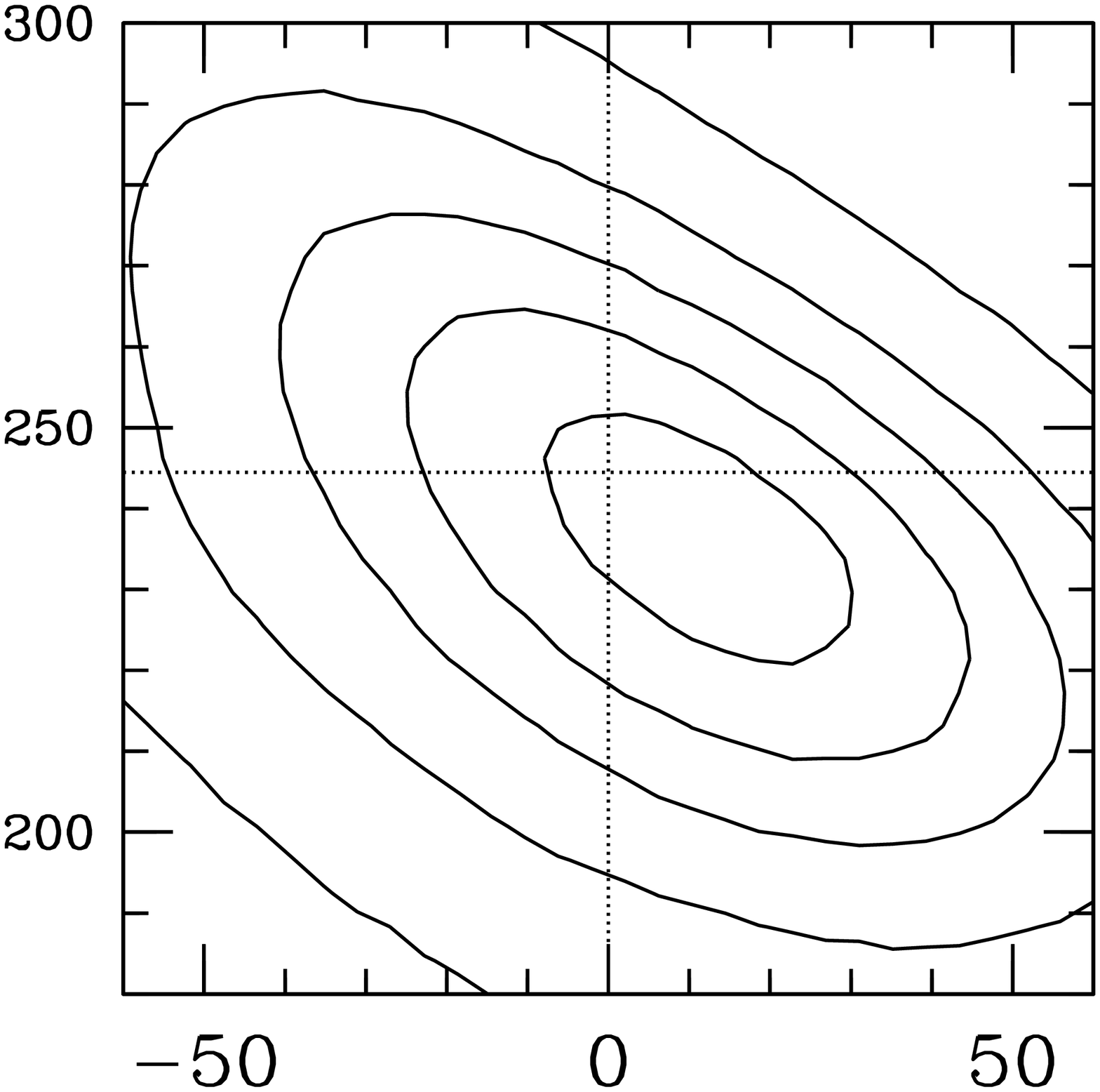}
    \includegraphics[width=.125\hsize]{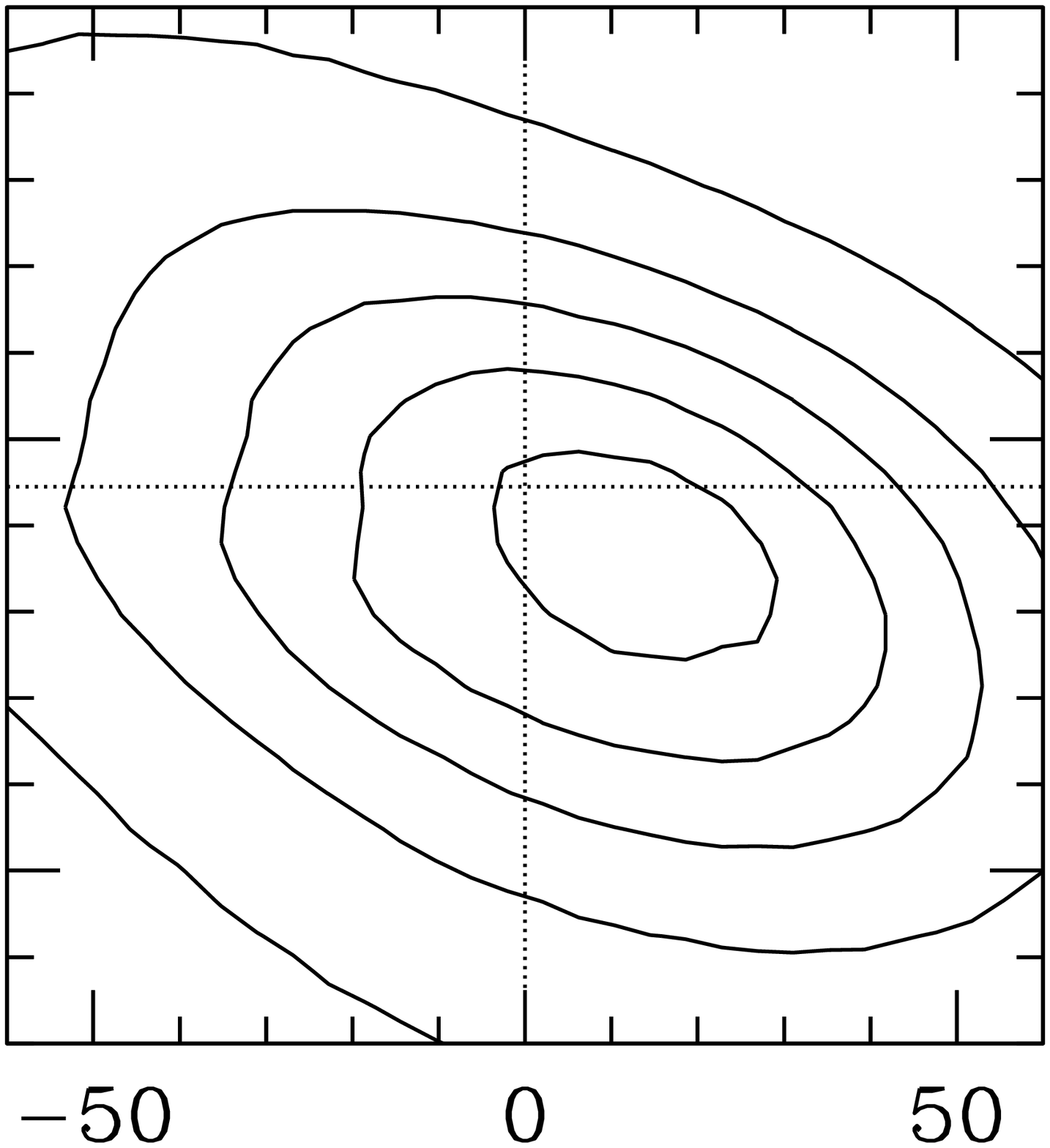}
    \includegraphics[width=.125\hsize]{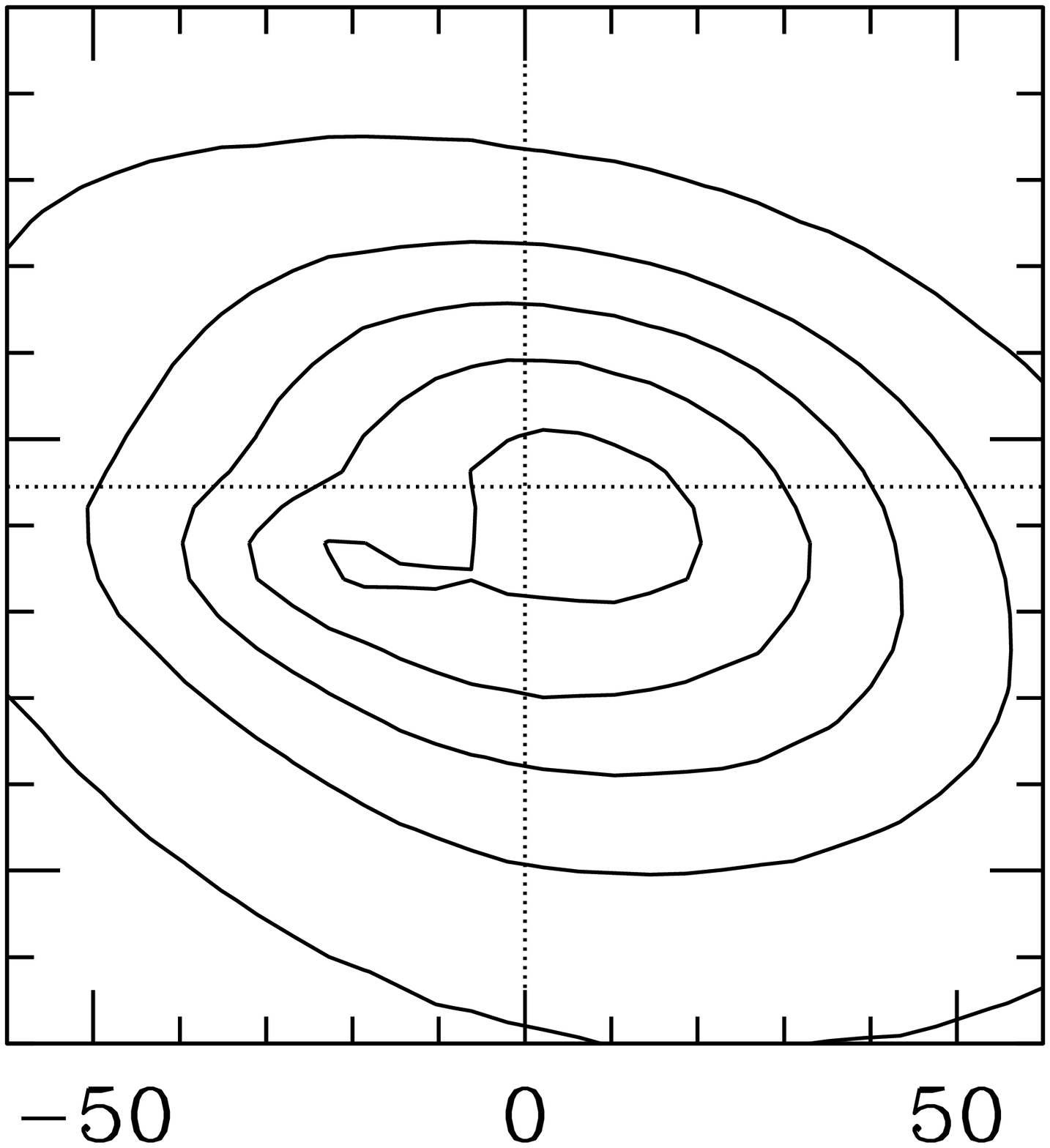}
    \includegraphics[width=.125\hsize]{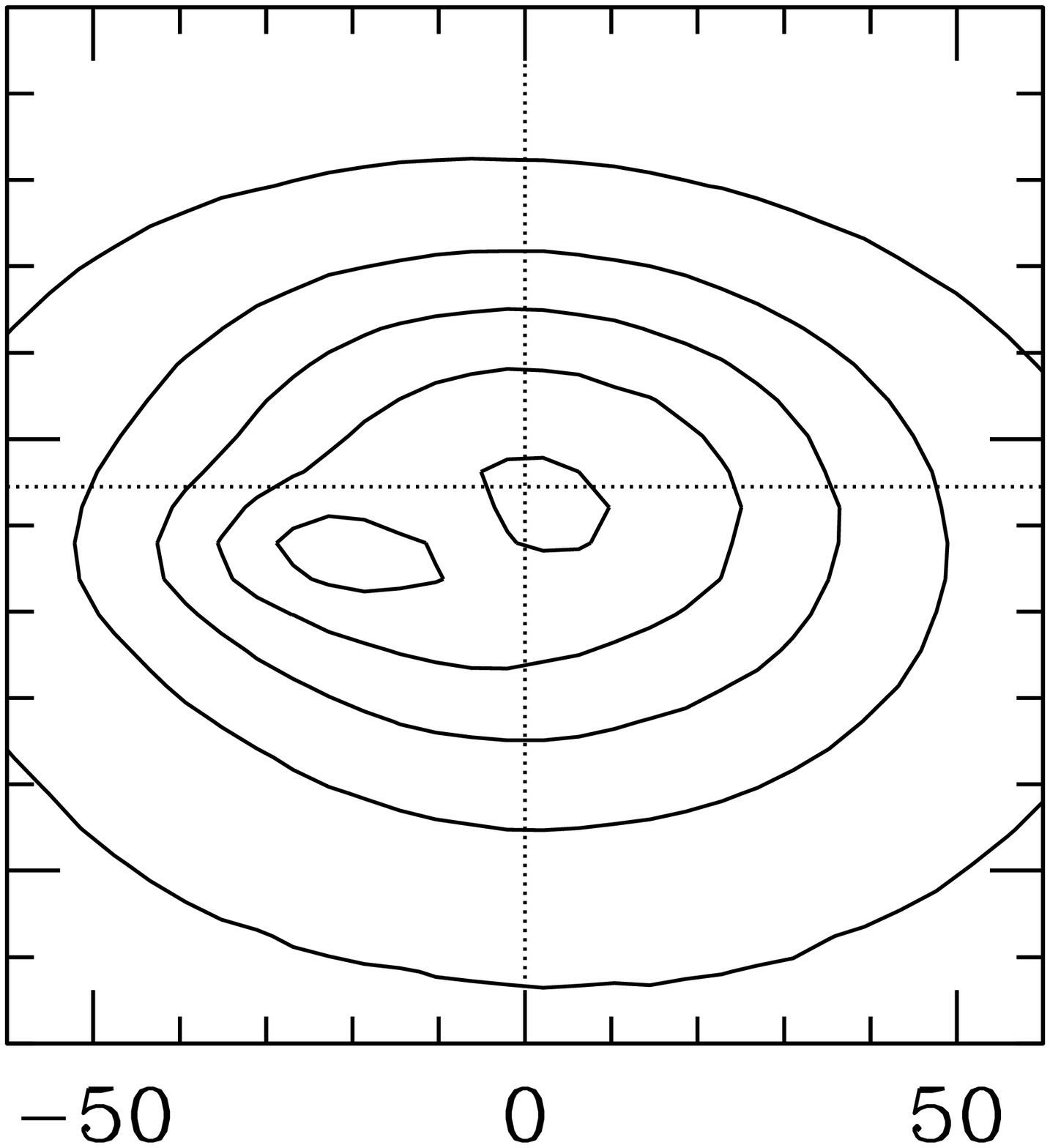}
    \includegraphics[width=.125\hsize]{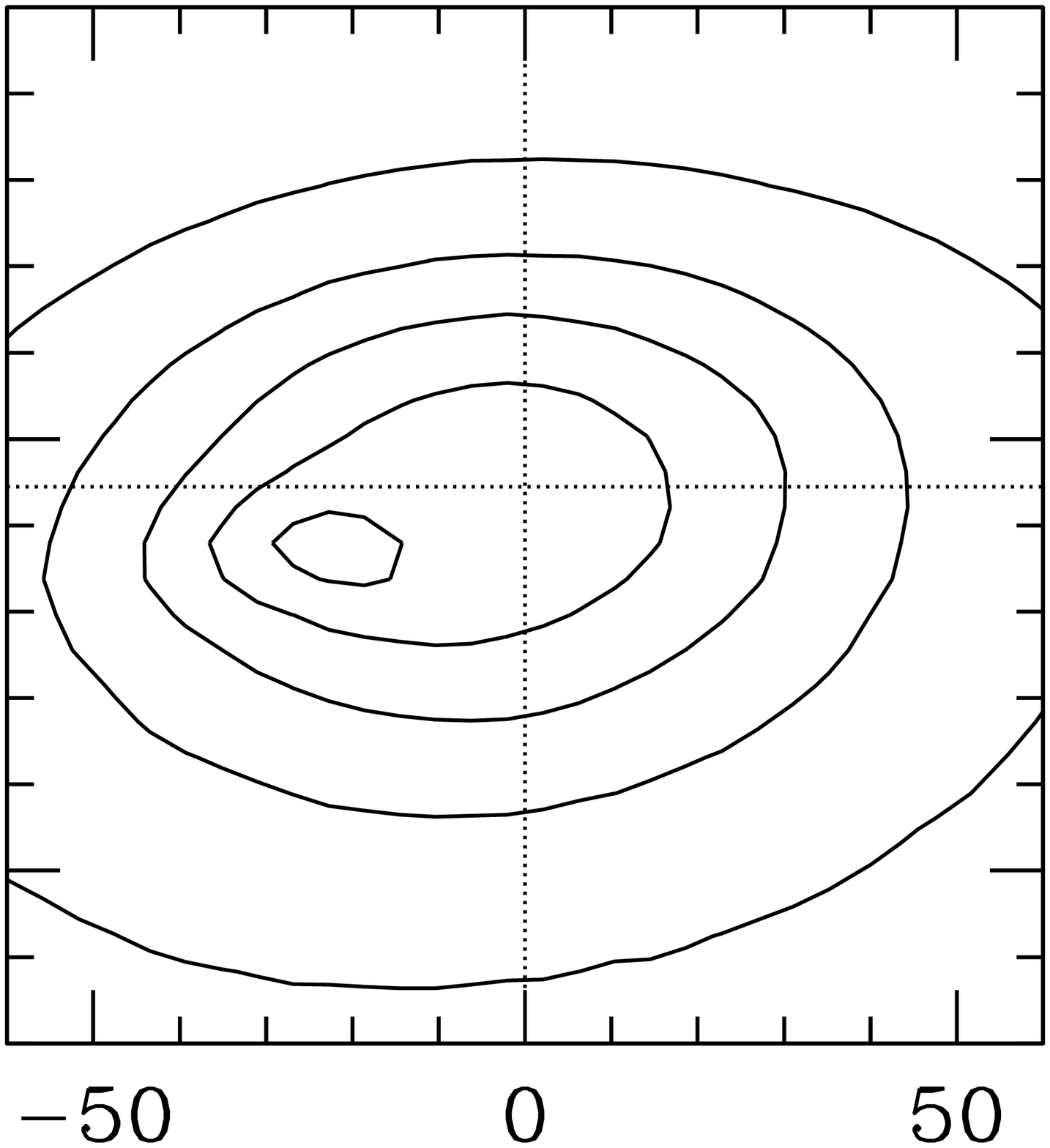}
    \includegraphics[width=.125\hsize]{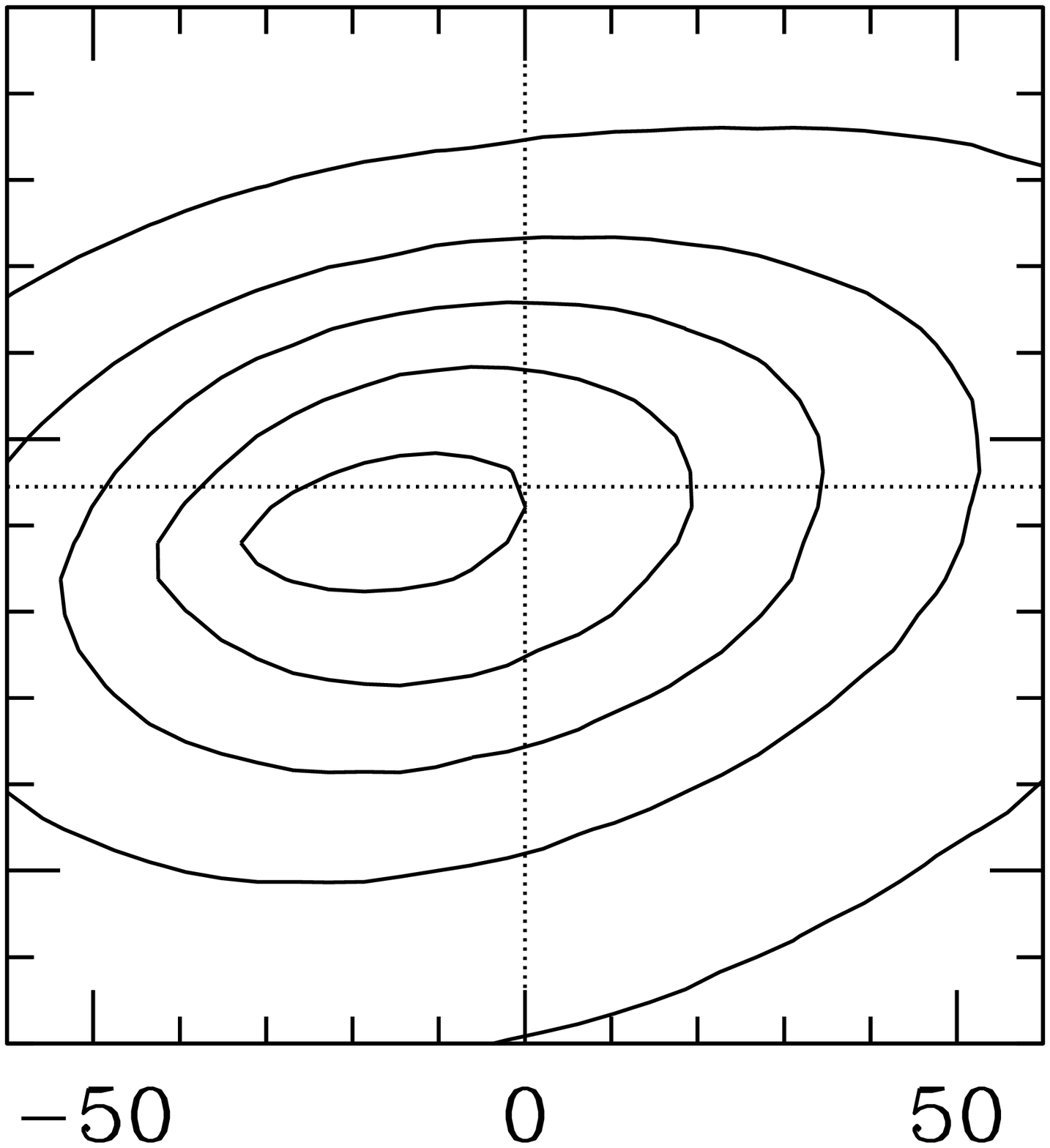}
    \includegraphics[width=.125\hsize]{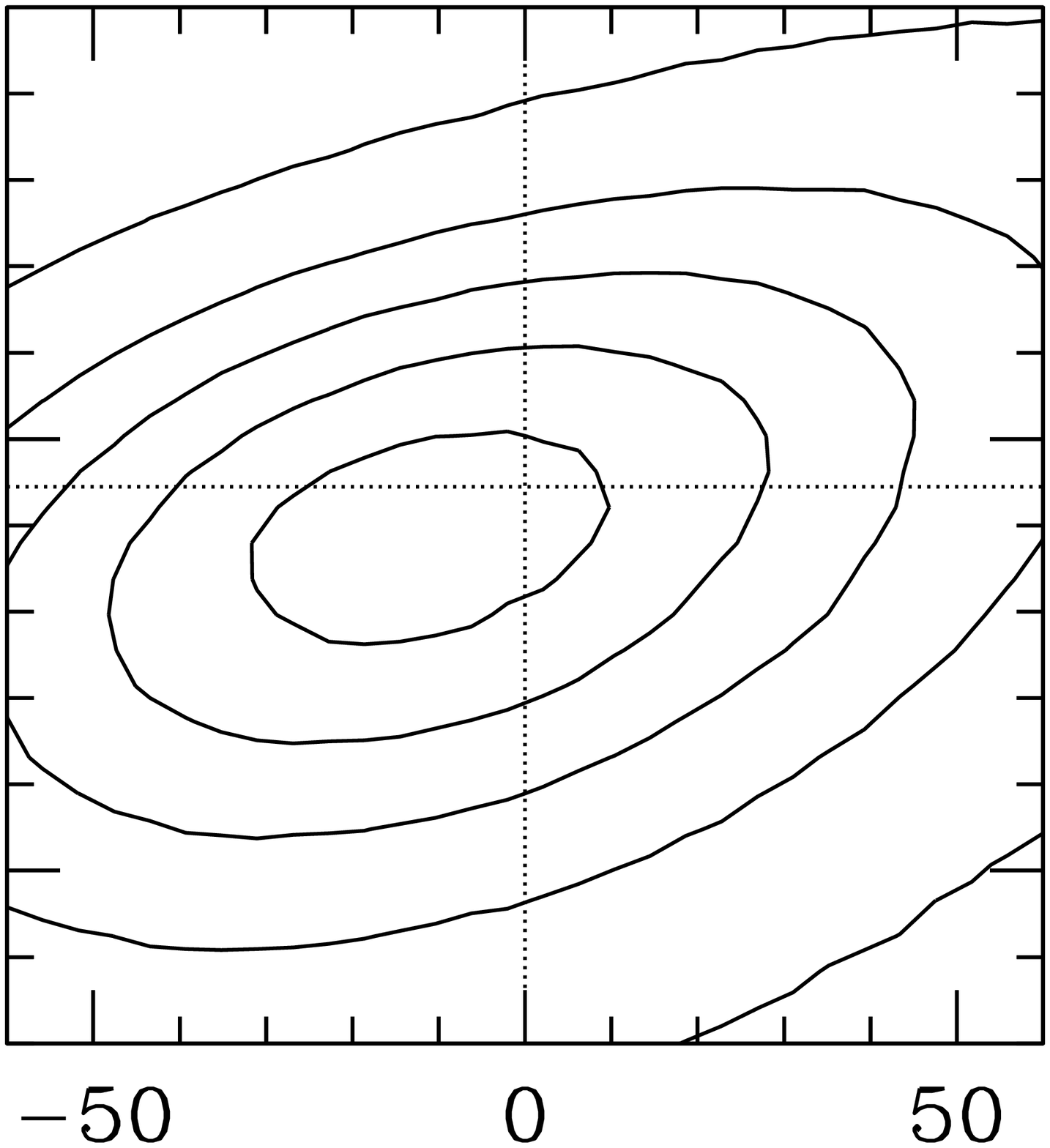}
  } \centerline{
    \includegraphics[width=.125\hsize]{HyadesPlots/OLR41_lab.eps}
    \includegraphics[width=.125\hsize]{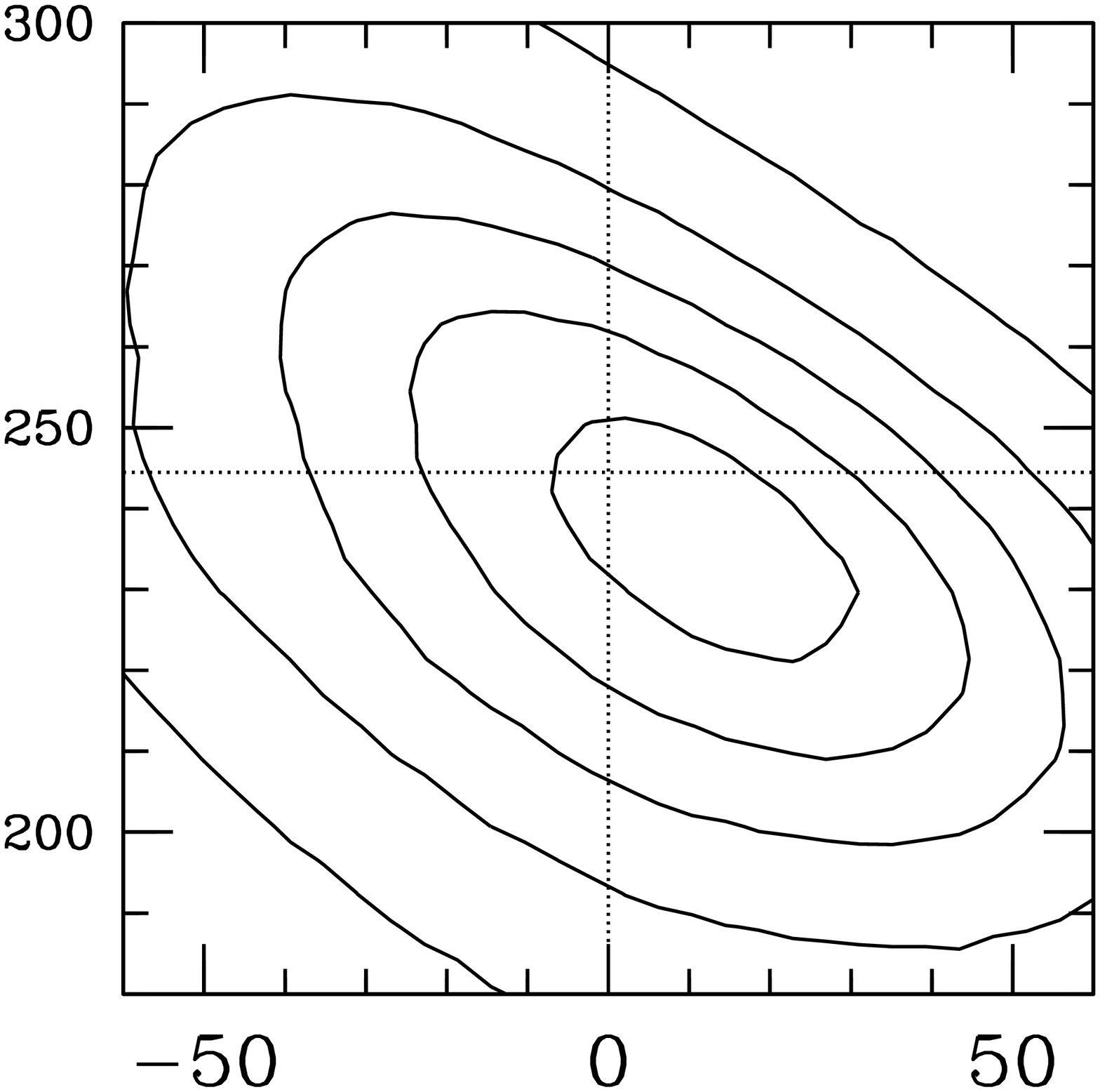}
    \includegraphics[width=.125\hsize]{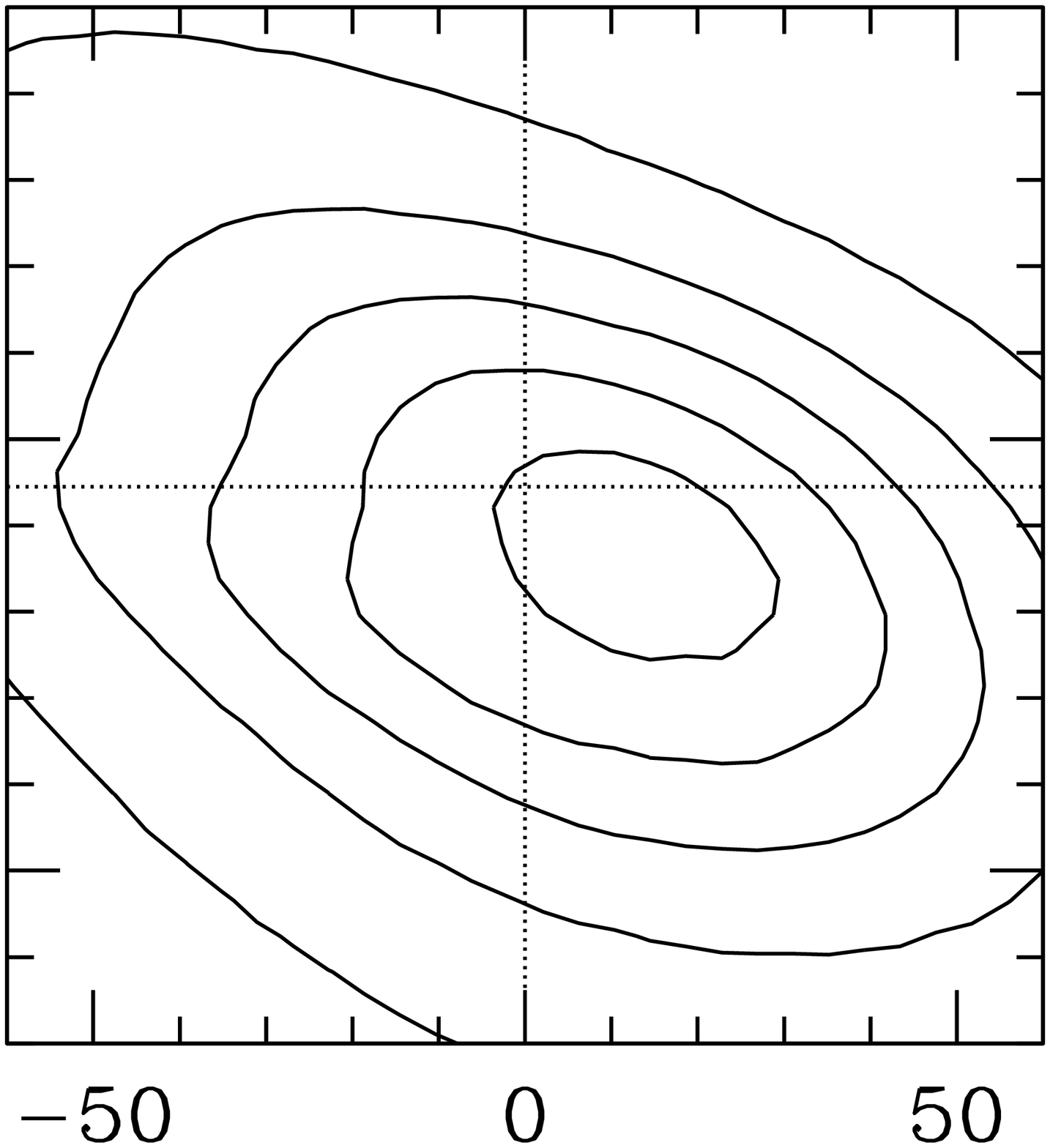}
    \includegraphics[width=.125\hsize]{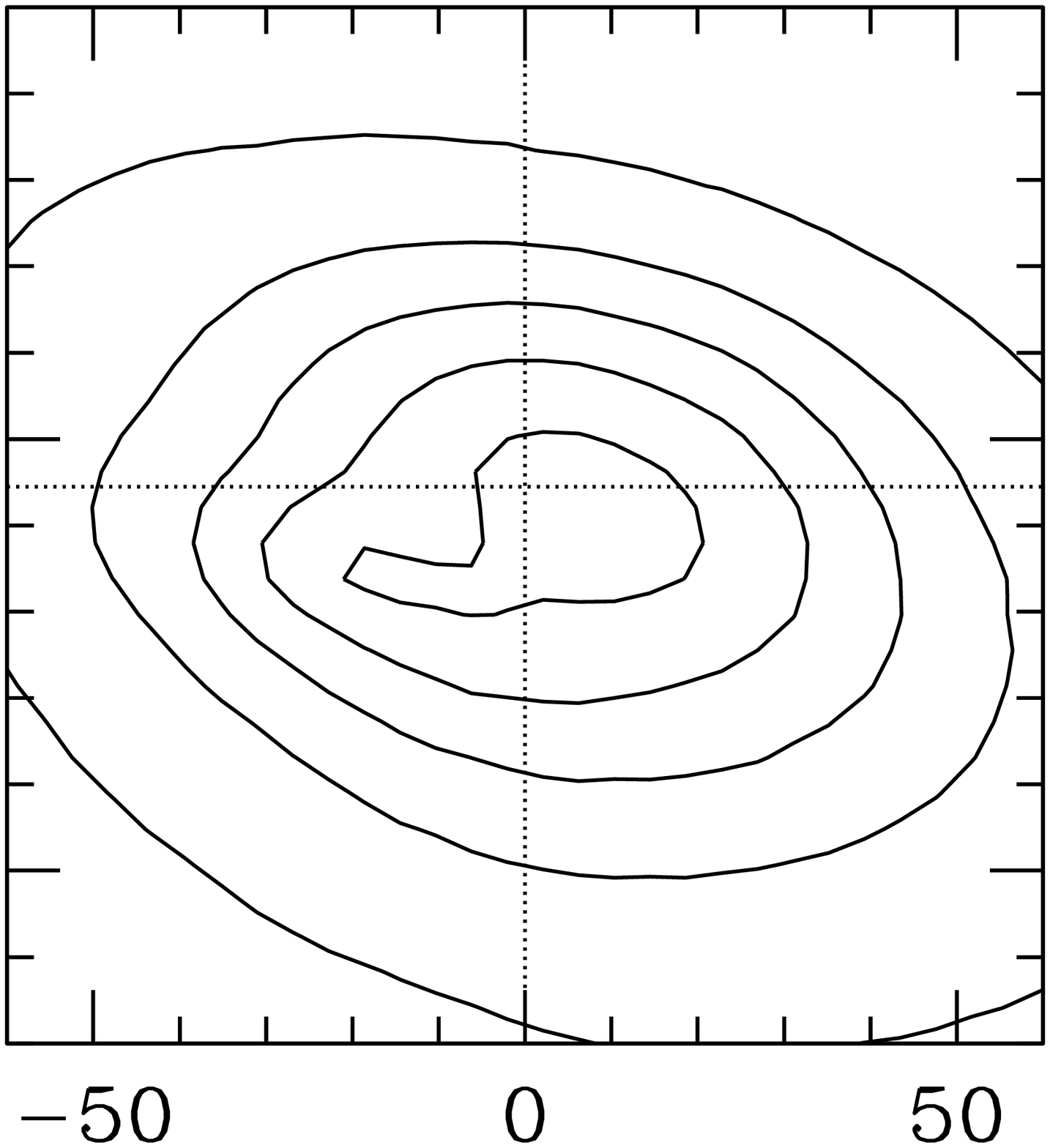}
    \includegraphics[width=.125\hsize]{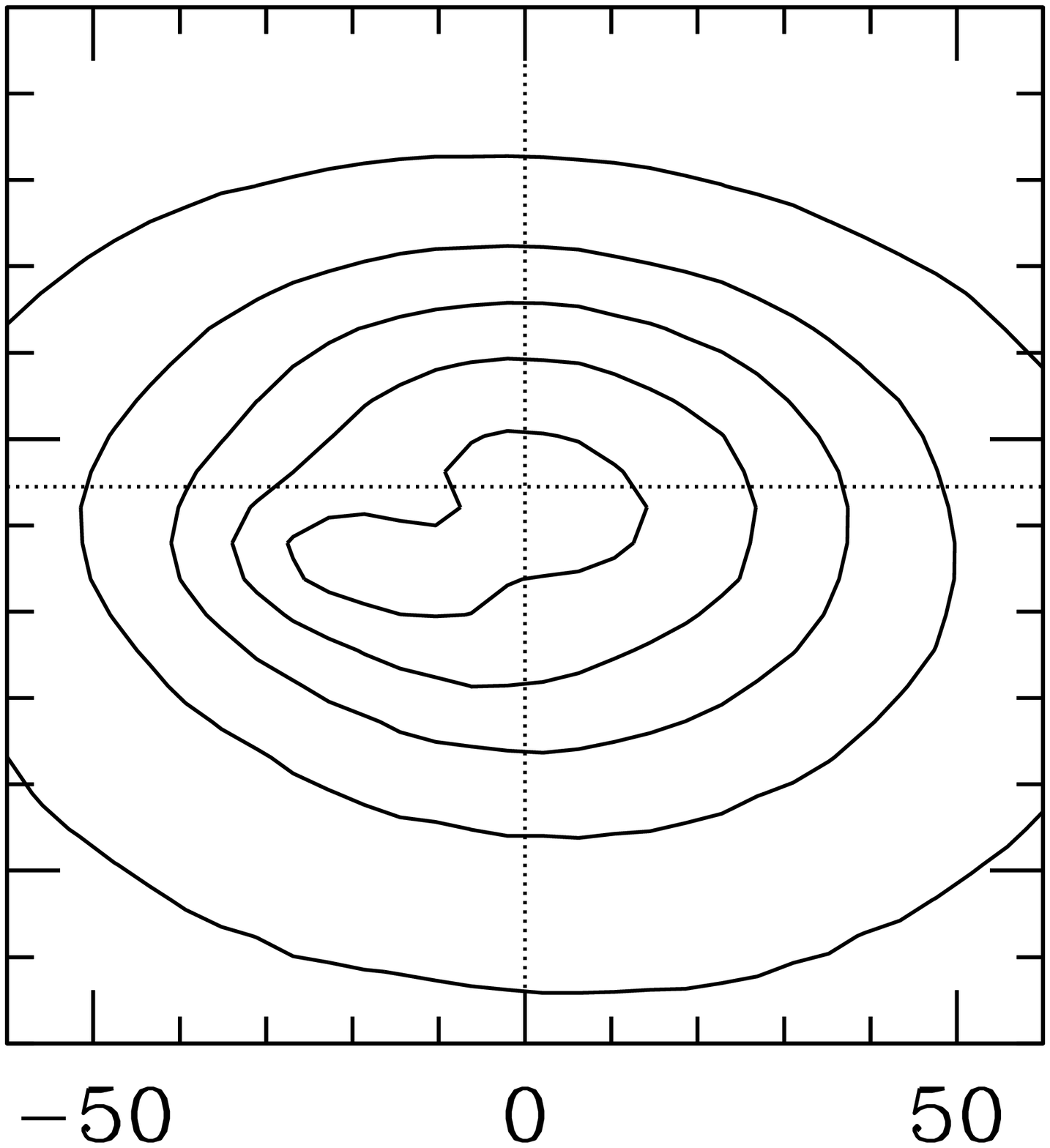}
    \includegraphics[width=.125\hsize]{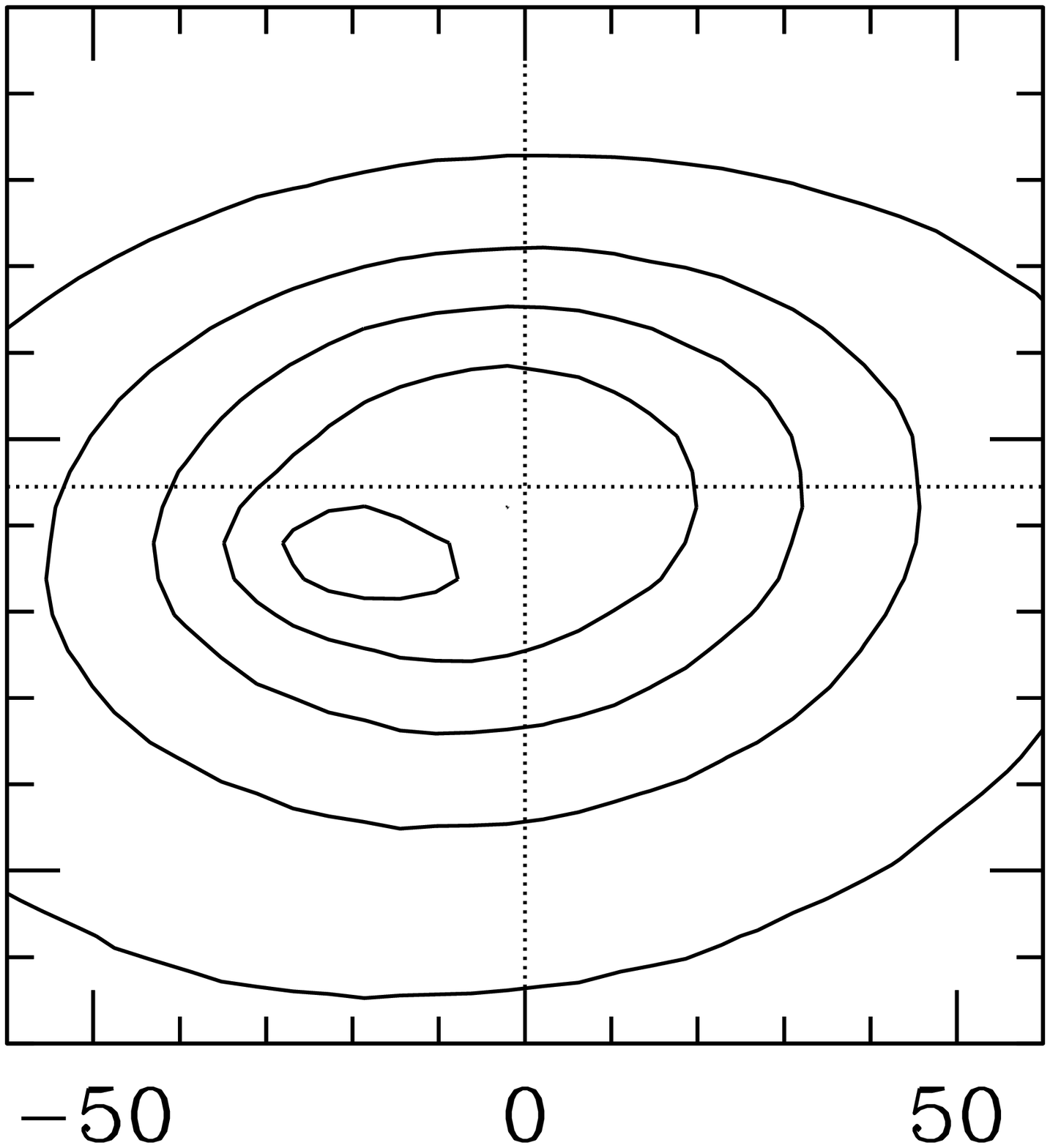}
    \includegraphics[width=.125\hsize]{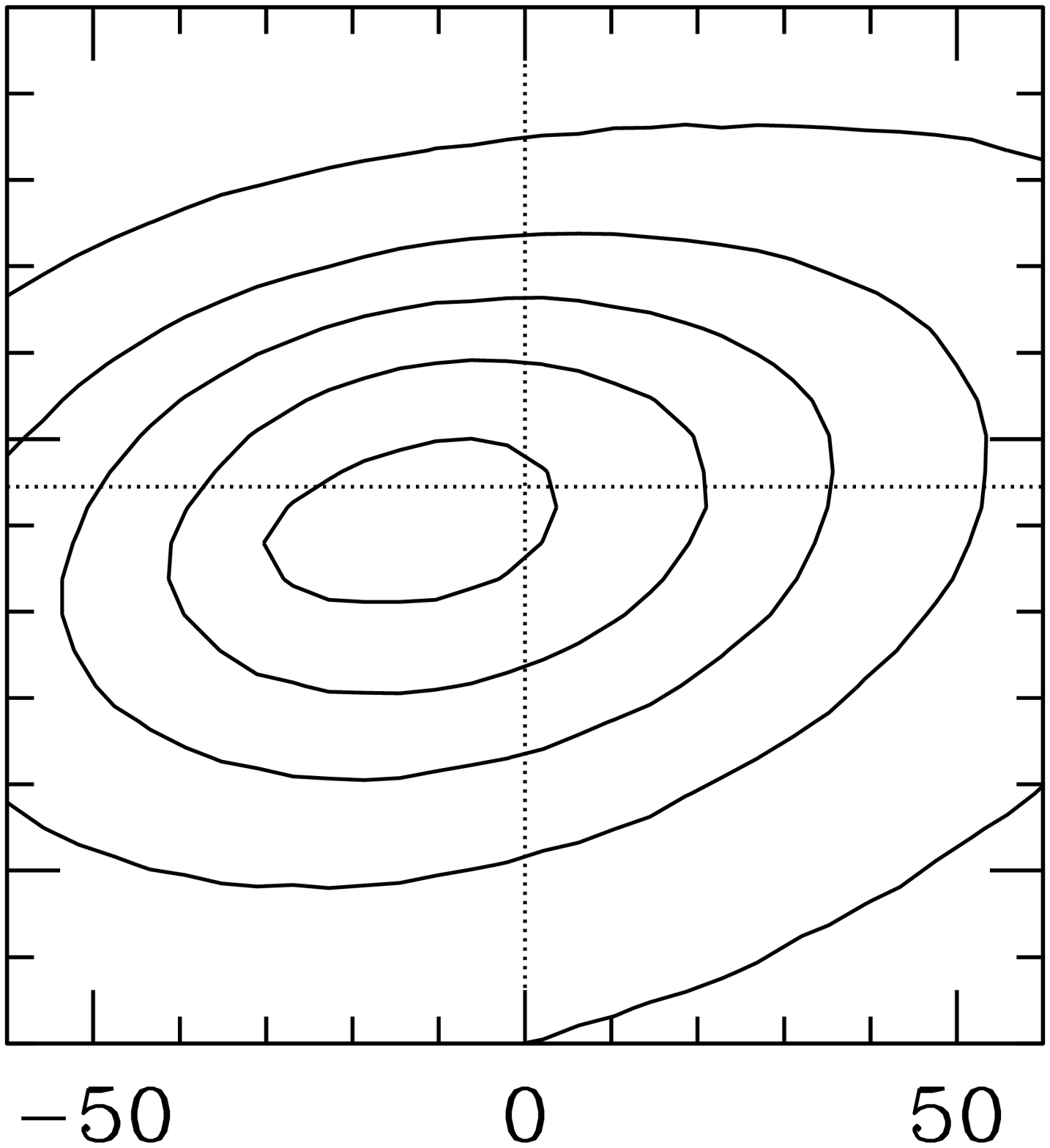}
    \includegraphics[width=.125\hsize]{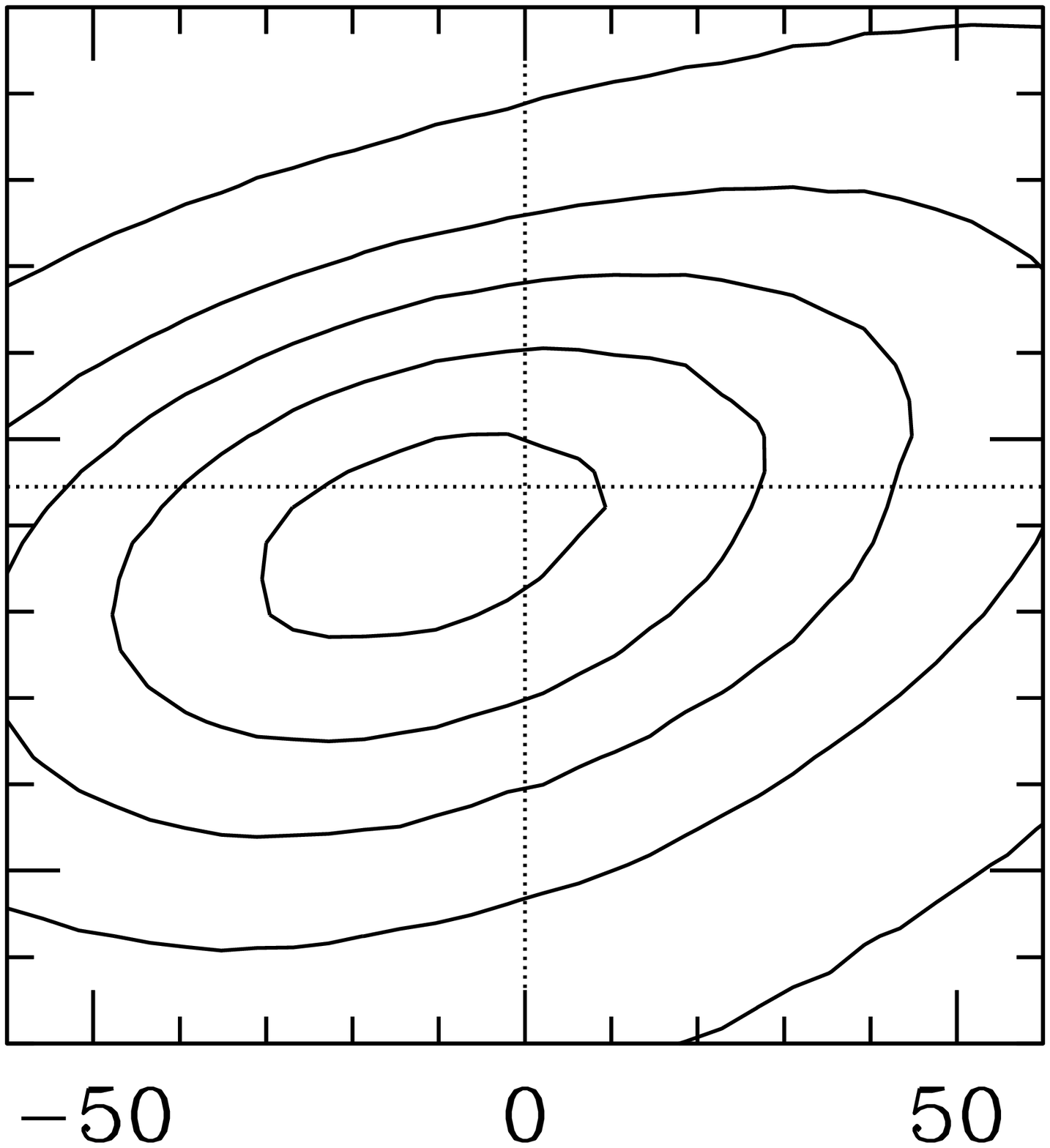}
  } \centerline{
    \includegraphics[width=.125\hsize]{HyadesPlots/ILR21_lab.eps}
    \includegraphics[width=.125\hsize]{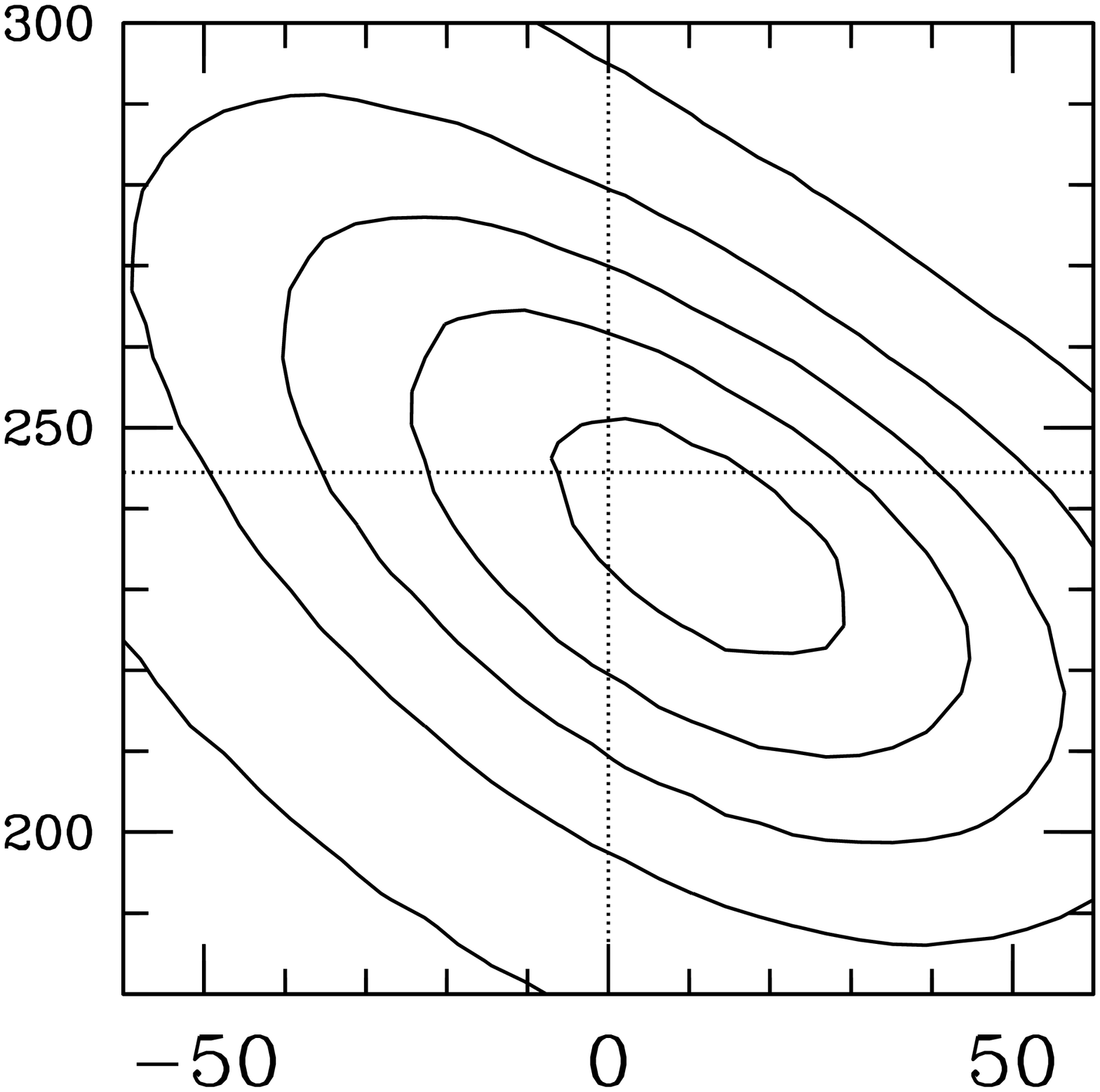}
    \includegraphics[width=.125\hsize]{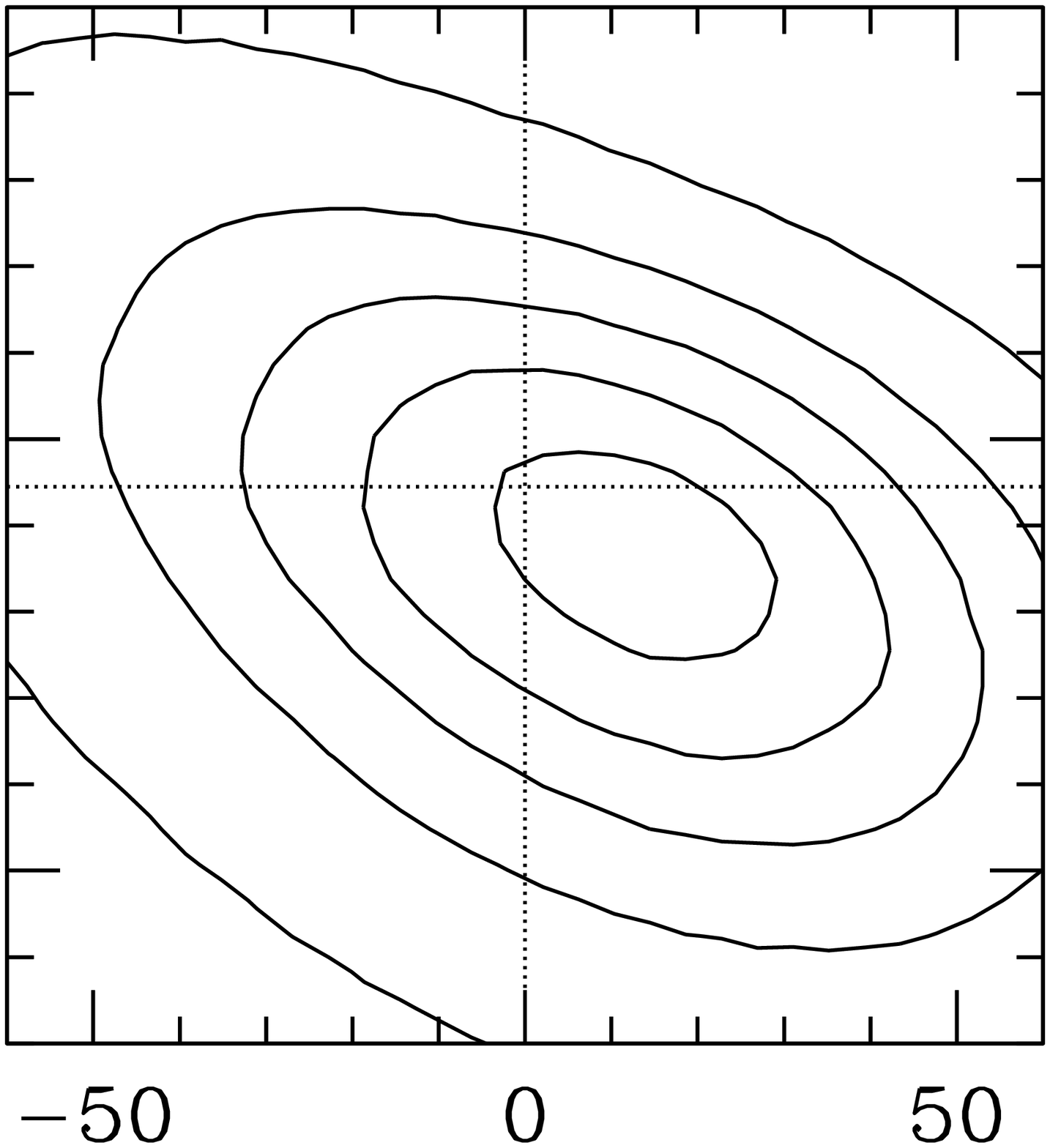}
    \includegraphics[width=.125\hsize]{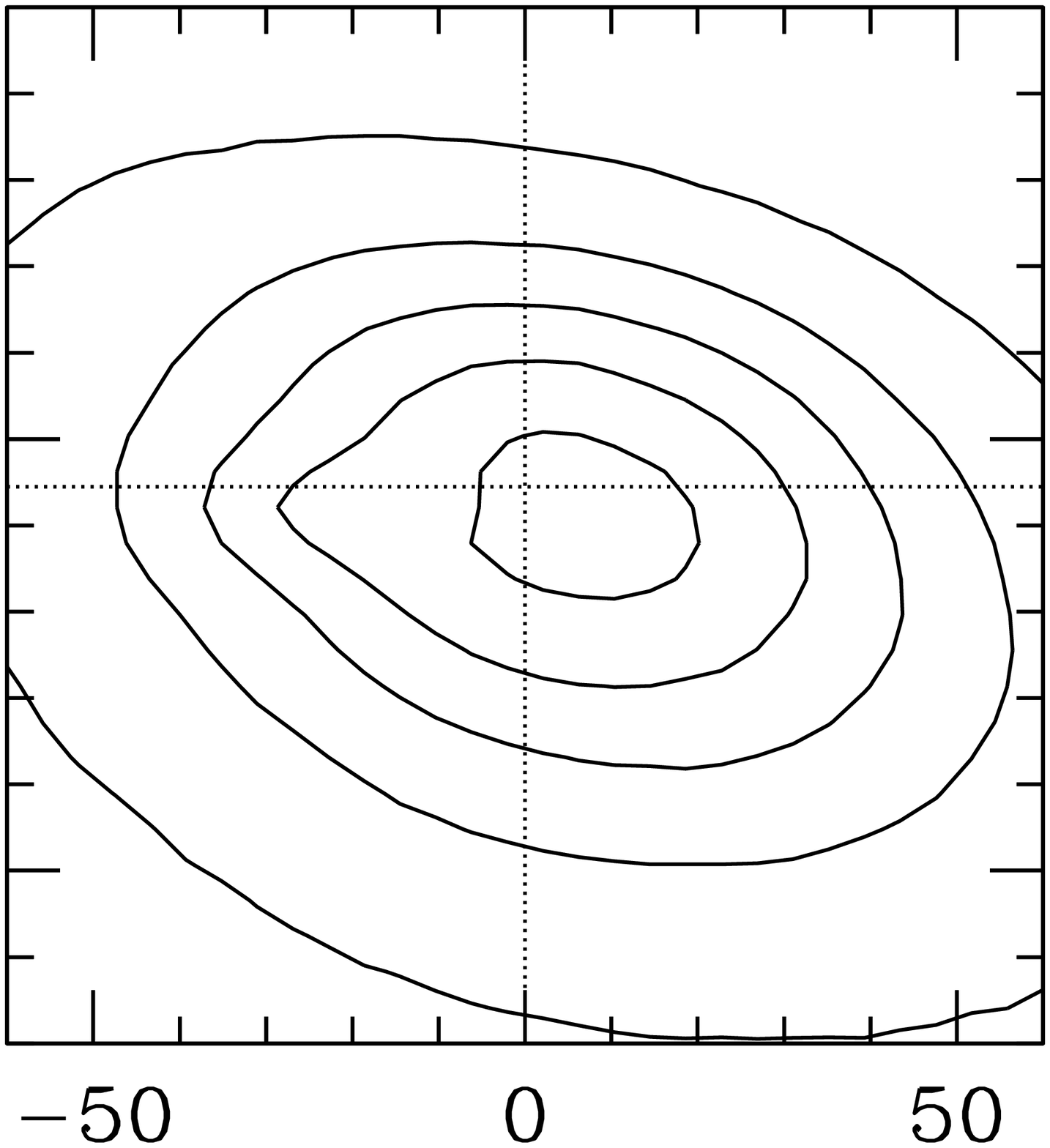}
    \includegraphics[width=.125\hsize]{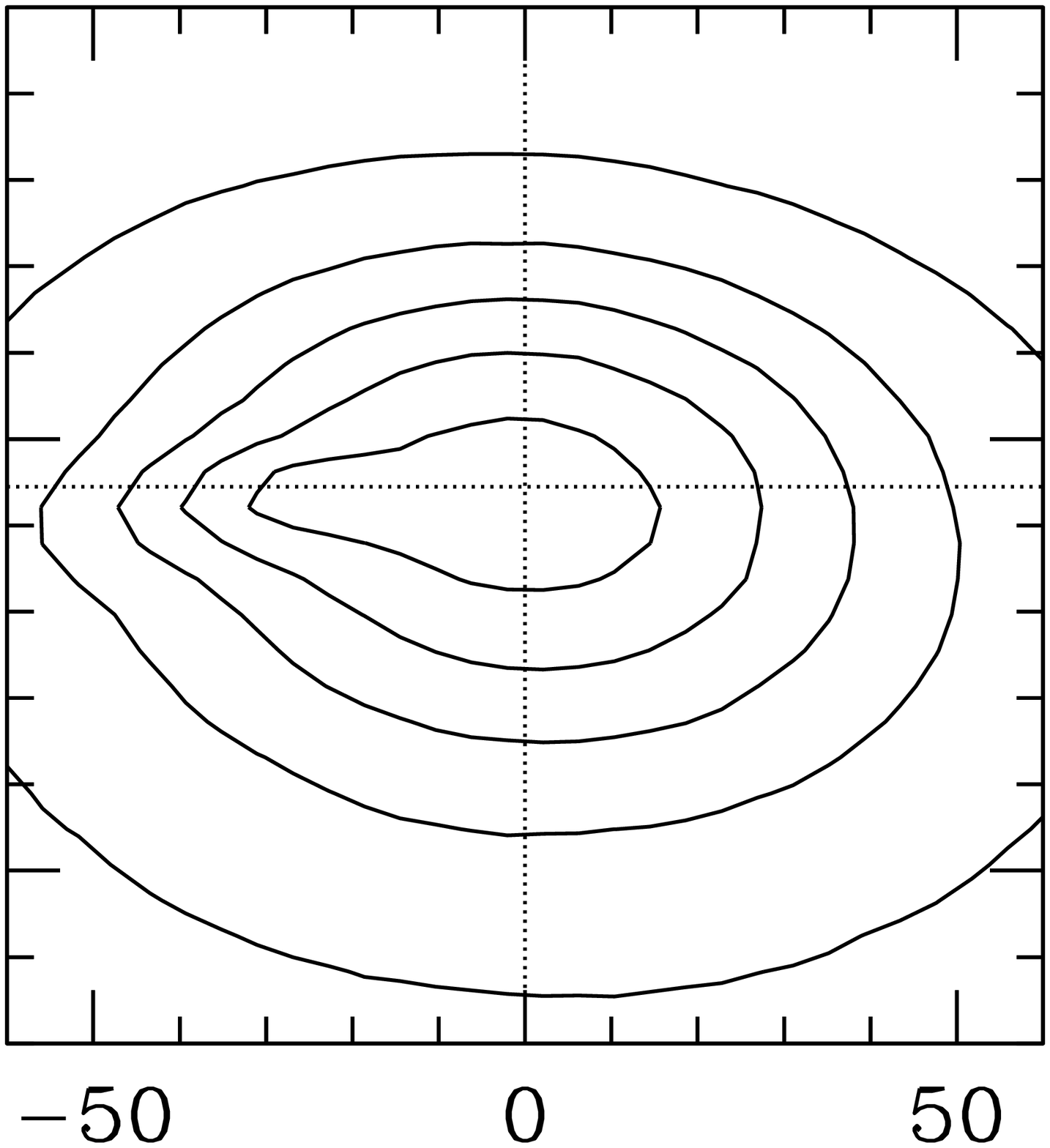}
    \includegraphics[width=.125\hsize]{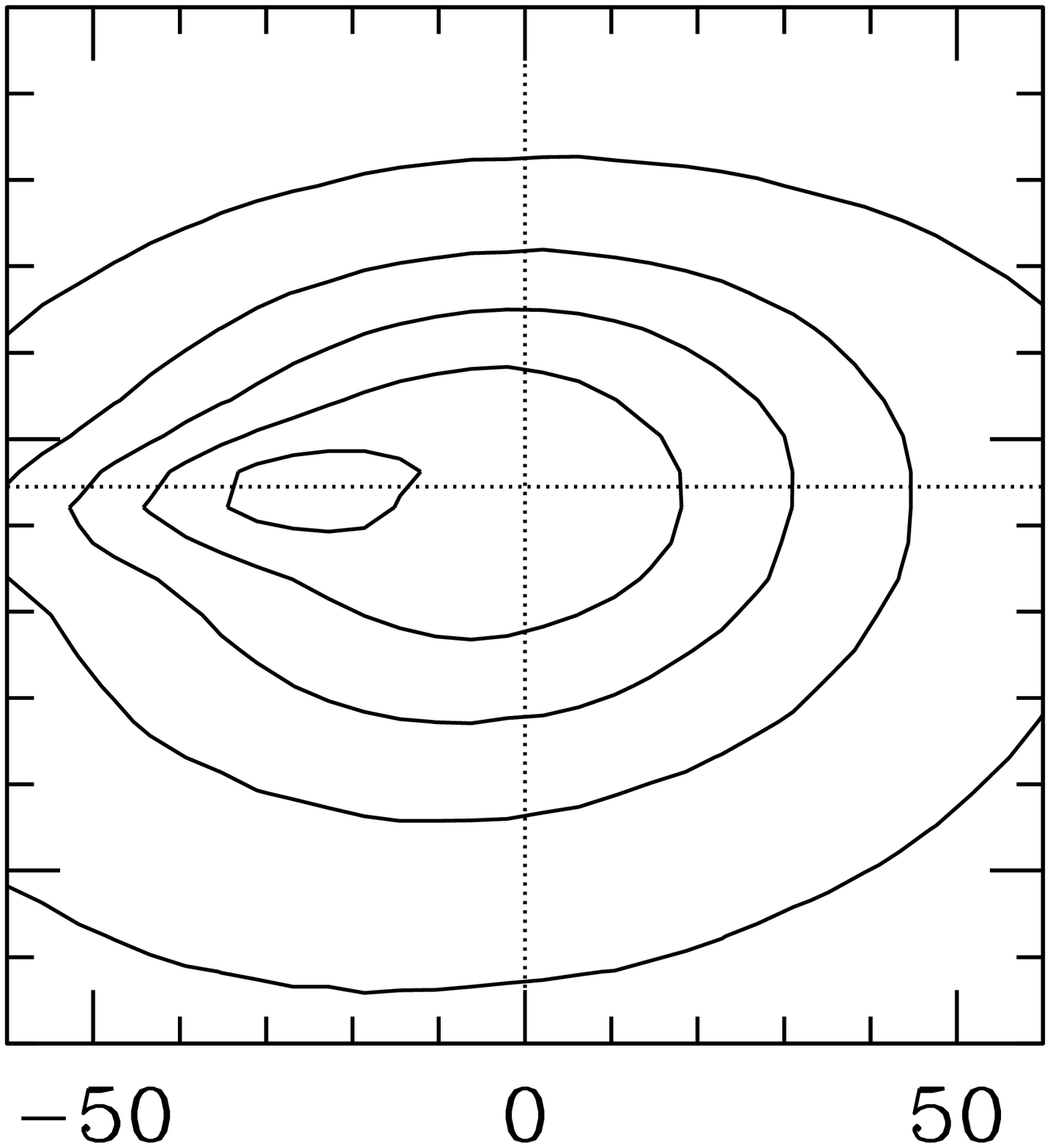}
    \includegraphics[width=.125\hsize]{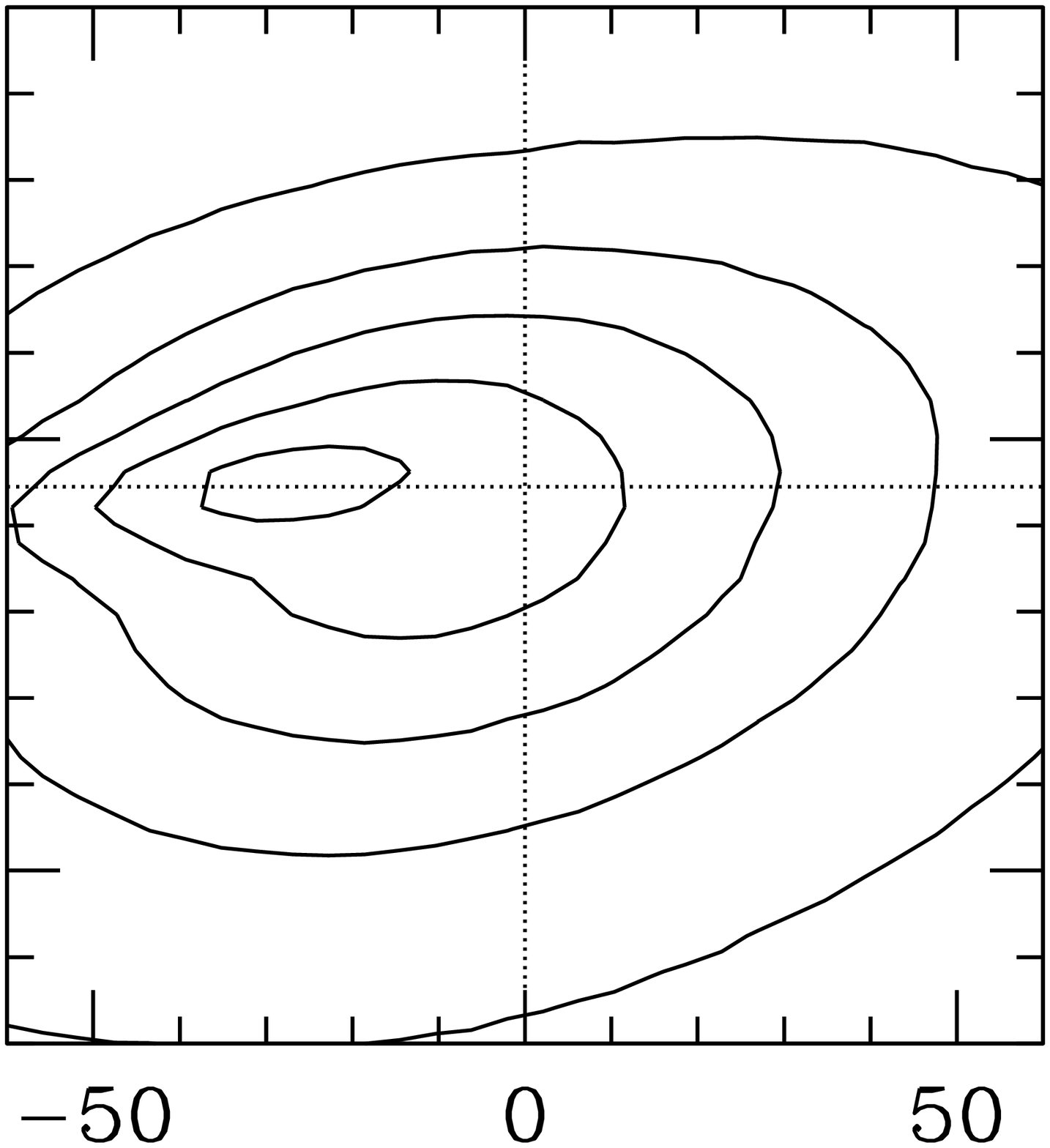}
    \includegraphics[width=.125\hsize]{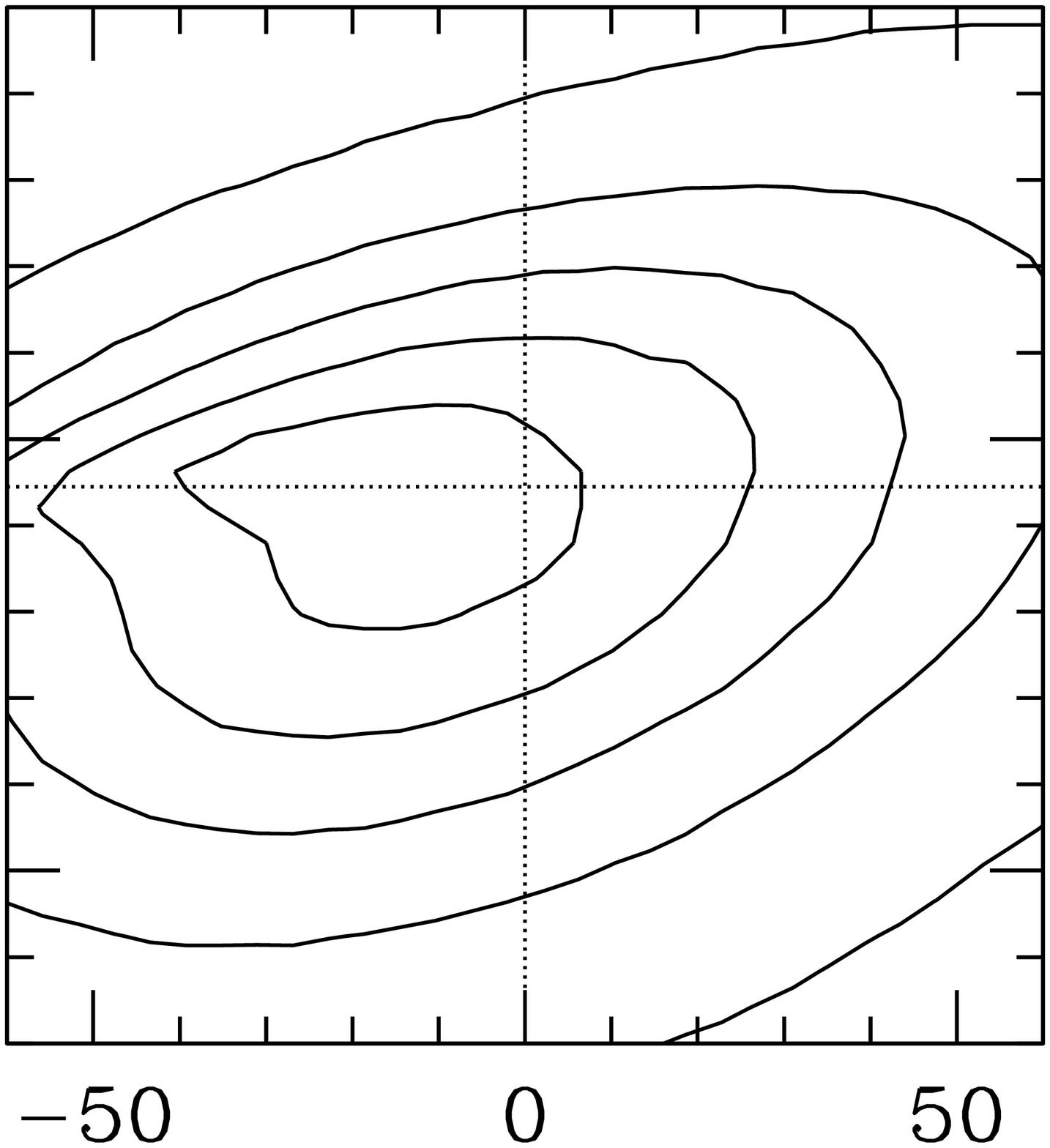}
  } \centerline{
    \includegraphics[width=.125\hsize]{HyadesPlots/ILR41_lab.eps}
    \includegraphics[width=.125\hsize]{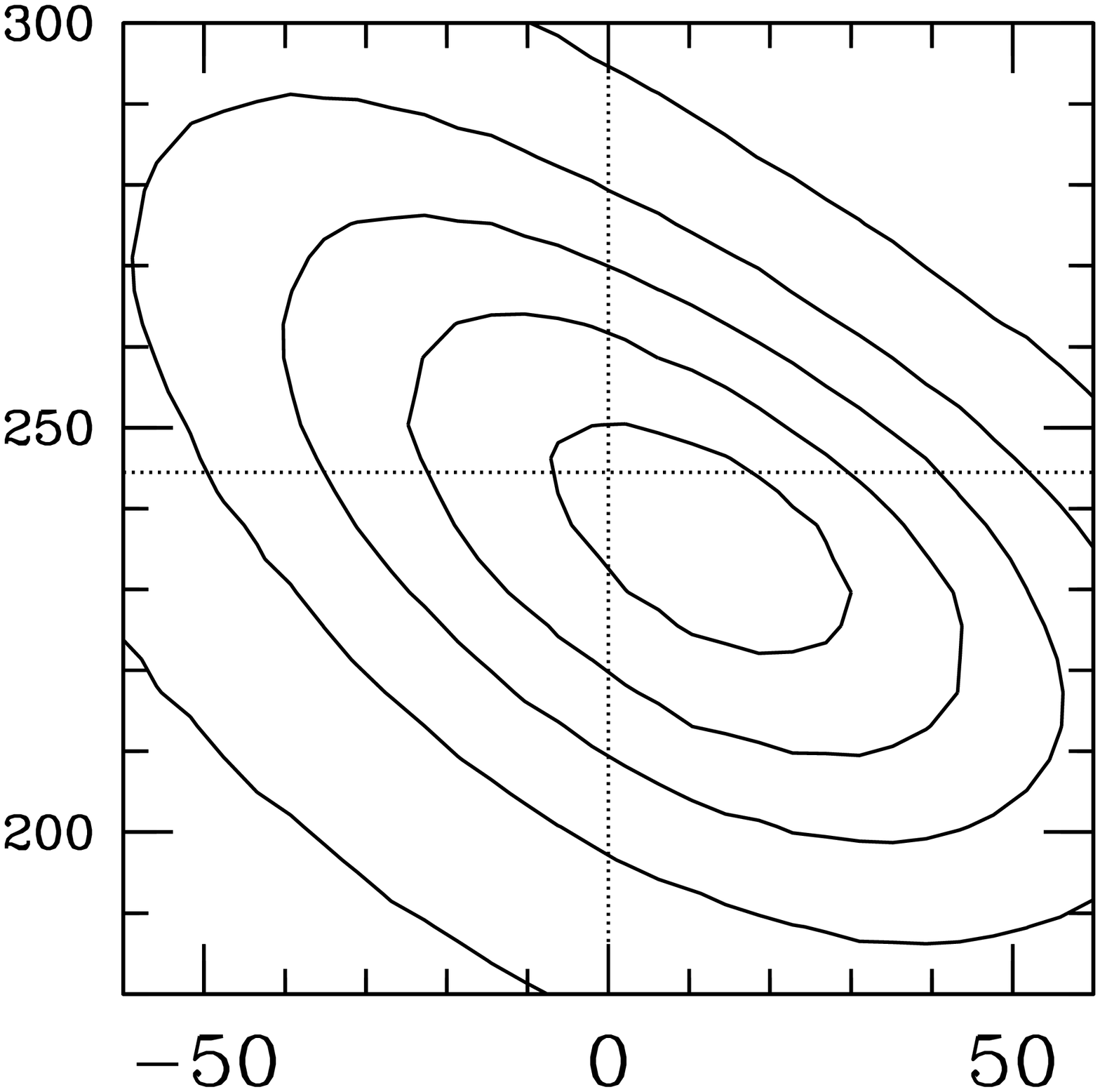}
    \includegraphics[width=.125\hsize]{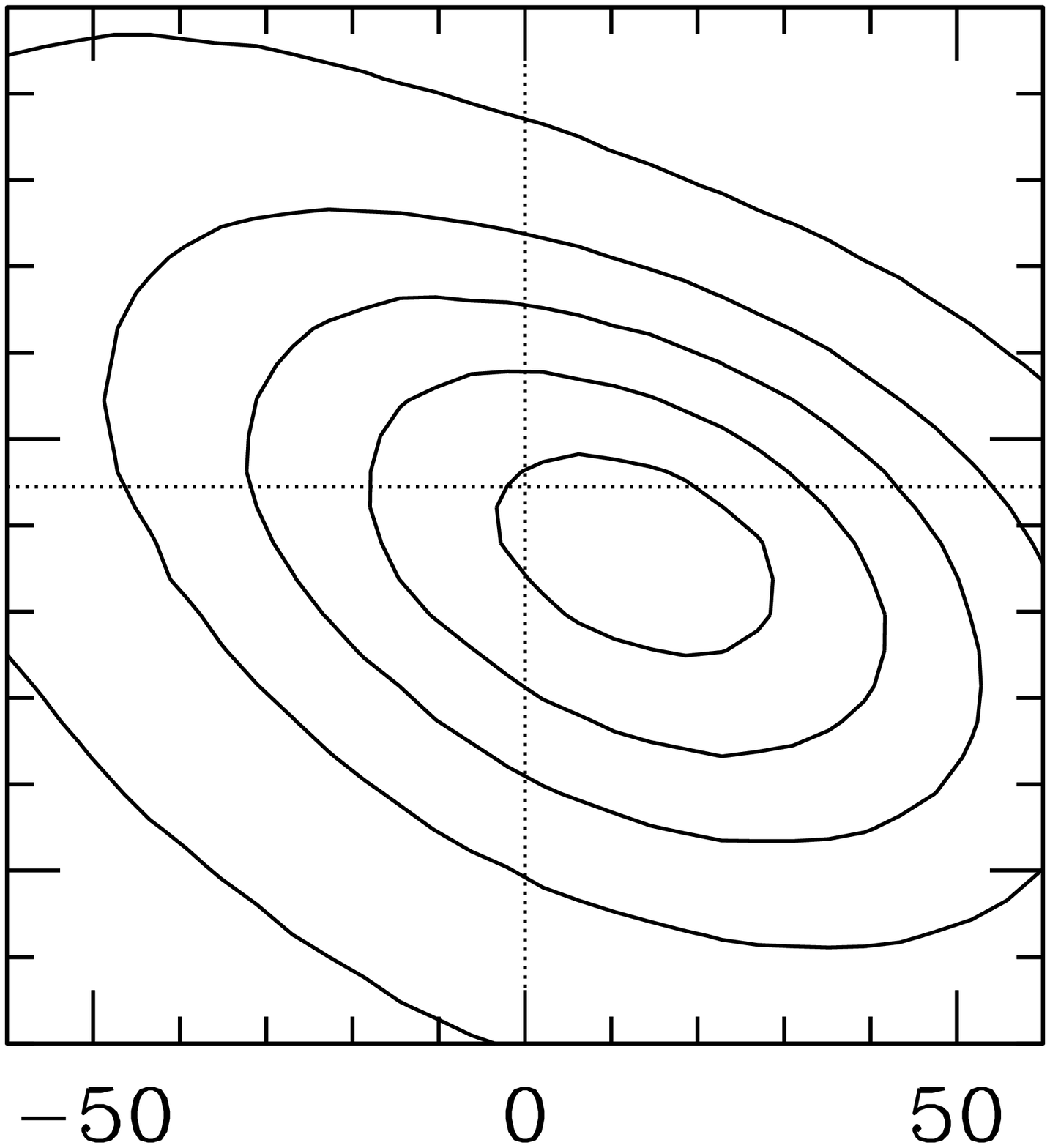}
    \includegraphics[width=.125\hsize]{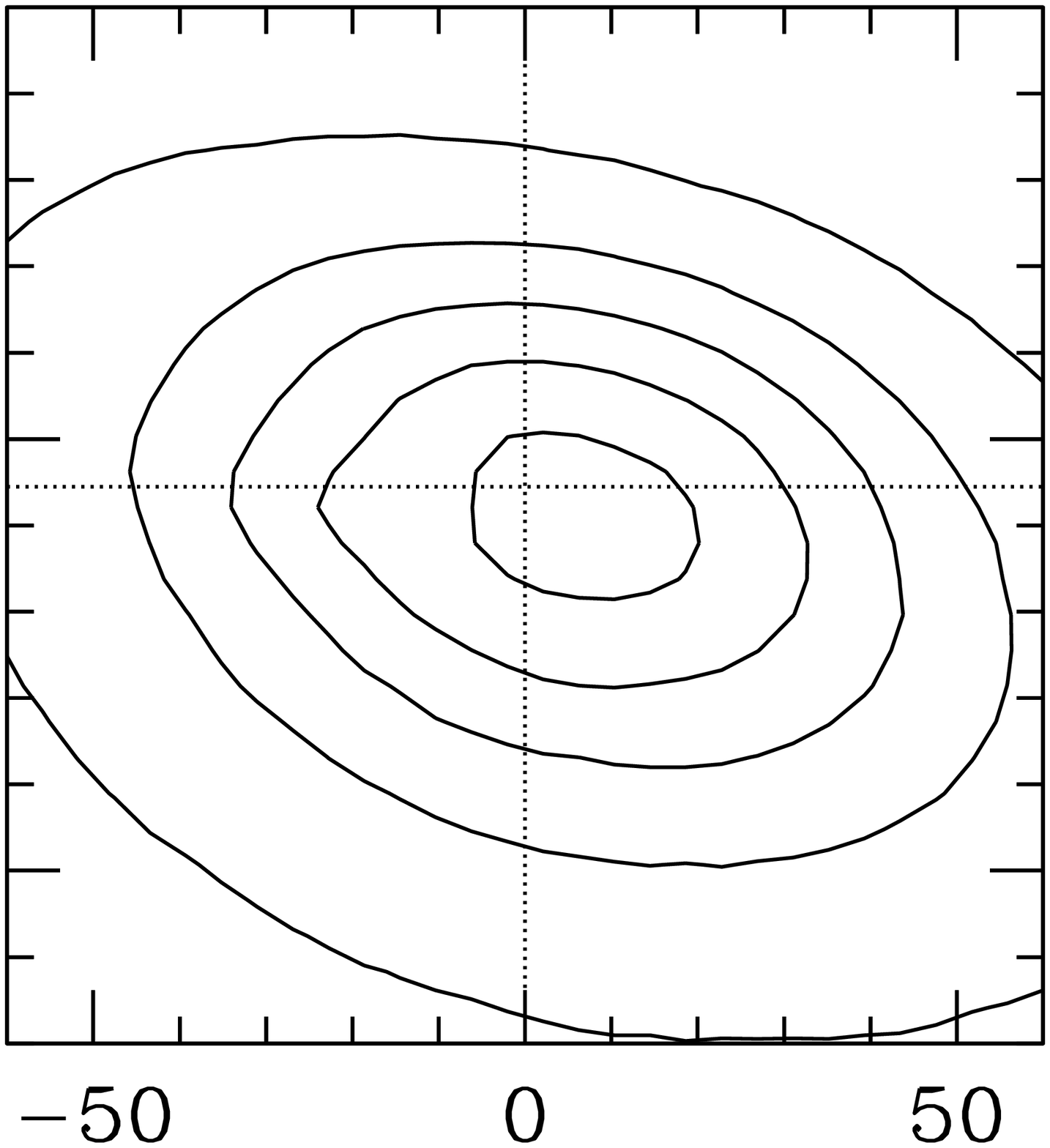}
    \includegraphics[width=.125\hsize]{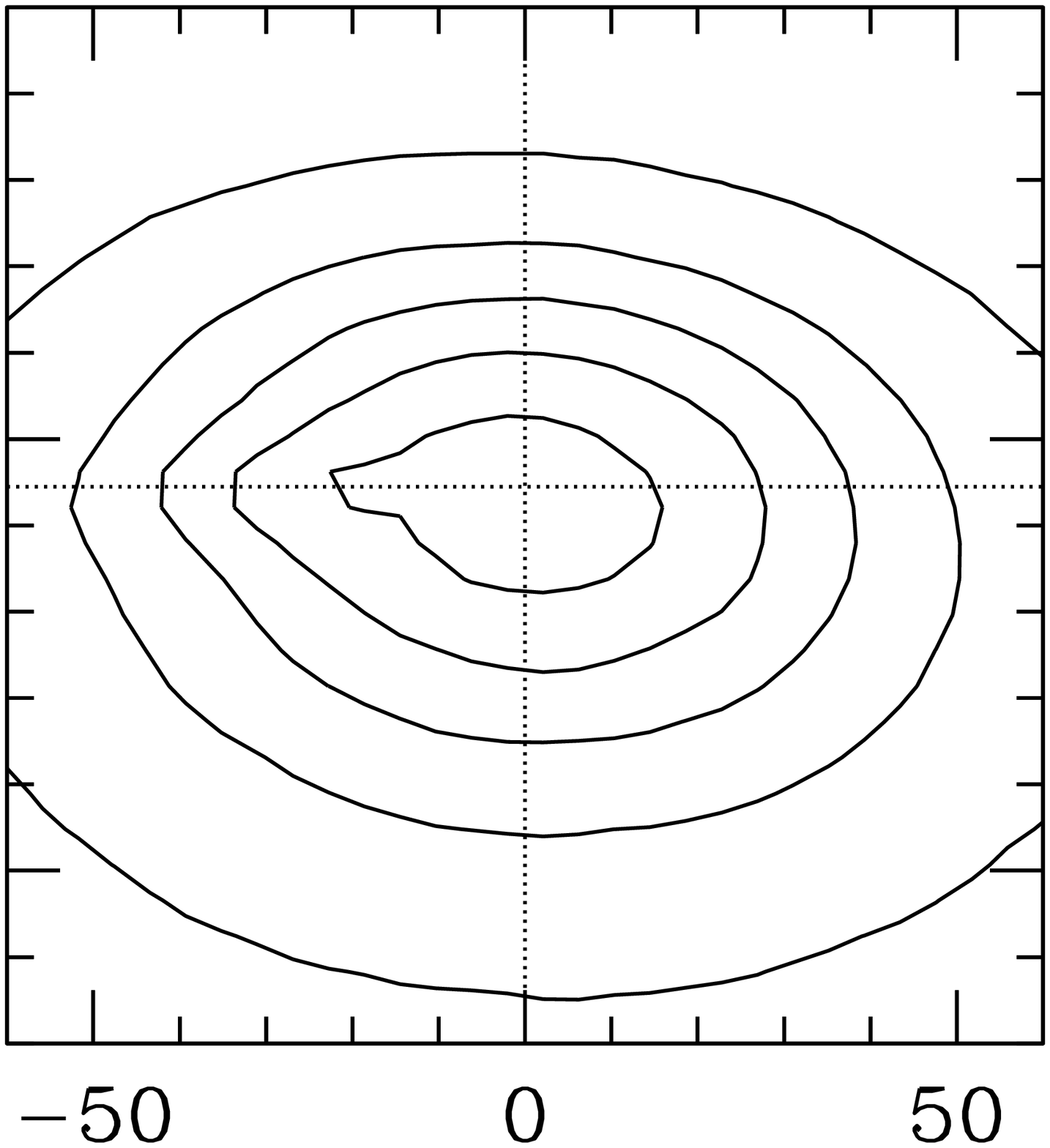}
    \includegraphics[width=.125\hsize]{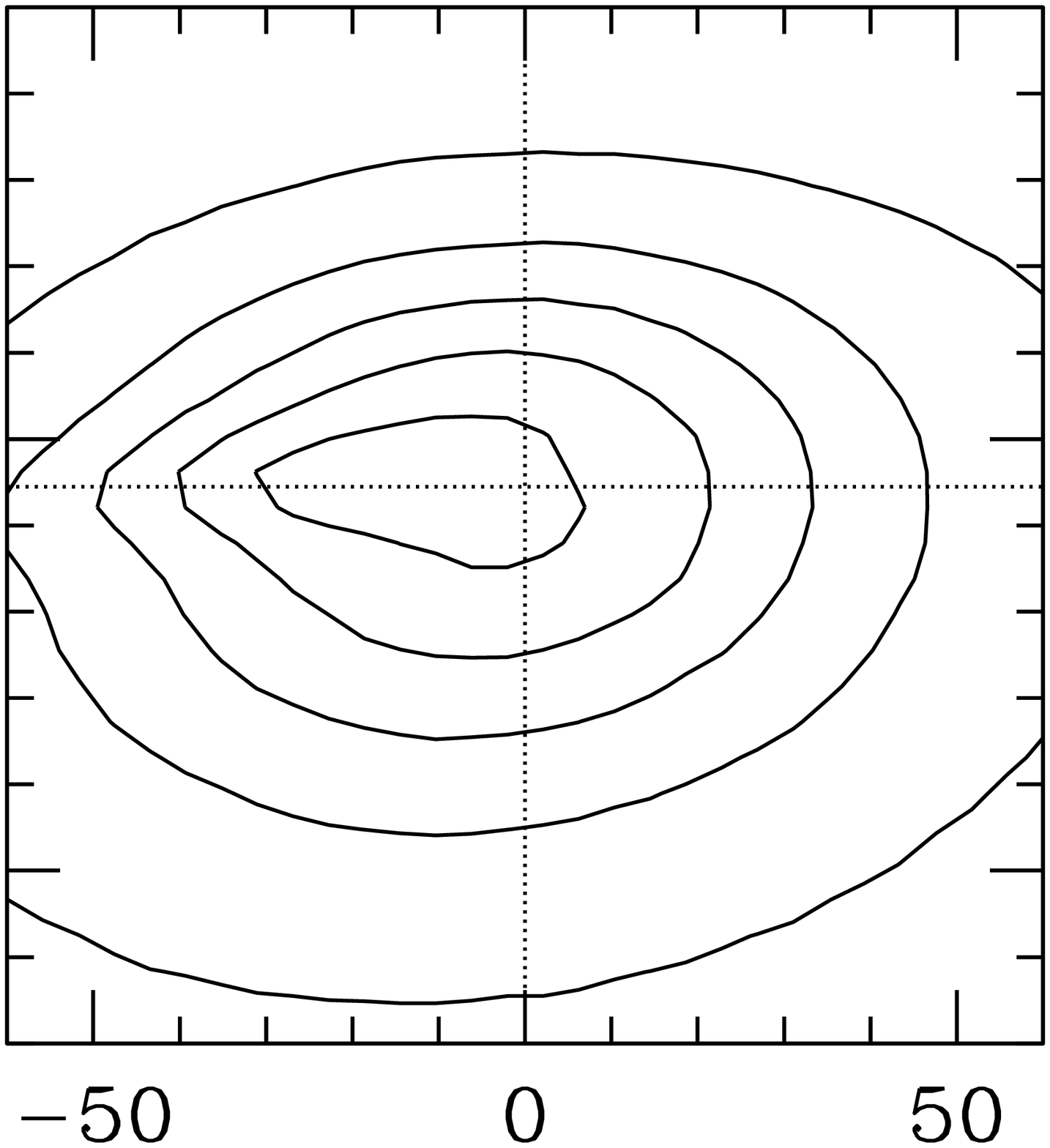}
    \includegraphics[width=.125\hsize]{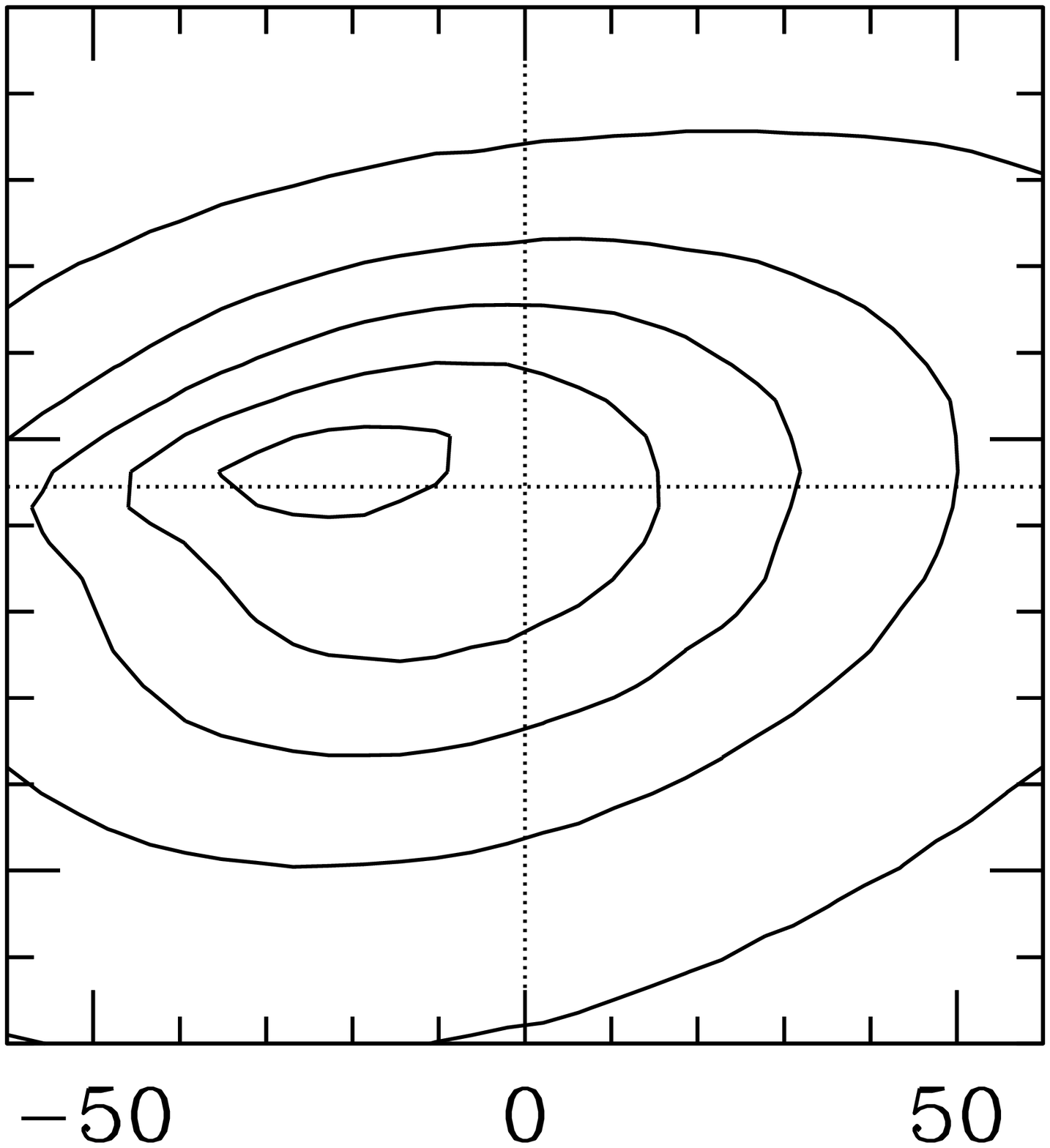}
    \includegraphics[width=.125\hsize]{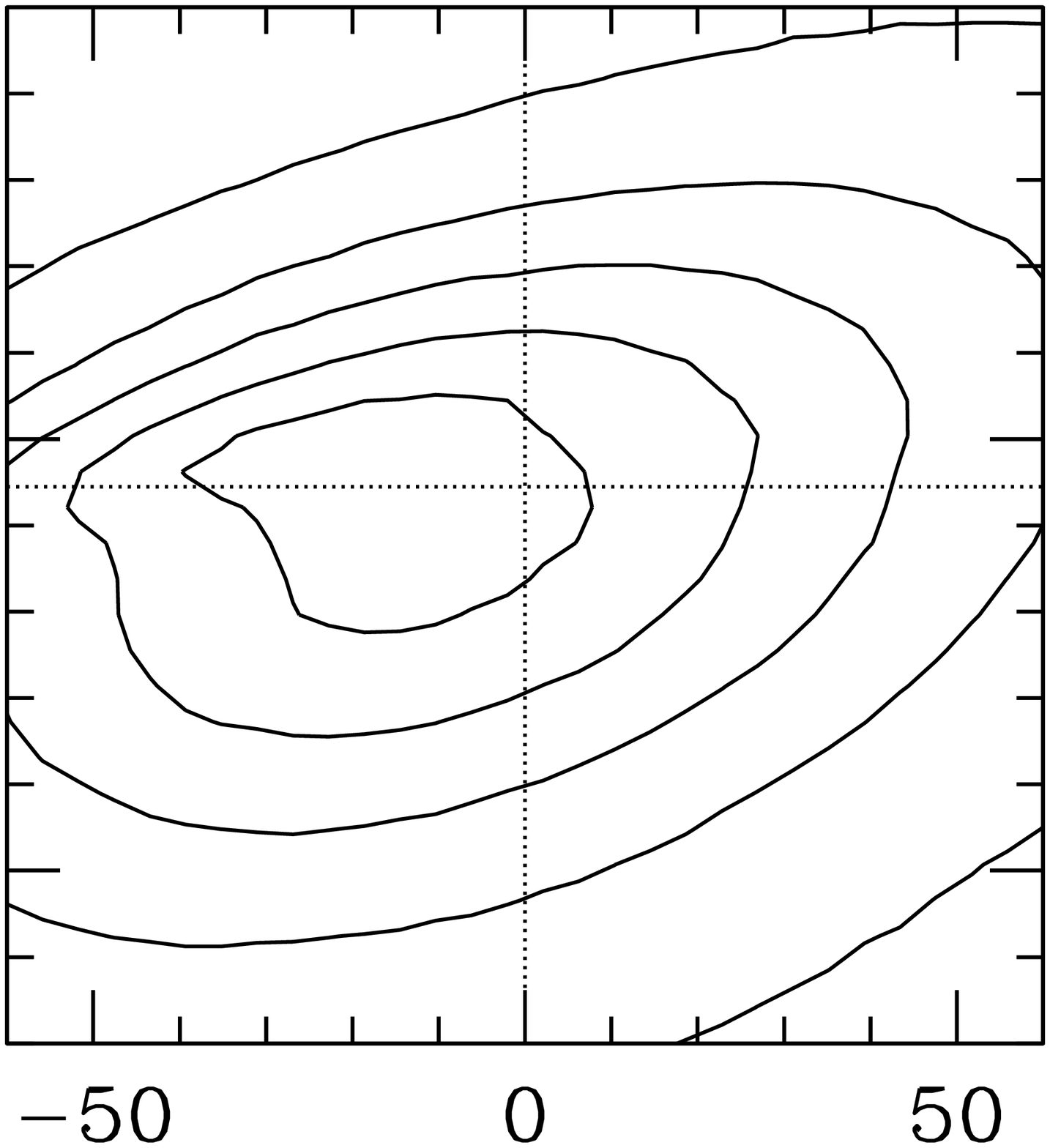}
  }
  \caption{
    Contour plots of the density in the ``observed'' $(-v_R)-v_\phi$ 
    plane of the stars in 
    the bins shown in Figure~\ref{fig:RAVEbins} for the four models,
    assuming observational uncertainties of $2\kms$ in radial
    velocity, $20$ per cent in distance, and $1\masyr$ in proper motion. 
    Again each plot covers the range $-60\kms < (-v_R) < 60\kms$, 
    $180\kms < v_\phi < 300\kms$.
\label{fig:RAVEe}
}
\end{figure*}

Isolating these signatures in real data is likely to be even harder
than Figures~\ref{fig:RAVE}\&~\ref{fig:RAVEe} 
would suggest, because there will be additional
noise due to the relatively small number of stars that will have been
observed in any given volume, and because the other stars will not be
smoothly distributed in velocity space, but will be in various
substructures, much as they are in the Solar neighbourhood. However,
sophisticated analysis techniques such as wavelet transforms
(A12 and references therein) may provide a means of finding even this
weak signal in noisy, background dominated, data. 

Currently available data, as analysed by A12, does 
show a feature consistent with any of these models at radii near 
$R_0$. It is shown in their Fig. 3, second panel (labelled $S_{R_\odot}$), 
with the feature  
labelled ``3'' as it is the third greatest overdensity in the plot. 

However, the panels showing the velocity distribution in bins at smaller 
or larger Galactocentric radii (labelled $S_{\mathrm{in}}$ and $S_{\mathrm{out}}$ 
respectively) do not show features that can be convincing associated with
any of these models. The feature labeled ``2'' in the $S_{\mathrm{in}}$ 
figure may appear to coincide with the overdensity expected from the ILR 
models shown here, but is at significantly higher $v_\phi$ 
($\sim15-25\kms$ faster than circular, as opposed to $\sim5\kms$ for the 
ILR models), and lacks the extension towards the left of the plot 
(i.e.~towards higher positive $v_R$) that is a distinctive feature of all 
of these models (and the Hyades in the Solar neighbourhood). While there 
appears to be some small overdensity at the position that the OLR models
would produce one in the $S_{\mathrm{out}}$ figure, it is so minor that it is 
unlabelled, and is of completely the wrong shape (the feature labeled ``4'' 
in this figure is at much lower $v_\phi$ than one would expect for the 
Hyades and, as noted by A12, is almost certainly the Hercules stream). 

A comparison of these models to figures showing the distribution of the
overdensities in velocity space in more closely spaced, 
overlapping bins taken from the RAVE 
data (Antoja, priv. comm.) does not provide any significant further evidence 
for either the ILR or OLR models. Given the substantial impact of 
observational errors illustrated by Figure~\ref{fig:RAVEe}, and the further 
problems associated with low number statistics and a non-smooth background, 
this cannot reasonably be taken as evidence that these models are all wrong, 
but rather as an illustration that more accurate 
data will be required to determine 
which, if any, of these models is correct.

\section{Conclusions}
In this paper we have taken a detailed look at four different models
for the Hyades, each of which accurately reproduce the signature of
the Hyades moving group in the Solar neighbourhood, but which differ
significantly from one another beyond it. The models represent stars
trapped at four different resonances, and are produced using the torus
fitting method to realize models described in angle-action variables.

We have shown that for each model we expect the Hyades to produce a
significant overdensity in velocity space around the Solar radius,
even $\sim1.5\kpc$ from the Sun in Galactic azimuth. However this
signature moves and fades with Galactocentric radius in a predictable
way that provides a clear way of determining which kind of resonance
is responsible. Most notably, if the Hyades are due to an ILR, they
tend to be found at larger Galactocentric radii ($R$) in the direction
of Galactic rotation, and smaller $R$ in the anti-rotation direction,
whereas the opposite is true if they are due to an OLR.

It is worth noting that, as shown by \cite{Se12}, there are good theoretical 
reasons to expect trapping at an ILR, and it can be seen in simulations. It 
is not obvious that one should expect to see similar at an OLR, and neither 
trapping at the OLR or the ultra-harmonic resonance is observed in 
\citeauthor{Se12}'s simulations. An additional concern is that, as noted 
by \cite{Se10}, a significant 
spiral pattern with co-rotation around $28\kpc$, as would 
be required to produce the Hyades through a 2:1 ILR, seems unlikely, which 
means that it would have to be due to some other perturbation, such as a 
rotating $m=2$ distortion of the halo. If the Hyades are the result of a 
an inner ultra-harmonic resonance with an $m=2$ perturbation (equivalent, 
in the models used in this paper, to a 4:1 ILR), 
this would be in keeping with the results of \cite{Siea12}, who 
used model with a two arm spiral to explain a gradient in mean Galactocentric 
radial velocity found in RAVE data, and the model proposed to explain some 
elements of the local velocity distribution by \cite{QuMi05}. It should be 
noted that the absence of resonant trapping at the ultra-harmonic resonance 
in the simulations of \cite{Se12} would seem to argue that if this is the 
result of resonance with a spiral with corotation at $\sim13\kpc$, the 
spiral would have to be 4-armed.

Presently available data (A12) suffers from low number statistics and
uncertain distances, which leads to a significant uncertainty in the
transverse velocity of the stars. This means we are not yet able to provide 
any significant extra insight as to which resonance is associated with the 
Hyades using observations from beyond the Solar neighbourhood.

With improved distance measurements and more observed stars,
this situation will improve, and these easily tuneable models provided
by the angle-action description of the phenomena will provide great
insight into the nature of the resonance and, therefore, the
perturbation to the Galaxy that is responsible.

\section*{Acknowledgments}
We're grateful to James Binney for a careful reading of a draft of this
paper, and to Teresa Antoja for providing detailed figures of the
velocity substructure found in the RAVE data. We thank the 
anonymous referee who provided a number of helpful suggestions and 
noticed one potentially terrible typographical error. This work is supported
by a grant from the Science and Technology Facilities Council.

\bibliographystyle{mn2e} \bibliography{new_refs}


\end{document}